\documentclass[multphys,vecphys]{svmult}

\usepackage{makeidx}         
\usepackage{graphicx}        
\usepackage{exscale,relsize}
\usepackage{epsf,color,graphics}
\usepackage{colordvi}
\usepackage{multicol}        
\usepackage[bottom]{footmisc}
\usepackage{amsmath}
\usepackage{amssymb}
\usepackage{txfonts,captcont,subfigure}
\usepackage{rotating}
\usepackage{natbib}      

\usepackage{natbib} 
\usepackage[below]{placeins}
\usepackage{url}

\def\etal{{\it et al}.}

\def\textindent#1{\indent{#1\enspace}\ignorespaces}
\def\itemitem{\par\indent \hangindent2\parindent \textindent}

\def\d3k{{\displaystyle {\rm d}{\bmit k} \over \displaystyle (2\pi)^3}}
\def\d3x{{d{\bf x}}}

\newcommand{\pscz} {PSC$z$}

\newcommand{\cgal}{\texttt{CGAL}\ }

\def\d3k{{\displaystyle {\rm d}{\bf k} \over \displaystyle (2\pi)^3}}
\def\hmpc{h^{-1} {\rm Mpc}}

\def\kms{\ {\rm km~s^{-1} }}

\font \bigtwo=cmbx10 scaled\magstep2
\font \bigthr=cmbx10 scaled\magstep3

\makeindex             

\begin{document}

\title*{\bigthr The Cosmic Web: Geometric Analysis}
\titlerunning{Geometric Analysis Cosmic Web} 
\author{Rien van de Weygaert \& Willem Schaap}
\institute{Kapteyn Astronomical Institute, University of Groningen,\\P.O. Box 800, 9700 AV Groningen,\\the Netherlands\\ 
\texttt{weygaert@astro.rug.nl}}
%
%
\maketitle

\begin{figure*}
\begin{center}
\vskip 0.0truecm
\mbox{\hskip 0.0truecm\includegraphics[height=14.0cm,angle=270.0]{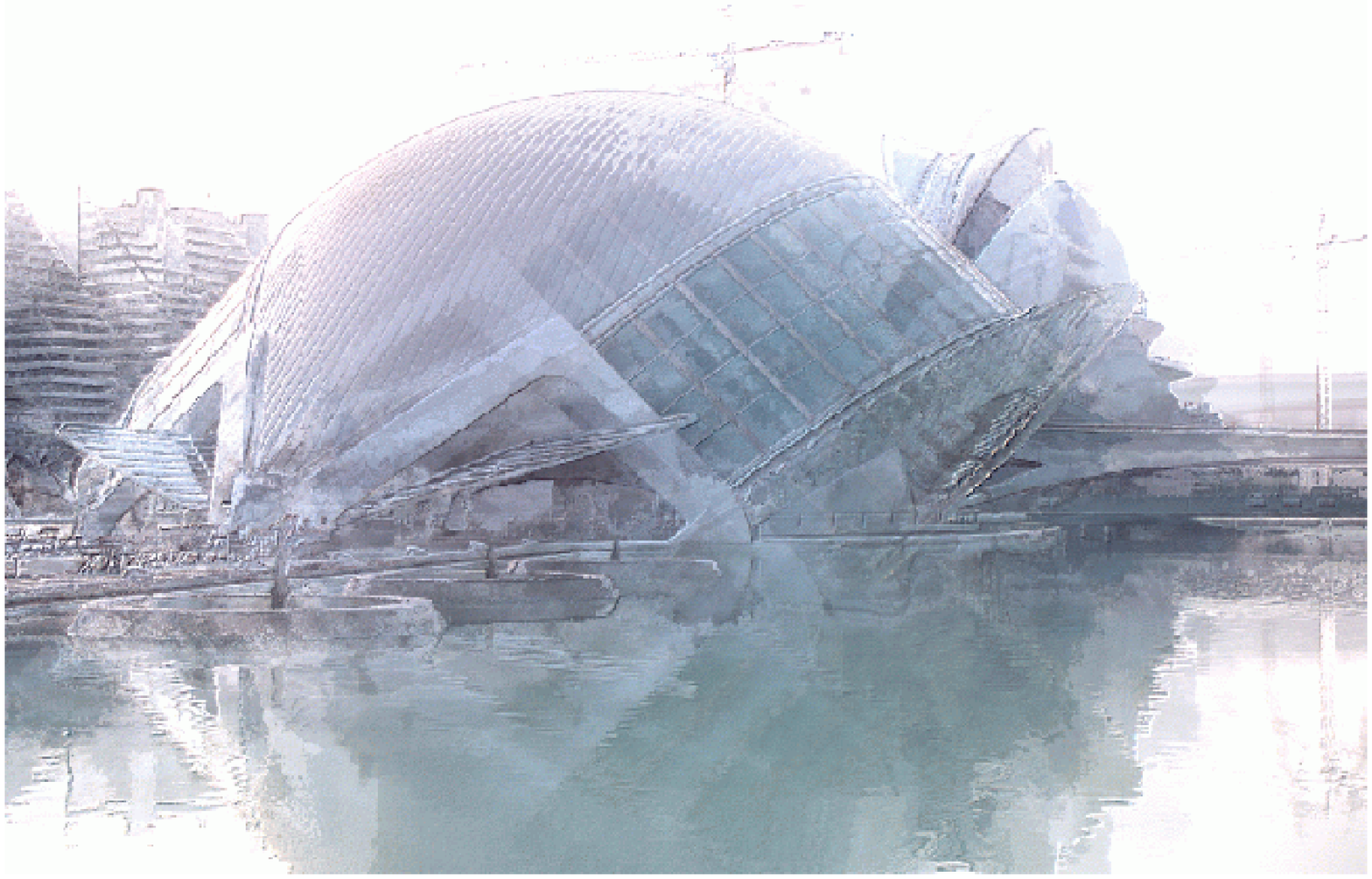}}
\vskip 0.0truecm
\label{fig:dtfepanel}
\end{center}
\end{figure*}

\vskip 0.5truecm
\begin{center}
\bigtwo Lectures at Summerschool\\
\ \\
``Data Analysis in Cosmology''\\
\ \\
Valencia, Spain (September 2004)
\end{center}

\vfill\eject

\section{Outline: Patterns in the Cosmic Web}
\noindent The spatial cosmic matter distribution on scales of a few up to more than a hundred 
Megaparsec displays a salient and pervasive foamlike pattern. Revealed through the painstaking efforts 
of redshift survey campaigns, it has completely revised our view of the matter distribution on these 
cosmological scales. The weblike spatial arrangement of galaxies and mass into elongated filaments, 
sheetlike walls and dense compact clusters, the existence of large near-empty void regions and the hierarchical nature 
of this mass distribution -- marked by substructure over a wide range of scales and densities -- are three 
major characteristics of we have come to know as the {\it Cosmic Web}. 

\begin{figure*}
\begin{center}
\vskip -0.5truecm
\mbox{\hskip -0.5truecm\includegraphics[width=20.0cm,height=13.0cm,angle=90.0]{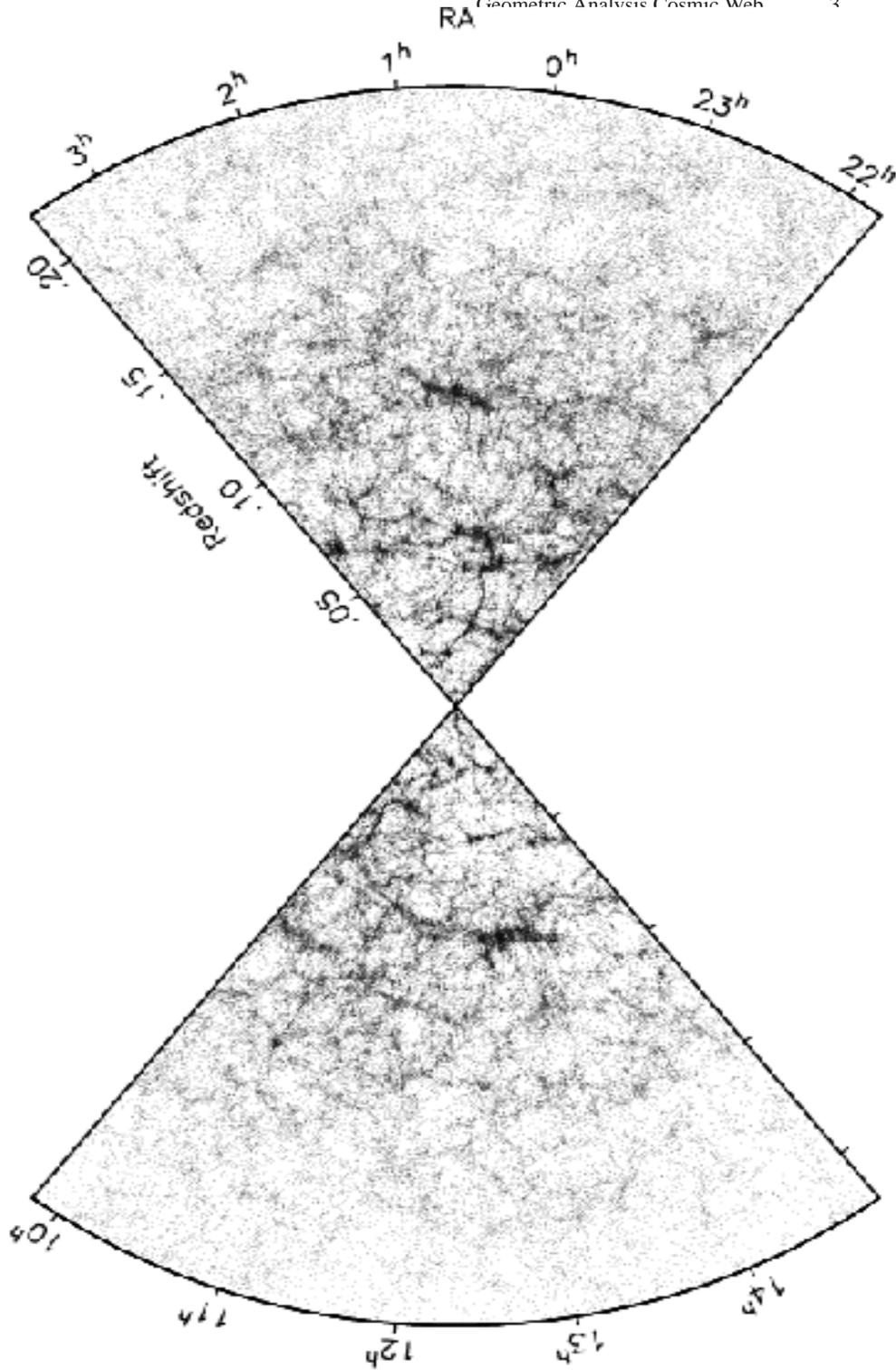}}
\caption{The galaxy distribution uncovered by the 2dF GALAXY REDSHIFT SURVEY. Depicted are the 
positions of 221414 galaxies in the final 2dFGRS catalogue. Clearly visible is the foamlike 
geometry of walls, filaments and massive compact clusters surrounding near-empty void regions. 
Image courtesy of M. Colless; see also Colless et al. (2003) and http://www2.aao.gov.au/2dFGRS/.}
\label{fig:2dfgaldist}
\end{center}
\end{figure*}
The vast Megaparsec cosmic web is undoubtedly one of the most striking examples of complex geometric patterns 
found in nature, and the largest in terms of sheer size. In a great many physical systems, the spatial organization 
of matter is one of the most readily observable manifestations of the forces and processes forming and moulding them. 
Richly structured morphologies are usually the consequence of the complex and nonlinear collective action of basic 
physical processes. Their rich morphology is therefore a rich source of information on the combination of physical forces at work and 
the conditions from which the systems evolved. In many branches of science the study of geometric patterns 
has therefore  developed into a major industry for exploring and uncovering the underlying physics 
\citep[see e.g.][]{balhaw1998}.

Computer simulations suggest that the observed cellular patterns are a prominent and natural 
aspect of cosmic structure formation through gravitational instability \citep{peebles80}, 
the standard paradigm for the emergence of structure in our Universe. Structure in the Universe is the result 
of the gravitational growth of tiny density perturbations and the accompanying tiny velocity perturbations in 
the primordial Universe. Supported by an impressive body of evidence, primarily those of temperature fluctuations 
in the cosmic microwave background \citep{smoot1992,bennett2003,spergel2006}, the character of the primordial random density 
and velocity perturbation field is that of a {\it homogeneous and isotropic spatial Gaussian process}. Such 
fields of primordial Gaussian perturbations in the gravitational potential are a natural product of an early 
inflationary phase of our Universe. 

The early linear phase of pure Gaussian density and velocity perturbations has been understood in great depth. 
This knowledge has been exploited extensively in extracting a truely impressive score of global cosmological 
parameters. Notwithstanding these successes, the more advanced phases of cosmic structure formation are still 
in need of substantially better understanding. Mildly nonlinear structures do contain a wealth of information 
on the emergence of cosmic structure at a stage features start to emerge as individually recognizable objects. 
The anisotropic filamentary and planar structures, the characteristic large underdense void regions and the 
hierarchical clustering of matter marking the weblike spatial geometry of the Megaparsec matter distribution 
are typical manifestations of mildly advanced gravitational structure formation. The existence of the intriguing 
foamlike patterns representative for this early nonlinear phase of evolution got revealed by major 
campaigns to map the galaxy distribution on Megaparsec scales revealed while ever larger computer N-body 
simulations demonstrated that such matter distributions are indeed typical manifestations of the gravitational 
clustering process. Nonetheless, despite the enormous progres true insight and physical understanding has 
remained limited. The lack of readily accessible symmetries and the strong nonlocal 
influences are a major impediment towards the development relevant analytical descriptions. The 
hierarchical nature of the gravitational clustering process forms an additional complication. 
While small structures materialize before they merge into large entities, each cosmic structure 
consists of various levels of substructure so that instead of readily recognizing one characteristic 
spatial scale we need to take into account a range of scales. Insight into the complex interplay of 
emerging structures throughout the Universe and at a range of spatial scales has been provided 
through a variety of analytical approximations. Computer simulations have provided us with a  
good impression of the complexities of the emerging matter distribution, but for the analysis 
of the resulting patterns and hierarchical substructure the toolbox of descriptive measures 
is still largely heuristic, ad-hoc and often biased in character.

\begin{figure*}
\begin{center}
\vskip 0.5truecm
\mbox{\hskip -1.0truecm\includegraphics[height=18.0cm]{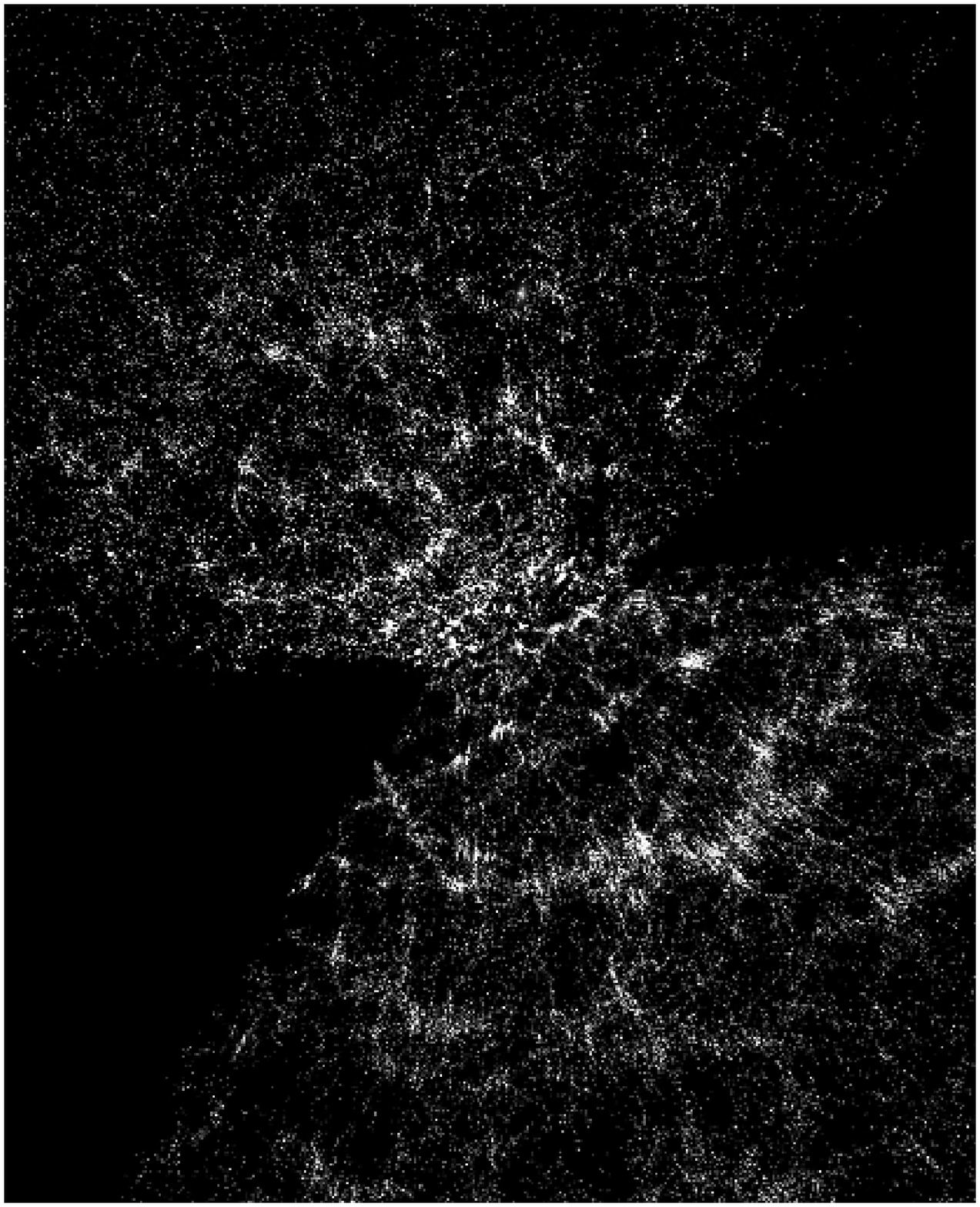}}
\end{center}
\end{figure*}
\begin{figure*}
\begin{center}
\vskip 0.0truecm
\caption{SDSS is the largest and most systematic sky survey in the history of astronomy. It is a 
combination of a sky survey in 5 optical bands of 25\% of the celestial (northern) sphere. Each 
image is recorded on CCDs in these 5 bands. On the basis of the images/colours and their brightness 
a million galaxies are subsequently selected for spectroscopic follow-up. The total sky area covered 
by SDSS is 8452 square degrees. Objects will be recorded to $m_{lim}=23.1$. In total the resulting 
atlas will contain 10$^8$ stars, 10$^8$ galaxies and 10$^5$ quasars. Spectra are taken of around 10$^6$ 
galaxies, 10$^5$ quasars and 10$^5$ unusual stars (in our Galaxy). Of the 5 public data releases 4 have been 
accomplished, ie. 6670 square degrees of images is publicly available, along with 806,400 spectra. 
In total, the sky survey is now completely done (107\%), the spectroscopic survey for 68\%. This image 
is taken from a movie made by Subbarao, Surendran \& Landsberg (see website: 
http://astro.uchicago.edu/cosmus/projects/sloangalaxies/). It depicts the resulting redshift 
distribution after the 3rd public data release. It concerns 5282 square degrees and contained 
528,640 spectra, of which 374,767 galaxies.}
\end{center}
\label{fig:sdssgaldist}
\end{figure*}

While cosmological theories are describing the development of structure in terms of 
continuous density and velocity fields, our knowledge stems mainly from discrete 
samplings of these fields. In the observational reality galaxies are the main tracers of the 
cosmic web and it is through measuring the redshift distribution of galaxies that we have 
been able to map its structure. Likewise, simulations of the evolving cosmic matter 
distribution are almost exclusively based upon N-body particle computer calculation, 
involving a discrete representation of the features we seek to study. 

Both the galaxy distribution as well as the particles in an N-body simulation are 
examples of spatial point processes in that they are (1) {\it discretely sampled}  
(2) have a {\it irregular spatial distribution}. The translation of {\it discretely sampled} and 
{\it spatially irregularly distributed} sampled objects into the related continuous fields is not 
necessarily a trivial procedure. The standard procedure is to use a filter to process the discrete 
samples into a representative reconstruction of the underlying continuous field. It is the design 
of the filter which determines the character of the reconstruction. 

Astronomical applications are usually based upon a set of user-defined filter functions. 
Nearly without exception the definition of these include pre-conceived knowledge about the features 
one is looking for. A telling example is the use of a Gaussian filter. This filter will 
suppress the presence of any structures on a scale smaller than the characteristic filter 
scale. Moreover, nearly always it is a spherically defined filter which tends to smooth out 
any existing anisotropies. Such procedures may be justified in situations in which 
we are particularly interested in objects of that size or in which physical 
understanding suggests the smoothing scale to be of particular significance. 
On the other hand, they may be crucially inept in situations of which we do not know 
in advance the properties of the matter distribution. The gravitational clustering process in 
the case of hierarchical cosmic structure formation scenarios is a particularly notorious 
case. As it includes structures over a vast range of scales and displays a rich palet 
of geometries and patterns any filter design tends to involve a discrimination against 
one or more -- and possibly interesting -- characteristics of the cosmic matter 
distribution it would be preferrable to define filter and reconstruction procedures 
that tend to be defined by the discrete point process itself. 

A variety of procedures that seek to define and employ more {\bf ``natural''} filters 
have been put forward in recent years. The scale of smoothing kernels can be adapted 
to the particle number density, yielding a density field that retains to a large extent the 
spatial information of the sampled density field. While such procedures may still have the 
disadvantage of a rigid user-defined filter function and filter geometry, a more sophisticated, 
versatile and particularly promising class of functions is that of wavelet-defined filters (see 
contribution B.J.T.~Jones). These can be used to locate contributions on a particular scale, and 
even to trace features of a given geometry. While its successes have been quite remarkable, the 
success of the application is still dependent on the particular class of employed wavelets. 

In this contribution we will describe in extensio the technique of the Delaunay Tessellation Field 
Estimator. DTFE is based upon the use of the Voronoi and Delaunay tessellations of a given spatial 
point distribution to form the basis of a natural, fully self-adaptive filter for 
discretely sampled fields in which the Delaunay tessellations are used as multidimensional 
interpolation intervals. The method has been defined, introduced and developed by \cite{schaapwey2000}
and forms an elaboration of the velocity interpolation scheme introduced by \cite{bernwey96}. 

Our focus on DTFE will go along with a concentration on the potential of spatial {\bf tessellations} as 
a means of estimating and interpolating discrete point samples into continuous field 
reconstructions. In particular we will concentrate on the potential of {\it Voronoi} 
and {\it Delaunay tessellations}. Both tessellations -- each others {\it dual} -- are fundamental 
concepts in the field of stochastic geometry. Exploiting the characteristics of these tessellations, 
we will show that the DTFE technique is capable of 
delineating the hierarchical and anisotropic nature of spatial point distributions and in outlining the presence 
and shape of voidlike regions. The spatial structure of the 
cosmic matter distribution is marked by precisely these , and precisely this potential has been 
the incentive for analyzing cosmic large scale structure. DTFE exploits three particular properties of 
Voronoi and Delaunay tessellations. The tessellations are very sensitive to the local point density, in 
that the volume of the tessellation cells is a strong function of the local (physical) 
density. The DTFE method uses this fact to define a local estimate of the density. Equally important is their 
sensitivity to the local geometry of the point distribution. This allows them to trace anisotropic 
features such as encountered in the cosmic web. Finally it uses the 
adaptive and minimum triangulation properties of Delaunay tessellations to use them as adaptive spatial 
interpolation intervals for irregular point distributions. In this it is the first order version of the 
{\it Natural Neighbour method} (NN method). The theoretical basis for the NN method, a generic smooth and 
local higher order spatial interpolation technique developed by experts in the field of computational 
geometry, has been worked out in great detail by \cite{sibson1980,sibson1981} and \cite{watson1992}.
As has been demonstrated by telling examples in geophysics \citep{braunsambridge1995} and solid mechanics 
\citep{sukumarphd1998} NN methods hold tremendous potential for grid-independent analysis and computations. 

Following the definition of the DTFE technique, we will present a systematic treatment of various virtues 
of relevance to the cosmic matter distribution. The related performance of DTFE will be illustrated by means 
of its success in analyzing computer simulations of cosmic structure formation as well as that of the galaxy 
distribution in large-scale redshift surveys such as the 2dFGRS and SDSS surveys. Following the 
definition and properties of DTFE, we will pay attention to extensions of the project. Higher order 
Natural Neighbour renditions of density and velocity fields involve improvements in terms of 
smoothness and discarding artefacts of the underlying tessellations. 

Following the determination of the ``raw'' DTFE produces density and velocity fields -- and/or other sampled 
fields, e.g. temperature values in SPH simulations -- the true potential of DTFE is realized in the subsequent 
stage in which the resulting DTFE fields get processed and analyzed. We will discuss an array of  
techniques which have been developed to extract information on aspects of the Megaparsec matter distribution. 
Straightforward processing involves simple filtering and the production of images from the reconstructed 
field. Also, we will shortly discuss the use of the tessellation fields towards defining new measures for 
the topology and geometry of the underlying matter distribution. The determination of the Minkowski 
functionals of isodensity surfaces, following the SURFGEN formalism \citep{sahni1998,jsheth2003,shandarin2004}, can be 
greatly facilitated on the basis of the Delaunay triangulation itself and the estimated DTFE density field. 
Extending the topological information contained in the Minkowski functionals leads us to the concept 
of {\it Betti} numbers and $\alpha${\it -shapes}, a filtration of the Delaunay complex of a dataset 
\citep{edelsbrunner1983,edelsbrunner1994,edelsbrunner2002}. With the associated persistence diagrams the $\alpha$-shapes encode 
the evolution of the Betti numbers. We have started to add to the arsenal of tools to quantify the patterns in the 
Megaparsec matter distribution \citep{vegter2007,eldering2006}. Particularly interesting are the recently developed elaborate and 
advanced techniques of {\it watershed void identification} \citep{platen2007} and 
{\it Multiscale Morphology Filter} \citep{aragonphd2007,aragonmmf2007}. These methods enable the unbiased identification 
and measurement of voids, walls, filaments and clusters in the galaxy distribution. 

Preceding the discussion on the DTFE and related tessellation techniques, we will first have 
to describe the structure, dynamics and formation of the {\it cosmic web}. The complex 
structure of the cosmic web, and the potential information it does contain, has been 
the ultimate reason behind the development of the DTFE. 

\section{Introduction: The Cosmic Web}
\label{sec:1}
\vskip -0.25truecm
Macroscopic patterns in nature are often due the collective action of basic, often even simple, 
physical processes. These may yield a surprising array of complex and genuinely unique physical 
manifestations. The macroscopic organization into complex spatial patterns is one of the most striking. 
The rich morphology of such systems and patterns represents a major source of information on the underlying 
physics. This has made them the subject of a major and promising area of inquiry.\\
\begin{figure*}
\begin{center}
\mbox{\hskip -0.5truecm\includegraphics[height=13.cm,angle=270.0]{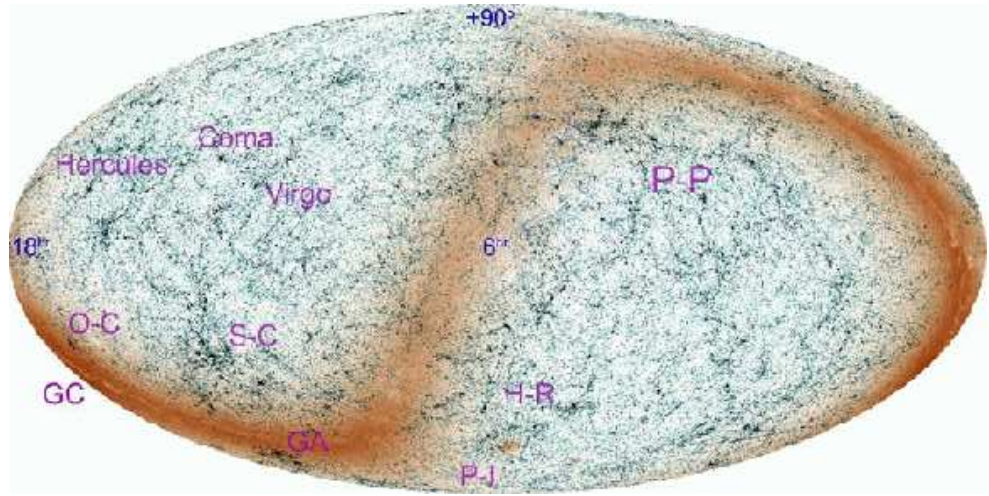}}
\vskip 0.25truecm
\caption{Equatorial view of the 2MASS galaxy catalog (6h RA at centre). The grey-scale represents the total 
integrated flux along the line of sight -- the nearest (and therefore brightest) galaxies produce a vivid contrast 
between the Local Supercluster (centre-left) and the more distant cosmic web. The dark band of the Milky Way clearly 
demonstrates where the galaxy catalog becomes incomplete due to source confusion. Some well known large-scale structures 
are indicated: P-P=Perseus-Pisces supercluster; H-R=Horologium-Reticulum supercluster; P-I=Pavo-Indus supercluster; 
GA=`Great Attractor'; GC=Galactic Centre; S-C=Shapley Concentration; O-C=Ophiuchus Cluster; Virgo, Coma, and 
Hercules=Virgo,Coma and Hercules superclusters. The Galactic `anti-centre' is front and centre, with the Orion and 
Taurus Giant Molecular Clouds forming the dark circular band near the centre. Image courtesy of J.H. Jarrett. 
Reproduced with permission from the Publications of the Astronomical Society of Australia 21(4): 396-403 (T.H. Jarrett). 
Copyright Astronomical Society of Australia 2004. Published by CSIRO PUBLISHING, Melbourne Australia.}
\label{fig:2massgal}
\end{center}
\end{figure*}
\subsection{Galaxies and the Cosmic Web}
\noindent One of the most striking examples of a physical system displaying a salient geometrical morphology, 
and the largest in terms of sheer size, is the Universe as a whole. The past few decades 
have revealed  that on scales of a few up to more than a hundred Megaparsec, the galaxies conglomerate 
into intriguing cellular or foamlike patterns that pervade throughout the observable cosmos. 
An initial hint of this cosmic web was seen in the view of the local Universe offered by the first CfA redshift slice 
\citep{lapparent1986}. In recent years this view has been expanded dramatically to the present grand 
vistas offered by the 100,000s of galaxies in the 2dF -- two-degree field -- Galaxy Redshift Survey, the 2dFGRS 
\citep{colless2003} and SDSS \citep[e.g.][]{tegmark2004} galaxy redshift surveys.
\footnote{See {\tt http://www.mso.anu.edu.au/2dFGRS/} and {\tt http://www.sdss.org/}}. 
\begin{figure*}
\begin{center}
\vskip -0.25truecm
\mbox{\hskip -0.5truecm\includegraphics[width=13.0cm]{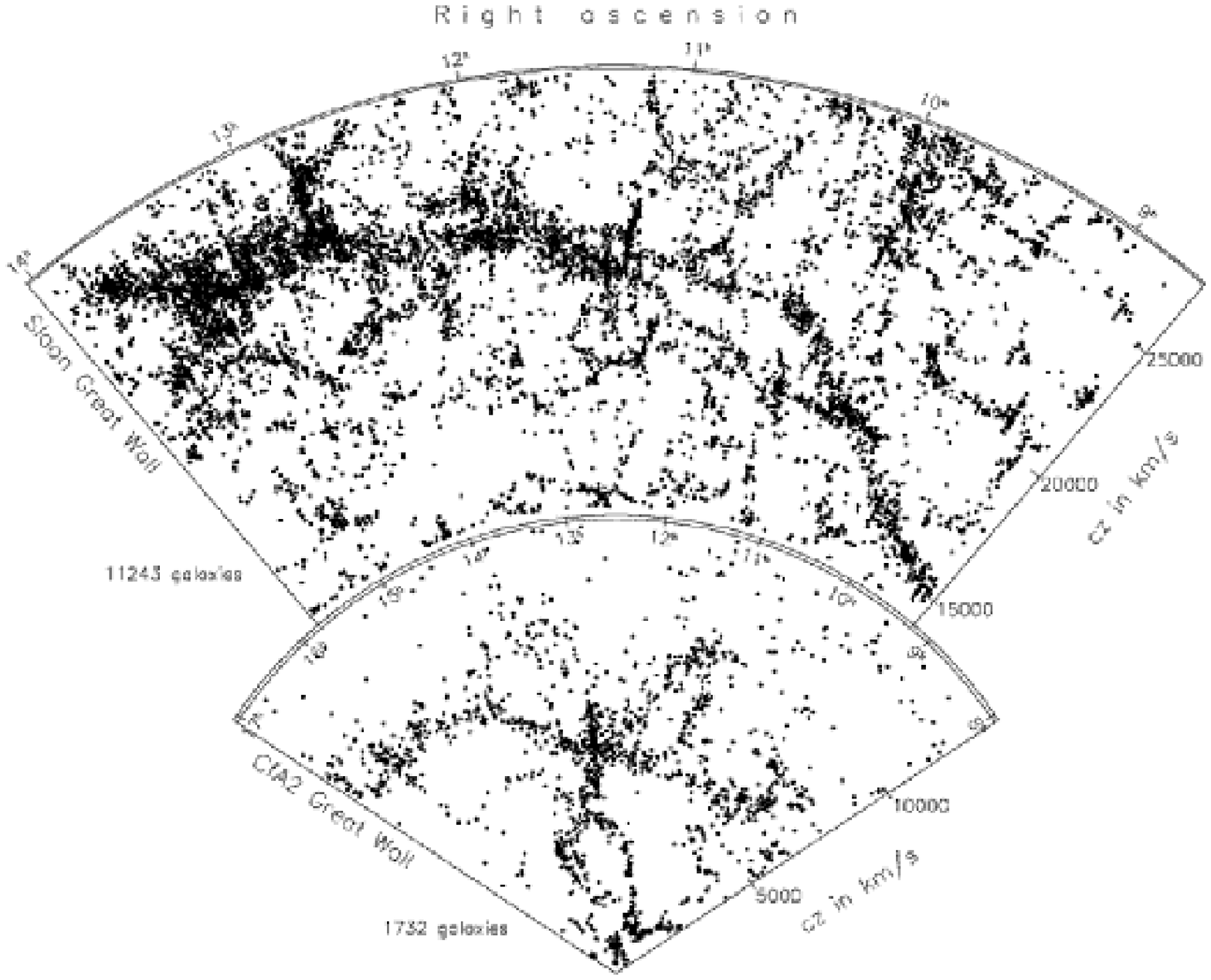}}
\caption{The CfA Great Wall (bottom slice, Geller \& Huchra 1989) compared with the Sloan Great Wall 
(top slice). Both structures represent the largest coherent structural in the galaxy redshift surveys 
in which they were detected, the CfA redshift survey and the SDSS redshift survey. The (CfA) 
Great Wall is a huge planar concentration of galaxies with dimensions that are estimated to be of 
the order of $60h^{-1} \times 170h^{-1} \times 5h^{-1}$ Mpc. Truely mindboggling is the Sloan Great Wall, 
a huge conglomerate of clusters and galaxies. With a size in the order of $400\hmpc$ it is at least three times 
larger than the CfA Great Wall. It remains to be seen whether it is a genuine physical structure or mainly a 
stochastic arrangement and enhancement, at a distance coinciding with the survey's maximum in the radial selection 
function. Image courtesy of M. Juri\'c, see also Gott et al. 2005. Reproduced by permission of the AAS.}
\vskip -0.5truecm
\label{fig:greatwall}
\end{center}
\end{figure*}
Galaxies are found in dense, compact clusters, in less dense filaments, and in sheetlike walls 
which surround vast, almost empty regions called {\it voids}.  This is most dramatically illustrated by the 
2dFGRS and SDSS maps. The published maps of the distribution of nearly 250,000 galaxies in two narrow ``slice'' 
regions on the sky yielded by the 2dFGRS surveys reveal a far from homogeneous distribution (fig.~\ref{fig:2dfgaldist}). 
Instead, we recognize a sponge-like arrangement, with galaxies aggregating in 
striking geometric patterns such as prominent filaments, vaguely detectable walls and dense compact clusters 
on the periphery of giant voids\footnote{It is important to realize that the interpretation of the 
Megaparsec galaxy distribution is based upon the tacit yet common assumption that it forms a a fair 
reflection of the underlying matter distribution. While there are various indications that 
this is indeed a reasonable approximation, as long as the intricate and complex 
process of the formation of galaxies has not been properly understood this should be 
considered as a plausible yet heuristic working hypothesis.}. The three-dimensional view emerging from the SDSS 
redshift survey provides an even more convincing image of the intricate patterns defined by the cosmic 
web (fig.~\ref{fig:sdssgaldist}). A careful assessment of the galaxy distribution in our immediate vicintiy reveals us how 
we ourselves are embedded and surrounded by beautifully delineated and surprisingly sharply defined weblike structures. In 
particular the all-sky nearby infrared 2MASS survey (see fig.~\ref{fig:2massgal}) provides us with a meticulously 
clear view of the web surrounding us \citep{jarr2004}. 
\begin{figure*}
\begin{center}
\vskip -0.5truecm
\mbox{\hskip -1.7truecm\includegraphics[width=12.5cm,angle=270.0]{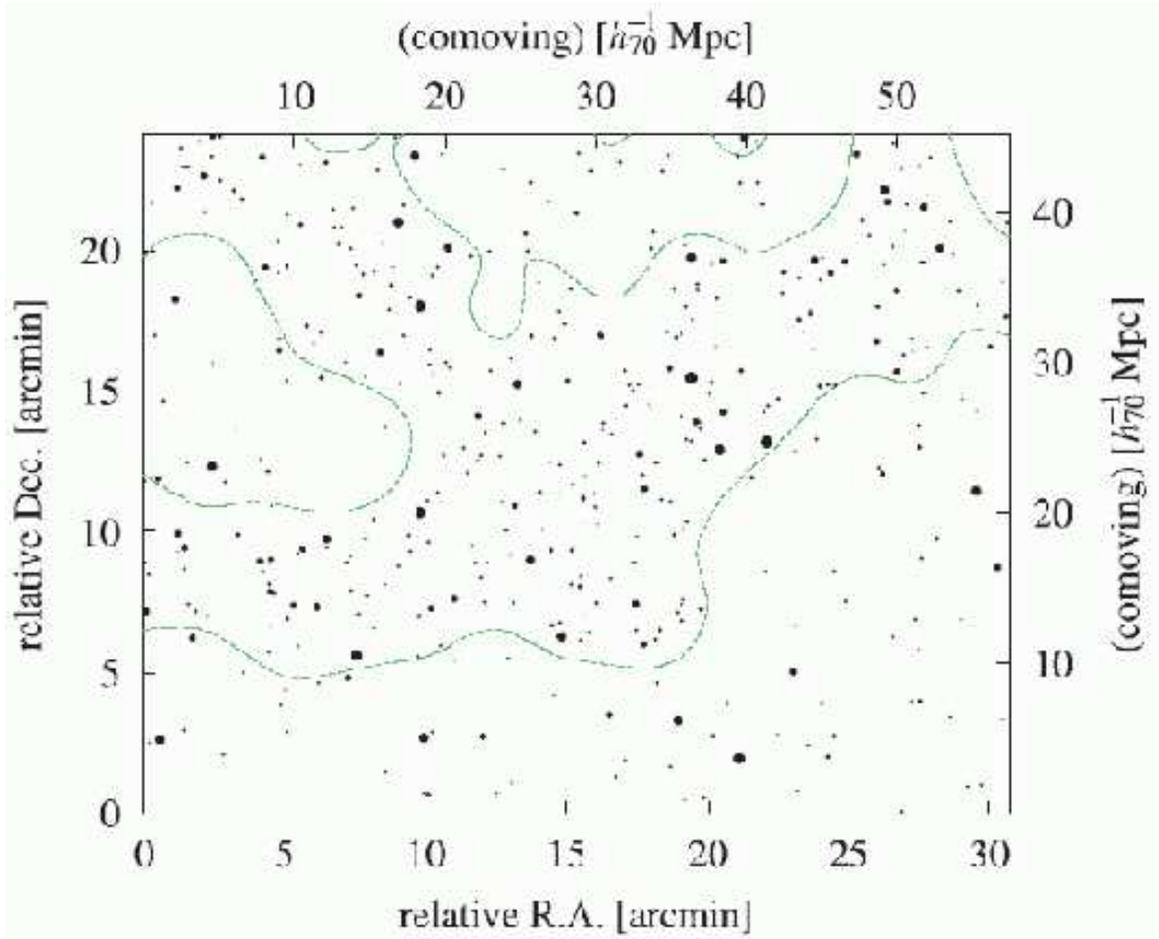}}
\caption{The cosmic web at high redshifts: a prominent weblike features at a redshift $z\sim 3.1$ found 
in a deep view obtained by the Subaru telescope. Large scale sky distribution of 283 strong Ly$\alpha$ emitters 
(black filled circles), the Ly$\alpha$ absorbers (red filled circles) and 
the extended Ly$\alpha$ emitters (blue open squares). The dashed lines indicate the high-density region of the strong 
Ly$\alpha$ emitters. From Hayashino et al. 2004. Reproduced by permission of the AAS.}
\vskip -0.5truecm
\label{fig:filsubaru}
\end{center}
\end{figure*}
The cosmic web is outlined by galaxies populating huge {\it filamentary} and {\it wall-like} 
structures, the sizes of the most conspicuous one frequently exceeding 100h$^{-1}$ Mpc. The 
closest and best studied of these massive anisotropic matter concentrations can be identified 
with known supercluster complexes, enormous structures comprising one or more rich clusters of 
galaxies and a plethora of more modestly sized clumps of galaxies. A prominent and representative 
nearby specimen is the Perseus-Pisces supercluster, a 5$h^{-1}$ wide ridge of at least 50$h^{-1}$ Mpc 
length, possibly extending out to a total length of 140$h^{-1}$ Mpc. While such giant elongated 
structures are amongst the most conspicuous features of the Megaparsec matter distribution, 
filamentary features are encountered over a range of scales and seem to represent a ubiquitous and 
universal state of concentration of matter. In addition to the presence of such filaments the galaxy 
distribution also contains vast planar assemblies. A striking local example is the {\it Great Wall}, a 
huge planar concentration of galaxies with dimensions that are estimated to be of the order of 
$60h^{-1} \times 170h^{-1} \times 5h^{-1}$ Mpc \citep{gellhuch1989}. In both the SDSS and 2dF surveys even 
more impressive planar complexes were recognized, with dimensions substantially in excess of those of 
the local Great Wall. At the moment, the socalled {\it SDSS Great Wall} appears to be the largest known 
structure in the Universe (see fig.~\ref{fig:greatwall}). Gradually galaxy surveys are opening the 
view onto the large scale distribution of galaxies at high redshifts. The Subaru survey has even managed 
to map out a huge filamentary feature at a redshift of $z\sim 3.1$, perhaps the strongest evidence for the 
existence of pronounced cosmic structure at early cosmic epochs (fig.~\ref{fig:filsubaru}).

\begin{figure*}
\begin{center}
\mbox{\hskip -0.45truecm\includegraphics[width=12.35cm]{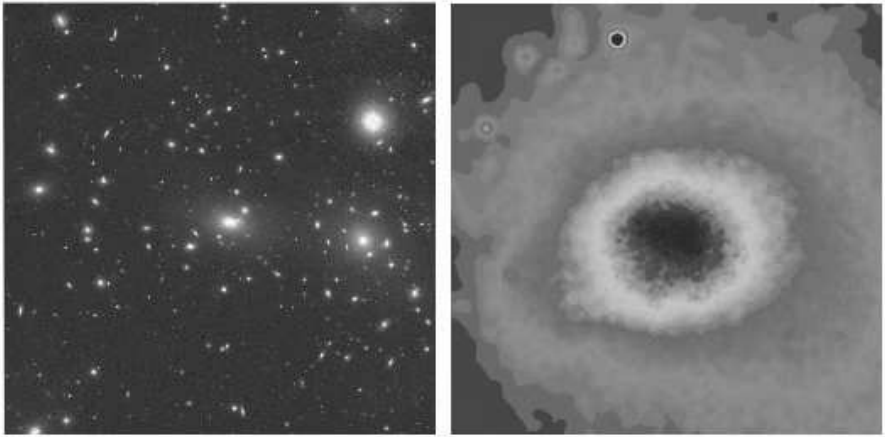}}
\caption{Comparison of optical and X-ray images of Coma cluster. Top: optical 
image. Image courtesy of O. L\'opez-Cruz. Lower: X-ray image (ROSAT).}
\end{center}
\label{fig:coma}
\end{figure*}

Of utmost significance for our inquiry into the issue of cosmic structure formation is the fact that the prominent 
structural components of the galaxy distribution -- clusters, filaments, walls and voids -- are not merely randomly 
and independently scattered features. On the contrary, they have arranged themselves in a seemingly 
highly organized and structured fashion, the {\it cosmic foam} or {\it cosmic web}. They are woven into an intriguing 
{\it foamlike} tapestry that permeates the whole of the explored Universe. The vast under-populated {\it void} regions in 
the galaxy distribution represent both contrasting as well as complementary spatial components to the surrounding {\it planar} 
and {\it filamentary} density enhancements. At the intersections of the latter we often find the most prominent density 
enhancements in our universe, the {\it clusters} of galaxies. \\

\subsection{Cosmic Nodes: Clusters}
\label{sec:clusters}
\begin{figure*}
\begin{center}
\vskip -0.25truecm
\mbox{\hskip -0.0truecm\includegraphics[width=11.8cm]{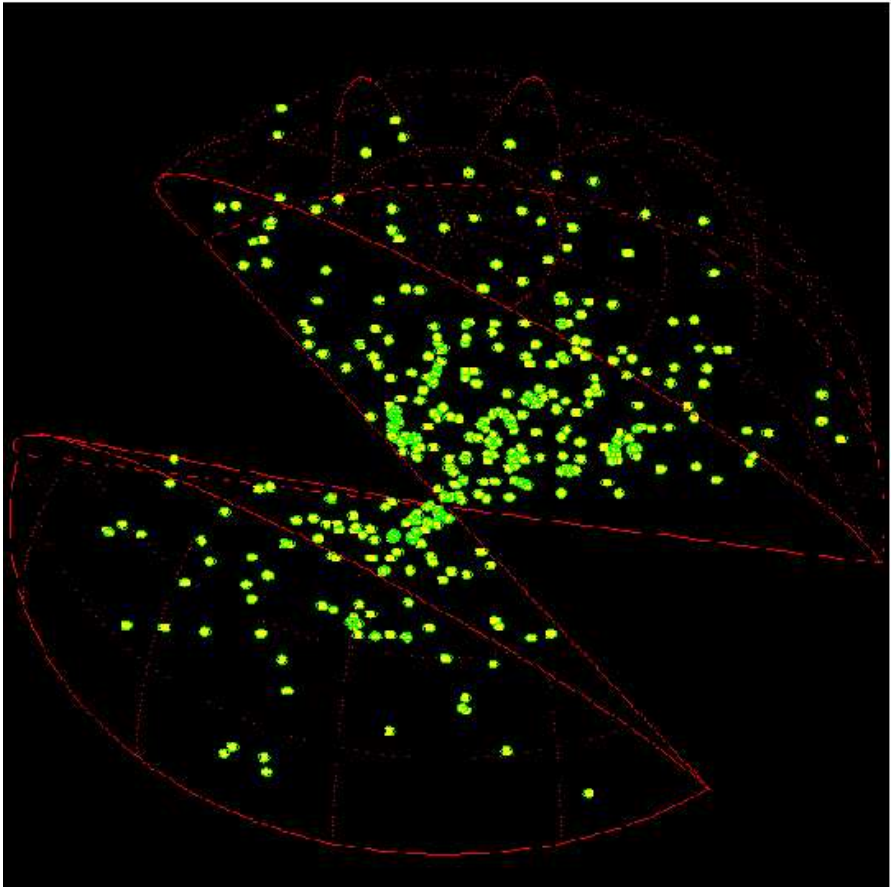}}
\caption{The spatial cluster distribution. The full volume of the X-ray REFLEX 
cluster survey within a distance of 600h$^{-1}$\hbox{Mpc}. The 
REFLEX galaxy cluster catalogue (B\"ohringer et al. 2001), 
contains all clusters brighter than an X-ray flux of $3\times 10^{-12} \hbox{erg} 
\hbox{s}^{-1} \hbox{cm}^{-2}$ over a large part of the southern sky. The missing part of the 
hemisphere delineates the region highly obscured by the Galaxy. 
Image courtesy of S. Borgani \& L. Guzzo, see also Borgani \& Guzzo (2001). Reproduced by permission 
of Nature.}
\vskip -0.25truecm
\end{center}
\label{fig:reflex}
\end{figure*}

Within and around these anisotropic features we find a variety of density condensations, ranging from 
modest groups of a few galaxies up to massive compact {\it galaxy clusters}. The latter stand out 
as the most massive, and most recently, fully collapsed and virialized objects in the Universe. 
Approximately $4\%$ of the mass in the Universe is assembled in rich clusters. They may be regarded 
as a particular population of cosmic structure beacons as they typically concentrate near the interstices 
of the cosmic web, {\it nodes} forming a recognizable tracer of the cosmic matter distribution 
\citep{borgguzz2001}. Clusters not only function as wonderful tracers of structure 
over scales of dozens up to hundred of Megaparsec (fig.~\ref{fig:reflex}) but also as useful probes for 
precision cosmology on the basis of their unique physical properties.

The richest clusters contain many thousands of galaxies within a relatively small volume of only a few 
Megaparsec size. For instance, in the nearby Virgo and Coma clusters more than a thousand galaxies have 
been identified within a radius of a mere 1.5$h^{-1}$ Mpc around their core (see fig.~\ref{fig:coma}). 
Clusters are first and foremost dense concentrations of dark matter, representing overdensities 
$\Delta \sim 1000$. In a sense galaxies and stars only form a minor constituent of clusters. 
The cluster galaxies are trapped and embedded in the deep gravitational wells of the dark matter. 
These are identified as a major source of X-ray emission, emerging from the diffuse extremely hot 
gas trapped in them. While it fell into the potential well, the gas got shock-heated to temperatures in 
excess of $T>10^7$ K, which results in intense X-ray emission due to the bremsstrahlung radiated by the 
electrons in the highly ionized intracluster gas. In a sense clusters may be seen as hot balls of X-ray 
radiating gas. The amount of intracluster gas in the cluster is comparable to that locked into stars, 
and stands for $\Omega_{ICM} \sim 0.0018$ \citep{fukupeeb2004}. The X-ray emission represents a particularly useful signature, 
an objective and clean measure of the potential well depth, directly related to the total mass of the cluster 
\citep[see e.g.][]{reiprich1999}. Through their X-ray brightness they can be seen out to large cosmic depths. 
The deep gravitational dark matter wells also strongly affects the path of passing photons. While the 
resulting strong lensing arcs form a spectacular manifestation, it has been the more moderate distortion of 
background galaxy images in the weak lensing regime \citep{kaiser1992,kaisersq1993} which has opened up 
a new window onto the Universe. The latter has provided a direct probe of the dark matter content of clusters 
and the large scale universe \citep[for a review see e.g.][]{mellier1999,refregier2003}. 

\begin{figure*}
\begin{center}
\vskip 0.25truecm
\mbox{\hskip -0.25truecm\includegraphics[height=12.5cm,angle=90.0]{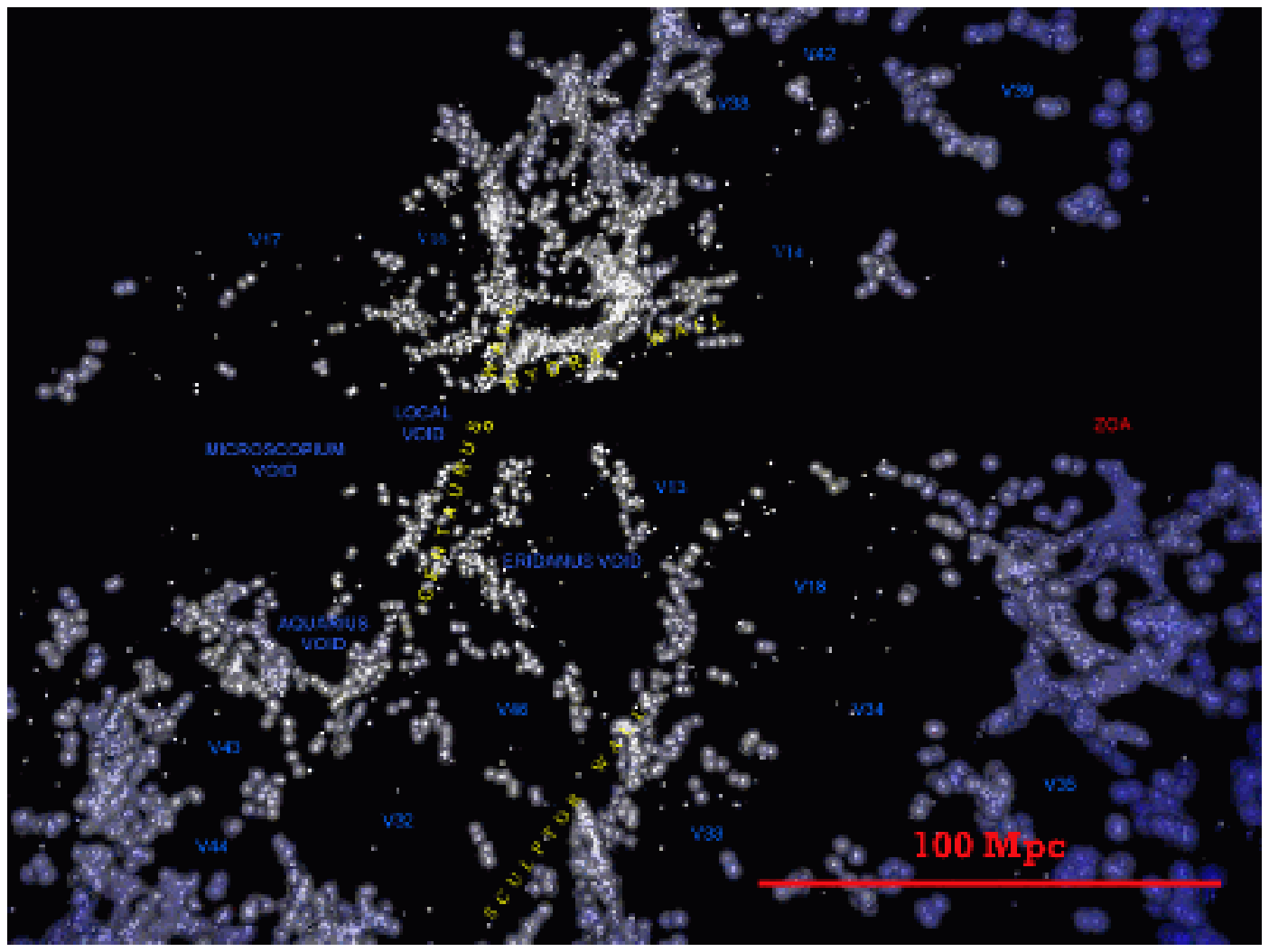}}
\vskip 0.25truecm
\caption{A region of the 6dF redshift survey marked by the presence of various major voids. The image concerns a 
3D rendering of the galaxy distribution in a 1000 km/s thick slice along the supergalactic SGX direction, 
at SGX=-2500 km/s. Image courtesy of A. Fairall.}
\end{center}
\label{fig:6dfvoid}
\end{figure*}
\subsection{Cosmic Depressions: the Voids}
\label{sec:webvoid}
Complementing this cosmic inventory leads to the existence of large {\it voids}, enormous regions with sizes in the range of 
$20-50h^{-1}$ Mpc that are practically devoid of any galaxy, usually roundish in shape and occupying the major 
share of space in the Universe. Forming an essential ingredient of the {\it Cosmic Web}, they are surrounded by elongated 
filaments, sheetlike walls and dense compact clusters.  

Voids have been known as a feature of galaxy surveys since the first surveys were compiled \citep{chincar75,gregthomp78,einasto1980}. Following the
discovery by \cite{kirshner1981,kirshner1987} of the most outstanding specimen, the Bo\"otes void, a hint of their central position within a weblike
arrangement came with the first CfA redshift slice \citep{lapparent1986}. This view has been expanded dramatically as maps of the spatial 
distribution of hundreds of thousands of galaxies in the 2dFGRS \citep{colless2003} and SDSS redshift survey \citep{abaz2003} became available, 
recently supplemented with a high-resolution study of voids in the nearby Universe based upon the 6dF survey \citep{heathjones2004,fairall2007}. 
The 2dFGRS and SDSS maps of figs.~\ref{fig:2dfgaldist} and ~\ref{fig:sdssgaldist}, and the void map of the 6dF survey in 
fig.~\ref{fig:6dfvoid} form telling illustrations.

Voids in the galaxy distribution account for about 95\% of the total volume \citep[see][]{kauffair1991,elad1996,elad1997,hoyvog2002,plionbas2002,
rojas2005}. The typical sizes of voids in the galaxy distribution depend on the galaxy population used to define the voids. Voids defined by 
galaxies brighter than a typical $L_*$ galaxy tend to have diameters of order $10-20h^{-1}$Mpc, but voids associated with rare luminous 
galaxies can be considerably larger; diameters in the range of $20h^{-1}-50h^{-1}$Mpc are not uncommon \citep[e.g.][]{hoyvog2002,plionbas2002}. 
These large sizes mean that only now we are beginning to probe a sufficiently large cosmological volume to allow meaningful statistics with voids 
to be done.  As a result, the observations are presently ahead of the theory.  

\section{Cosmic Structure Formation:\\
\ \ \ \ from Primordial Noise to Cosmic Web}
The fundamental cosmological importance of the {\it Cosmic Web} is that it comprises  
features on a typical scale of tens of Megaparsec, scales at which the Universe still 
resides in a state of moderate dynamical evolution. Structures have only freshly emerged 
from the almost homogeneous pristine Universe and have not yet evolved beyond 
recognition. Therefore they still retain a direct link to the matter distribution in 
the primordial Universe and thus still contain a wealth of direct information on 
the cosmic structure formation process. 

In our exploration of the cosmic web and the development of appropriate tools towards the 
analysis of its structure, morphology and dynamics we start from the the assumption that 
the cosmic web is traced by a population of discrete objects, either galaxies in the real 
observational world or particles in that of computer simulations. The key issue will be to 
reconstruct the underlying continuous density and velocity field, retaining the geometry and 
morphology of the weblike structures in all its detail. 

In this we will pursue the view that filaments are the basic elements of the cosmic web, 
the key features around which most matter will gradually assemble and the channels along which 
matter is transported towards the highest density knots within the network, the clusters of galaxies. 
Likewise we will emphasize the crucial role of the voids -- the large underdense and expanding 
regions occupying most of space -- in the spatial organization of the various structural elements 
in the cosmic web. One might even argue that it are the voids which should be seen as the key 
ingredients of the cosmic matter distribution. This tends to form the basis for geometrical 
models of the Megaparsec scale matter distribution, with the Voronoi model as its main 
representative \citep[see e.g][]{weygaertphd1991,weygaert2002,weygaert2007a,weygaert2007b}\footnote{These Voronoi 
models are spatial models for cellular/weblike galaxy distributions, not to be confused with the
application of Voronoi tessellations in DTFE and the tessellation based methods towards spatial 
interpolation and reconstruction}. 

\begin{figure*} 
  \vskip 0.25truecm
  \center
     \mbox{\hskip -0.25truecm\includegraphics[height=13.5cm,angle=90.0]{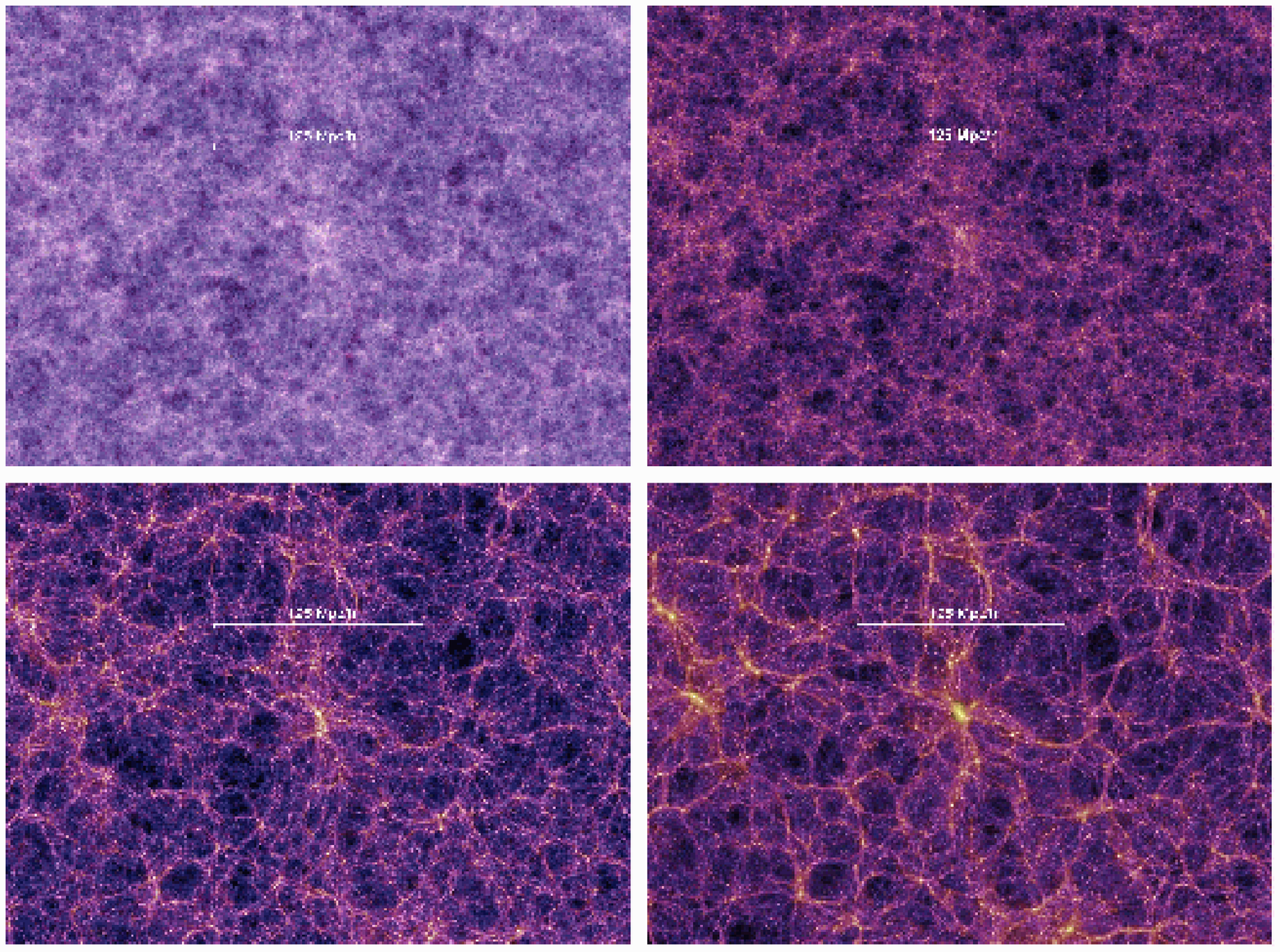}}
\end{figure*}
\begin{figure*}
\caption{The Cosmic Web in a box: a set of four time slices from the Millennium simulation 
of the $\Lambda$CDM model. The frames show the projected (dark) matter distribution in slices of 
thickness $15\hmpc$, extracted at $z=8.55, z=5.72, z=1.39$ and $z=0$. These redshifts correspond 
to cosmic times of 600 Myr, 1 Gyr, 4.7 Gyr and 13.6 Gyr after the Big Bang. The four frames have a size 
of $125\hmpc$. The evolving mass distribution reveals the major 
characteristics of gravitational clustering: the formation of an intricate filamentary web, the 
hierarchical buildup of ever more massive mass concentrations and the evacuation of large underdense 
voids. Image courtesy of V. Springel \& Virgo consortium, also see Springel et al. 2005.}
\label{fig:millennium}
\end{figure*}

\subsection{Gravitational Instability}
The standard paradigm of cosmic structure formation is that of gravititional instability 
scenarios \citep{peebles80,zeld1970}. Structure in the Universe is the result of the gravitational growth of 
tiny density perturbations and the accompanying tiny velocity perturbations in the primordial Universe. 
Supported by an impressive body of evidence, primarily that of temperature fluctuations in the cosmic 
microwave background \citep{smoot1992,bennett2003,spergel2006}, the character of the primordial random density and 
velocity perturbation field is that of a {\it homogeneous and isotropic spatial Gaussian process}. 
Such fields of primordial Gaussian perturbations in the gravitational potential are a natural product 
of an early inflationary phase of our Universe. 

The formation and moulding of structure is a result to the gravitational growth of 
the primordial density- and velocity perturbations. Gravity in slightly overdense regions 
will be somewhat stronger than the global average gravitational deceleration, as will be 
the influence they exert over their immediate surroundings. In these regions the 
slow-down of the initial cosmic expansion is correspondingly stronger and, when 
the region is sufficiently overdense it may even come to a halt, turn around  
and start to contract. If or as long as pressure forces are not sufficient to counteract 
the infall, the overdensity will grow without bound, assemble more and more matter 
by accretion of matter from its surroundings, and ultimately fully collapse to form 
a gravitationally bound and virialized object. In this way the primordial overdensity 
finally emerges as an individual recognizable denizen of our Universe, their precise 
nature (galaxy, cluster, etc.) and physical conditions determined by the scale, mass 
and surroundings of the initial fluctuation. 

\subsection{Nonlinear Clustering}
Once the gravitational clustering process has progressed beyond the initial linear growth 
phase we see the emergence of complex patterns and structures in the density field. 
Highly illustrative of the intricacies of the structure formation process is that of the 
state-of-the-art N-body computer simulation, the Millennium simulation 
by \citep[][]{springmillen2005}. Figure~\ref{fig:millennium} shows four 
time frames out of this massive 10$^{10}$ particle simulation of a $\Lambda CDM$ matter 
distribution in a $500\hmpc$ box. The time frames correspond to redshifts $z=8.55$, 
$z=5.72$, $z=1.39$ and $z=0$ (ie. at epochs 600 Myr, 1 Gyr, 4.7 Gyr and 13.6 Gyr after 
the Big Bang). The earlies time frame is close to that of the condensation of the 
first stars and galaxies at the end of the {\it Dark Ages} and the reionization of the 
gaseous IGM by their radiation. The frames contain the Dark Matter 
particle distribution in a $15\hmpc$ thick slice of a $125\hmpc$ region centered on 
the central massive cluster of the simulation. 

The four frames provide a beautiful picture of the unfolding Cosmic Web, starting from a 
field of mildly undulating density fluctations towards that of a pronounced and intricate 
filigree of filamentary features, dented by dense compact clumps at the nodes of the network. 
Clearly visible is the hierarchical nature in which the filamentary network builds up. At first consisting of 
a multitude of small scale edges, they quickly merge into a few massive elongated 
channels. 

Large N-body simulations like the Millennium simulation and the many others currently 
available all reveal a few ``universal'' characteristics of the (mildly) nonlinear 
cosmic matter distribution. Three key characteristics of the Megaparsec universe 
stand out:\\

\begin{center}
{\vbox{\Large{ 
\begin{itemize}
\item[$\bullet$] Hierarchical clustering
\item[$\bullet$] Weblike spatial geometry
\item[$\bullet$] Voids
\end{itemize}}}}
\end{center}
\ \\
{\it Hierarchical} clustering implies that the first objects to condense first are small and that 
ever larger structures form through the gradual merging of smaller structures. Usually an object 
forms through the accretion of all matter and the fusion of all substructure within its realm, including 
that of the small-scale objects which had condensed out at an earlier stage. The {\it second} fundamental 
aspect is that of {\it anisotropic gravitational collapse}. Aspherical overdensities, on any scale and in 
any scenario, will contract such that they become increasingly anisotropic. At first they turn into a flattened 
{\it pancake}, rapidly followed by contraction into an elongated filament and possibly, dependent on scale, 
total collapse into a galaxy or a cluster may follow. This tendency to collapse anisotropically 
finds its origin in the intrinsic primordial flattening of the overdensity, augmented by the anisotropy of 
the gravitational force field induced by the external matter distribution (i.e. by tidal forces). It is 
evidently the major agent in shaping the weblike cosmic geometry. The {\it third} manifest feature of 
the Megaparsec Universe is the marked and dominant presence of large roundish underdense regions, the 
{\it voids}. They form in and around density troughs in the primordial density field. Because of their lower interior gravity 
they will expand faster than the rest of the Universe, while their internal matter density rapidly decreases as 
matter evacuates their interior. They evolve in the nearly 
empty void regions with sharply defined boundaries marked by filaments and walls. Their essential role in the 
organization of the cosmic matter distribution got recognized early after their discovery. 
Recently, their emergence and evolution has been explained within the context of hierarchical gravitational scenarios 
\cite{shethwey2004}. 

The challenge for any viable analysis tool is to trace, highlight and measure each of these aspects 
of the cosmic web. Ideally it should be able to do so without resorting to user-defined parameters or functions, 
and without affecting any of the other essential characteristics. We will argue in this contributation that 
the {\it DTFE} method, a linear version of {\it natural neighbour} interpolation, is indeed able to deal 
with all three aspects (see fig.~\ref{fig:hierarchydtfe}).

%
%
\section{Spatial Structure and Pattern Analysis}
\label{sec:statspatial}
Many attempts to describe, let alone identify, the features and components of the Cosmic Web have been of a mainly heuristic nature. 
There is a variety of statistical measures characterizing specific aspects of the large scale matter distribution \citep[for an 
extensive review see][]{martinez2002}. For completeness and comparison, we list briefly a selection of methods for structure characterisation 
and finding.  It is perhaps interesting to note two things about this list: 
\begin{enumerate}
\item[a)] each of the methods tends to be specific to one particular structural entity 
\item[b)] there are no explicit wall-finders. 
\end{enumerate}
This emphasises an important aspect of our Scale Space approach: it provides a uniform approach to finding Blobs, Filaments and Walls 
as individual objects that can be catalogued and studied.

\subsection{Structure from higher moments}
The clustering of galaxies and matter is most commonly described in terms of a hierarchy of correlation functions \citep{peebles80}. 
The two-point correlation function -- and the equivalent power spectrum, its Fourier transform \citep{peacockdodss1994,tegmark2004} 
-- remains the mainstay of cosmological clustering analysis and has a solid physical basis. However, the nontrivial and nonlinear 
patterns of the cosmic web are mostly a result of the phase correlations in the cosmic matter distribution \citep{rydengram1991,
chiang2000,pcoles2000}. While this information is contained in the moments of cell counts \citep{peebles80,lapparent1991,gaztanaga1992} and, 
more formally so, in the full hierarchy of M-point correlation functions $\xi_M$, except for the lowest orders their measurement has proven 
to be practically unfeasible \citep{peebles80,szapudi1998,jones2005}. Problem remains that these higher order correlation functions 
do not readily translate into a characterization of identifiable features in the cosmic web. 

The Void probability Function \citep{white1979,lachieze1992} provides a characterisation of the ''voidness'' of the Universe in terms of a function 
that combined information from many higher moments of the point distribution.  But, again, this did not provide any identification of 
individual voids. 

\subsection{Topological methods}
The shape of the local matter distribution may be traced on the basis of an analysis of the statistical properties of its inertial moments 
\citep{babul1992,vishniac1995,basilakos2001}. These concepts are closely related to the full characterization of the topology of the matter 
distribution in terms of four Minkowski functionals \citep{mecke1994,schmalzing1999}. They are solidly based on the theory of spatial statistics 
and also have the great advantage of being known analytically in the case of Gaussian random fields. In particular, the \textit{genus} of the 
density field has received substantial attention as a strongly discriminating factor between intrinsically different spatial patterns 
\citep{gott1986,hoyle2002}. 

The Minkowski functionals provide global characterisations of structure. An attempt to extend its scope towards providing locally defined 
topological measures of the density field has been developed in the SURFGEN project defined by Sahni and Shandarin and their coworkers 
\citep{sahni1998,jsheth2003,shandarin2004}. The main problem remains the user-defined, and thus potentially biased, nature of the continuous 
density field inferred from the sample of discrete objects. The usual filtering techniques suppress substructure on a scale smaller than the 
filter radius, introduce artificial topological features in sparsely sampled regions and diminish the flattened or elongated morphology of the 
spatial patterns. Quite possibly the introduction of more advanced geometry based methods to trace the density field may prove a major advance 
towards solving this problem. \citet{martinez2005} have generalized the use of Minkowski Functionals by calculating their values in variously 
smoothed volume limited subsamples of the 2dF catalogue. 

\subsection{Cluster and Filament finding}
In the context of analyzing distributions of galaxies we can think of cluster finding algorithms.  There we might define a cluster as an 
aggregate of neighbouring galaxies sharing some localised part of velocity space.  Algorithms like HOP attempt to do this.  However, there 
are always issues arising such as how to deal with substructure: that perhaps comes down to the definition of what a cluster is. Nearly 
always coherent structures are identified on the basis of particle positions alone. Velocity space data is often not used since there is 
no prior prejudice as to what the velocity space should look like. 

The connectedness of elongated supercluster structures in the cosmic matter distribution was first probed by means of percolation analysis, 
introduced and emphasized by Zel'dovich and coworkers \citep{zeldovich1982}, while a related graph-theoretical construct, the minimum 
spanning tree 
of the galaxy distribution, was extensively probed and analysed by Bhavsar and collaborators \citep{barrow1985,colberg2007} in an attempt 
to develop an objective measure of filamentarity. 

Finding filaments joining neighbouring clusters has been tackled, using quite different techniques, by \citet{colberg2005} and by 
\citet{pimbblet2005}.  More general filament finders have been put forward by a number of authors.  Skeleton analysis of the density field 
\citep{novikov2006} describes continuous density fields by relating density field gradients to density maxima and saddle points. This is 
computationally intensive but quite effective, though it does depend on the artefacts in the reconstruction of the continuous density field.

\citet{stoica2005} use a generalization of the classical Candy model to locate and catalogue filaments in galaxy surveys.  This 
appraoch has the advantage that it works directly with the original point process and does not require the creation of a continuous density 
field. However, it is very computationally intensive.

A recently developed method, the Multiscale Morphology Filter \citep{aragonphd2007,aragonmmf2007} (see sect.~\ref{sec:mmf}), seeks to identify different 
morphological features over a range of scales. Scale space analysis looks for structures of a mathematically specified type in a hierarchical, 
scale independent, manner.  It is presumed that the specific structural characteristic is quantified by some appropriate parameter (e.g.: density, 
eccentricity, direction, curvature components).  The data is filtered to produce a hierarchy of maps having different resolutions, and at each 
point, the dominant parameter value is selected from the hierarchy to construct the scale independent map.  While this sounds relatively 
straightforward, in practice a number 
of things are required to execute the process. There must be an unambiguous definition of the structure-defining characteristic. The 
implementation of \cite{aragonmmf2007} uses the principal components of the local curvature of the density field at each point as a morphology 
type indicator.  This requires that the density be defined at all points of a grid, and so there must be a method for going from a discrete 
point set to a grid sampled continuous density field. This is done using the DTFE methodology since that does minimal damage to the structural 
morphology of the density field (see sect.~\ref{sec:dtfe}).

\subsection{Void Finding}
Voids are distinctive and striking features of the cosmic web, yet finding them systematically in surveys and simulations has proved rather 
difficult. There have been extensive searches for voids in galaxy catalogues and in numerical simulations (see sect.~\ref{sec:webvoid}). 
Identifying voids and tracing their outline within the complex spatial geometry of the Cosmic Web appears to be a nontrivial
issue. The fact that voids are almost empty of galaxies means that the sampling density plays a key role in determining what is or is 
not a void \citep{schmidt2001}. There is not an unequivocal definition of what a void is and as a result there is considerable disagreement 
on the precise outline of such a region \citep[see e.g.][]{shandfeld2006}. 

Moreover, void finders are often predicated on building void structures out of cubic cells \citep{kauffair1991} or out of spheres 
\citep[e.g.][]{patiri2006a}.  Such methods attempt to synthesize voids from the intersection of cubic or spherical 
elements and do so with varying degrees of success.  Because of the vague and different definitions, and the range of different 
interests in voids, there is a plethora of void identification procedures \citep{kauffair1991,elad1997,aikmah1998,hoyvog2002,arbmul2002, 
plionbas2002,patiri2006a,colberg2005b,shandfeld2006}. The ``voidfinder'' algorithm of \cite{elad1997} has been at the basis of most 
voidfinding methods. However, this succesfull approach will not be able to analyze complex spatial configurations in which voids may 
have arbitrary shapes and contain a range and variety of substructures. The \textit{Void Finder Comparison Project} of \citet{colpear2007} 
will clarify many of these issues. 

The watershed-based WVF algorithm of \citet[][see sect.~\ref{sec:watershed}]{platen2007} aims to avoid issues of both sampling density 
and shape. This new and objective voidfinding formalism has been specifically designed to dissect in a selfconsistent
manner the multiscale character of the void network and the weblike features marking its boundaries.  The {\it Watershed Void Finder}
(WVF) is based on the watershed algorithm \citep{beulan1979,meyerbeucher1990,beumey1993}. This is a concept from the field of 
{\it mathematical morphology} and {\it image analysis}. The WVF is defined with respect to the DTFE density field of a discrete point 
distribution \citep{schaapwey2000}, 
assuring optimal sensitivity to the morphology of spatial structures and an unbiased probe over the full range of substructure in the mass
distribution. Because the WVF void finder does not impose a priori constraints on the size, morphology and shape of a voids it has the
potential to analyze the intricacies of an evolving void hierarchy. 

\section{Structural Reconstruction}
In the real world it is impossible to get exhaustive values of data at every desired point of space. Also 
astronomical observations, physical experiments and computer simulations often produce discretely sampled data 
sets in two, three or more dimensions. This may involve the value of some physical quantity measured at an irregularly 
distributed set of reference points. Also cosmological theories describe the development of structure in terms 
of continuous (dark matter) density and velocity fields while to a large extent our knowledge stems from a 
discrete sampling of these fields.  

In the observational reality galaxies are the main tracers of the cosmic web and it is mainly through 
measuring of the redshift distribution of galaxies that we have been able to map its structure. Another 
example is that of the related study of cosmic flows in the nearby Universe, based upon the measured peculiar 
velocities of a sample of galaxies located within this cosmic volume. Likewise, simulations of the evolving 
cosmic matter distribution are almost exclusively based upon N-body particle computer calculation, involving 
a discrete representation of the features we seek to study. Both the galaxy distribution as well as the particles 
in an N-body simulation are examples of {\it spatial point processes} in that they are 
\begin{itemize}
\item[-] {\it discretely sampled} 
\item[-] have an {\it irregular spatial distribution}.
\end{itemize}
\noindent 
The principal task for any formalism seeking to process the discretely sampled field is to optimally retain or 
extract the required information. Dependent on the purpose of a study, various different strategies may be 
followed. One strategy is to distill various statistical measures, or other sufficiently descriptive cosmological 
measures, characterizing specific aspects of the large scale matter distribution 
\citep[see][also see sect.~\ref{sec:statspatial}]{martinez2002}. 
In essence this involves the compression of the available information into a restricted set of parameters or functions, with the 
intention to compare or relate these to theoretical predictions. The alternative is to translate the {\it discretely sampled} and 
{\it spatially irregularly distributed} sampled objects into related continuous fields. While demanding in itself, it is complicated by the 
highly inhomogeneous nature of the sample point distribution. The translation is a far from trivial procedure. If 
forms the subject of an extensive literature in computer science, visualization and applied sciences. An interesting 
comparison and application of a few different techniques is shown in fig.~\ref{fig:landscapint}. It shows how 
some methods to be discussed in the following sections fare when applied to the reconstruction of Martian 
landscapes from measurements by the MOLA instrument on the Mars Global Surveyor \citep{abramov2004}.  
\begin{figure*}
\begin{center}
\mbox{\hskip -0.15truecm\includegraphics[height=19.5cm]{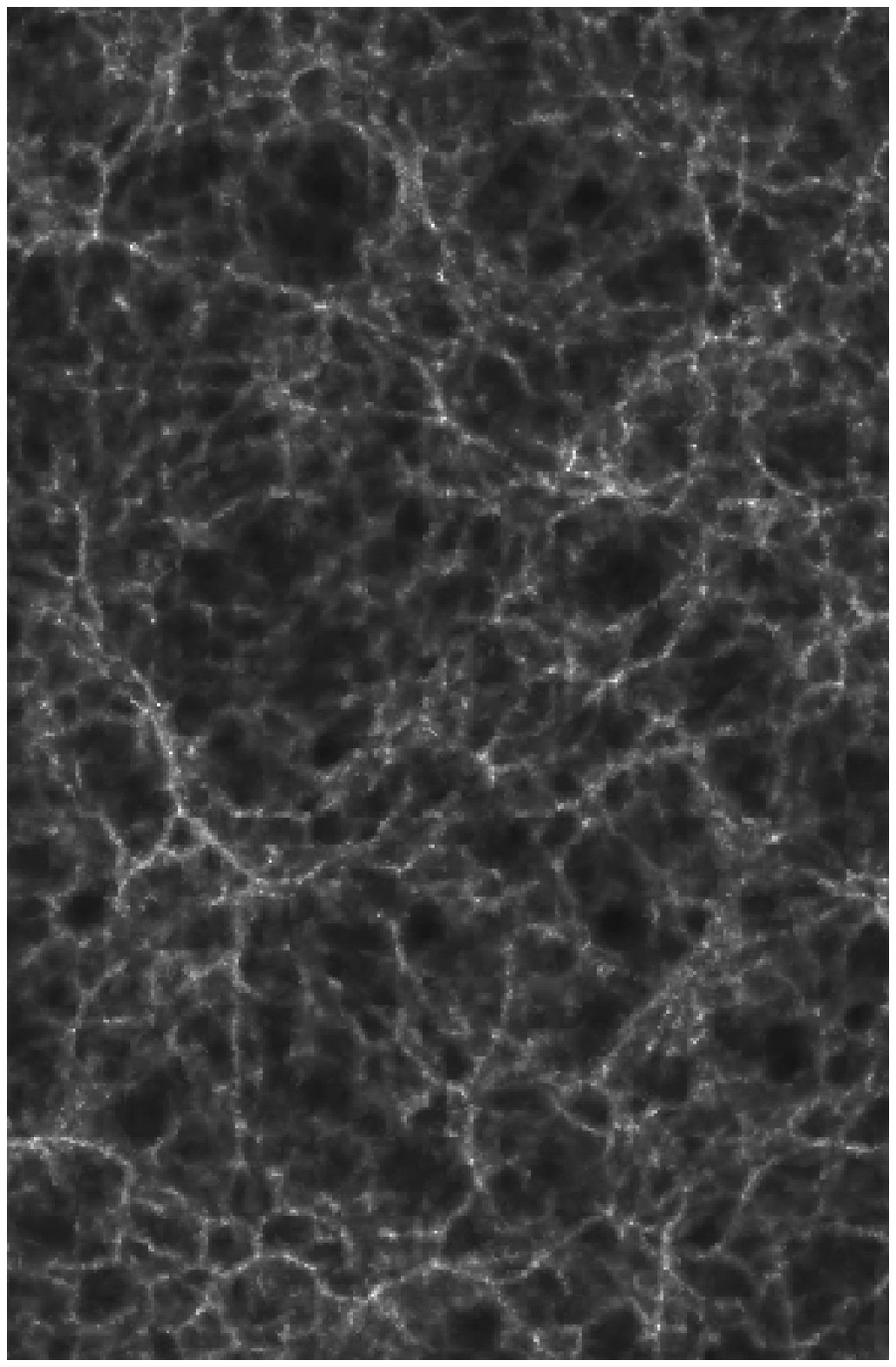}}
\end{center}
\end{figure*}
\begin{figure*}[t]
\caption{DTFE processed image of the Cosmic Web: GIF N-body simulation of structure formation in a $\Lambda$CDM 
cosmology. Part of the central X-slice. Simulation courtesy: J. Colberg.}
\label{fig:gifdtfeweb}
\end{figure*}
\subsection{Spatial Data: Filtering and Interpolation}
Instead of direct statistical inference from datasets one can seek to reconstruct the underlying continuous field(s). 
For a meaningful analysis and interpretation of spatial data it is indeed often imperative and/or preferrable 
to use methods of parameterization and interpolation to obtain estimates of the related field values throughout 
the sample volume. The {\it reconstructed} continuous field may subsequently be processed in order to yield 
a variety of interesting parameters. 

It involves issues of {\it smoothing} and {\it spatial interpolation} of the measured data over the 
sample volume, of considerable importance and interest in many different branches of science. 
Interpolation is fundamental to graphing, analysing and understanding of spatial data. 
Key references on the involved problems and solutions include those by \cite{ripley1981,watson1992,cressie1993}. While 
of considerable importance for astronomical purposes, many available methods escaped attention. A systematic 
treatment and discussion within the astronomical context is the study by \cite{rybickipress1992},
who focussed on linear systems as they developed various statistical procedures related to linear prediction and 
optimal filtering, commonly known as Wiener filtering. An extensive, systematic and more general survey of 
available mathematical methods can be found in a set of publications by \cite{lombardi2001,lombardi2002,lombardi2003}. 

A particular class of spatial point distributions is the one in which the point process forms a representative 
reflection of an underlying smooth and continuous density/intensity field. The spatial distribution of the 
points itself may then be used to infer the density field. This forms the basis for the interpretation and analysis 
of the large scale distribution of galaxies in galaxy redshift surveys. The number density of galaxies in redshift survey 
maps and N-body particles in computer simulations is supposed to be proportional to the underlying matter density. 

\begin{figure} 
  \center
     \mbox{\hskip -0.5truecm\includegraphics[width=12.5cm]{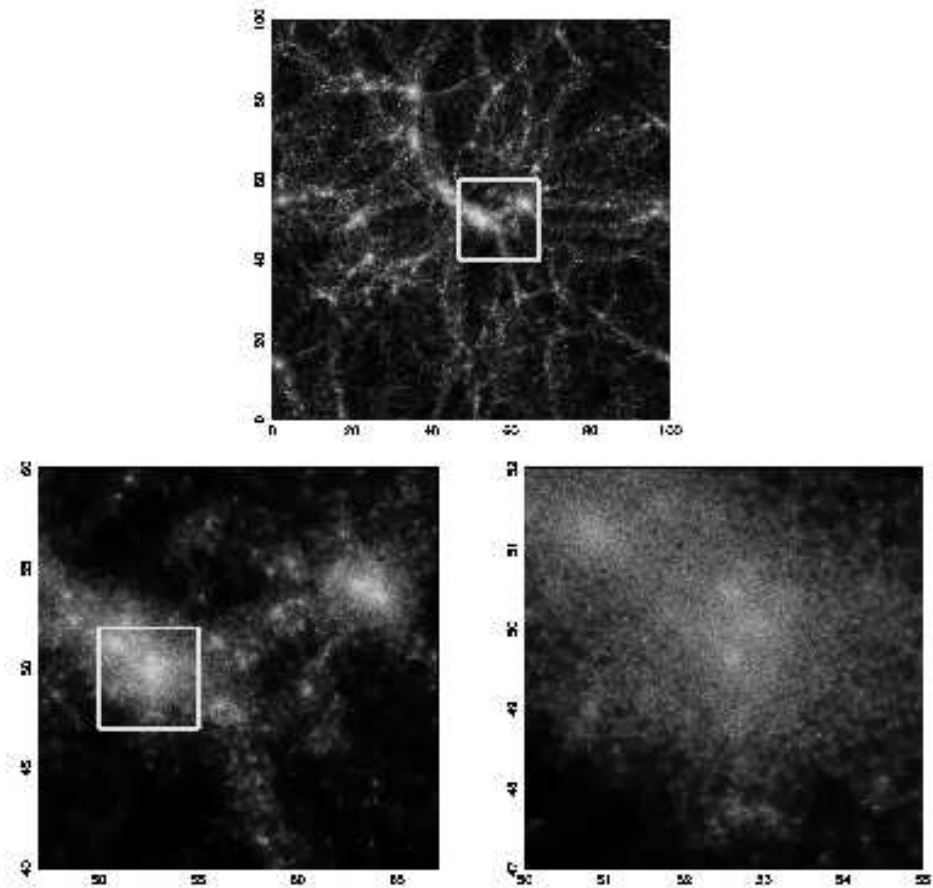}}
  \vskip -0.2truecm
  \caption{Cosmic density field illustrating the large dynamic range 
  which is present in the large scale matter distribution. In the left-hand
  frame the density field in a 10$h^{-1}$Mpc wide slice through a
  cosmological simulation is depicted. In the subsequent frames
  zoom-ins focusing on a particular structure are shown. On all
  depicted scales structures are present.}
\label{fig:hierarchydtfe}
\end{figure}
\subsection{Local Interpolation: Natural Neighbour Methods}
The complex spatial geometry and large density variations marking the cosmic web ideally should be analyzed by a 
technique which would (1) not lose information against the backdrop of a highly inhomogeneous spatial resolution and 
(2) which is capable of tracing hierarchically structured and anisotropic spatial patterns in an entirely objective 
fashion. Nearly all existing techniques for analyzing galaxy redshift surveys or numerical simulations of cosmic structure 
formation have important shortcomings with respect to how they treat the weblike geometry of the large scale matter 
distribution and trace the cosmic matter distribution over a wide range of densities. The limited available mathematical 
machinery has often been a major obstacle in exploiting the potentially large information content of the cosmic web. 

The various aspects characterizing the complex and nontrivial spatial structure of the cosmic web have proven to be notoriously difficult 
to quantify and describe. For the analysis of weblike patterns the toolbox of descriptive measures is still largely ad-hoc  
and is usually biased towards preconceived notions of their morphology and scale. None of the conventional, nor even 
specifically designed, measures of the spatial matter distribution have succeeded in describing all relevant features 
of the cosmic web. Even while they may succeed in quantifying one particular key aspect it usually excludes the ability 
to do so with other characteristics.

For many applications a `local' interpolation and reconstruction method appears to provide the preferred path. In this case 
the values of the variable at any point depend only on the data in its neighbourhood \citep[see e.g][]{sbm1995}. Local schemes 
usually involve a discretization of the region into an adaptive mesh. In data interpolation this usually represents a more realistic 
approach, and generically it also tends to be independent of specific model assumptions while they are very suited for 
numerical modeling applications. When the points have a regular distribution many local methods are available for smooth 
interpolation in multidimensional space. Smooth, local methods also exist for some specific irregular point distributions. 
A telling example are the `locally constrained' point distributions employed in applications of the finite element method. 

In this review we specifically concentrate on a wide class of tessellation-based {\it multidimensional} and entirely {\it local} 
interpolation procedures, commonly known as {\it Natural Neighbour Interpolation} \citep[ch. 6]{watson1992,braunsambridge1995,
sukumarphd1998,okabe2000}. The local {\it nutural neighbour} methods are based upon the {\it Voronoi and Delaunay tessellations} of 
the point sample, basic concepts from the field of stochastic and computational geometry \citep[see][and references 
therein]{okabe2000}. These spatial volume-covering divisions of space into mutually disjunct triangular (2-D) or tetrahedral (3-D) cells 
adapt to the local {\it density} and the local {\it geometry} of the point distribution (see fig.~\ref{fig:gifdtfeweb} and 
fig.~\ref{fig:hierarchydtfe}). The natural neighbour interpolation schemes exploit these virtues and thus adapt automatically and in 
an entirely natural fashion to changes in the density or the 
geometry of the distribution of sampling points. For the particular requirements posed by astronomical and cosmological datasets, for which it 
is not uncommon to involve millions of points, we have developed a linear first-order version of {\it Natural Neighbour Interpolation}, the 
{\it Delaunay Tessellation Field Estimator} \citep[DTFE,][]{bernwey96,schaapwey2000,schaapphd2007}. Instead of involving user-defined filters which 
are based upon artificial smoothing kernels the main virtue of natural neighbour methods is that they are intrinsically {\it self-adaptive}  
and involve filtering kernels which are defined by the {\it local density and geometry} of the point process or object distribution. 

\begin{figure} 
  \centering
    \mbox{\hskip -0.1truecm\includegraphics[height=20.0cm]{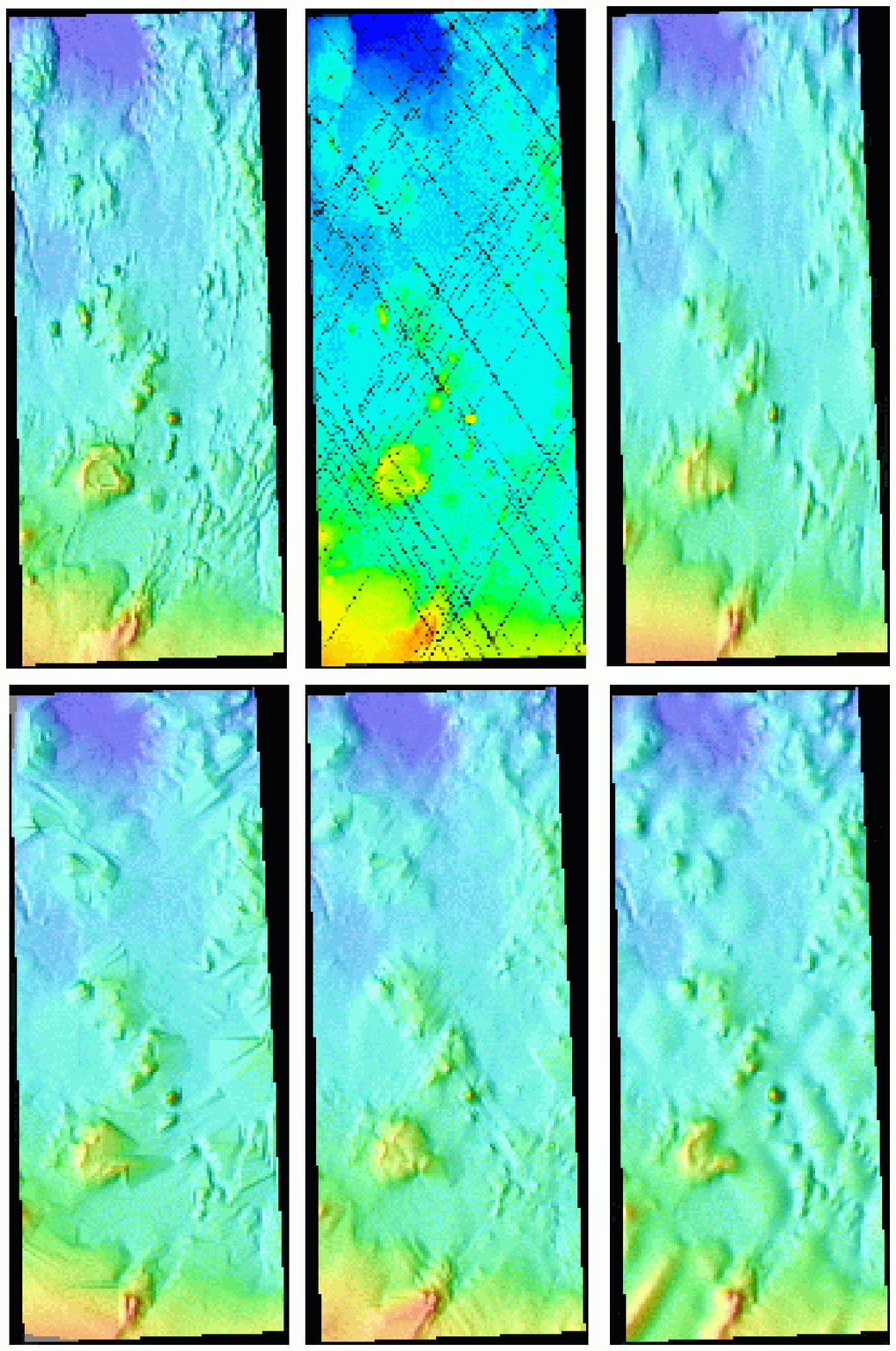}}
    \vskip -0.5truecm
\end{figure} 
\begin{figure}
\caption{Interpolation of simulated Mars Orbiter Laster Altimeter (MOLA) data acquired from a 
digital elevation model of the Jokulsa and Fjollum region of Iceland, in anticipation of 
landscape reconstructions of terrain on planet Mars. The ``original'' highly resolved image of the terrain 
is shown in the top left frame. The comparison concerns data that were measured at the track points indicated 
in the top centre image. The medium resolution ($209 \times 492$) interpolations are: natural neighbour (top right), 
linear (DTFE) interpolation (bottom left), nearest neighbour interpolation (bottom centre) and 
spline interpolation (bottom right). Evidently, the nn-neighbour map is the most natural looking 
reconstruction. Image courtesy of O. Abramov and A. McEwen, figure modified from Abramov \& McEwen 2004}
\label{fig:landscapint} 
\end{figure}

\subsection{Meshless methods}
\noindent The Natural Neighbour schemes, including DTFE, are mesh-based. With current technology this
is computationally feasible as long as the domain of the matter density field
has at most three dimensions. However, also interesting would be to extend attention to 
six-dimensional phase-space, incorporating not only the location of
galaxies/particles but also their velocities. This will double the number of
dimensions and makes mesh-based methods a real challenge.  While the
application of DTFE to the analysis of the phase-space of dark haloes has been
shown to lead to very good results \citep{arad2004}, 
studies by \cite{ascalbin2005} and \cite{sharma2006} argued it would be far more
efficient and reasonably accurate to resort to the simpler construct of a {\it
k-d tree} \citep[][for an early astronomical implementation]{bentley1975,frbent1977,bentfr1978,weygaert1987}.
While \cite{ascalbin2005,sharma2006} do not take into
account that phase-space is not a simple metric space but a symplectic one, it may indeed be a real
challenge to develop mesh-based methods for the analysis of structure in
phase-space.  Although algorithms for the computation of higher dimensional
Voronoi and Delaunay tesselations have been implemented (e.g., in \cgal), the
high running time and memory use make further processing computationally
infeasible.

Meshless spatial interpolation and approximation methods for datasets in spaces 
of dimensions greater than three may therefore provide the alternative of choice. 
There are a variety of meshless multivariate data interpolation schemes. 
Examples are Shepard's interpolant \citep{shepard1968}, moving least squares 
approximants \citep{lancaster1981}, or Hardy's multiquadrics \citep{hardy1971}. 

\subsubsection{Spline interpolation}
\noindent Spline interpolation \citep{schoenberg46a,schoenberg46b} is based on 
interpolating between sampling points by means of higher-order polynomials. The
coefficients of the polynomial are determined `slightly' non-locally,
such that a global smoothness in the interpolated function is
guaranteed up to some order of derivative. The order of the
interpolating polynomials is arbitrary, but in practice cubic splines
are most widely used. Cubic splines produce an interpolated function
that is continuous through the second derivative. To obtain a cubic
spline interpolation for a dataset of $N+1$ points $N$ separate cubics
are needed. Each of these cubics should have each end point match up
exactly with the end points to either side.  At the location of these
points the two adjacent cubics should also have equal first and second
derivatives. A full mathematical derivation can be found in e.g. 
\citet{geraldwheat1999,numrecipes}. 

Spline interpolation is a widely used procedure. Equalising the
derivatives has the effect of making the resulting interpolation
appear smooth and visually pleasing. For this reason splines are for
example frequently used in graphical applications.  Splines can
provide extremely accurate results when the original sample rate is
notably greater than the frequency of fluctuation in the data. Splines
however cannot deal very well with large gaps in the dataset. Because
the gap between two points is represented by a cubic, these result in
peaks or troughs in the interpolation. Also, splines are rather 
artificially defined constructs.

\subsubsection{Radial Basis Functions}
\noindent One of the most promisings schemes may be that of Radial Basis Functions, 
\citep[RBFs, see e.g.][]{powell1992,arnold1991,wendland2005}. RBFs may be used to determine a 
smooth density field interpolating three-dimensional spatial and four-dimensional spatio-temporal data sets, or
even data sets in six-dimensional phase space.  In the first step of this
approach the implicit function is computed as a linear combination of
translates of a single radial basis function.  This function is determined by
the geometric constraint that the input sample points belong to its zero set.
If the input is a density map, the geometric constraint boils down to the
implicit function interpolating the densities at the input points (and some
additional constraints preventing the construction of the zero function).
The construction of Radial Basis Functions with suitable interpolation
properties is discussed in \cite{sw-ccrbf-01}, while an early review of the mathematical
problems related to RBF-interpolation may be found in \cite{powell1992}. 
A nice overview of the state of the art in scattered data modeling using Radial 
Basis Functions may be obtained from the surveys \cite{b-airf-01,i-sdmur-02,lf-sdts-99}.

In practice variational implicit surfaces, based on Radial Basis Functions which minimize 
some global energy or curvature functional, turn out to be very flexible~\citep{dts-rsvru-02,
to-stuvi-99}: they are adaptive to curvature
variations, can be used for enhancement of fine detail and sharp features
that are missed or smoothed out by other implicit techniques, and can overcome
noise in the input data since they are approximating rather than interpolating.
Especially the use of parameter dependent or anisotropic radial basis
functions allows for graceful treatment of sharp features and provide multiple
orders of smoothness~\citep{to-rsuab-01}.

\section{Spatial Tessellations}
\begin{figure*}
  \vskip -0.0truecm
  \mbox{\hskip -0.0truecm\includegraphics[width=11.8cm]{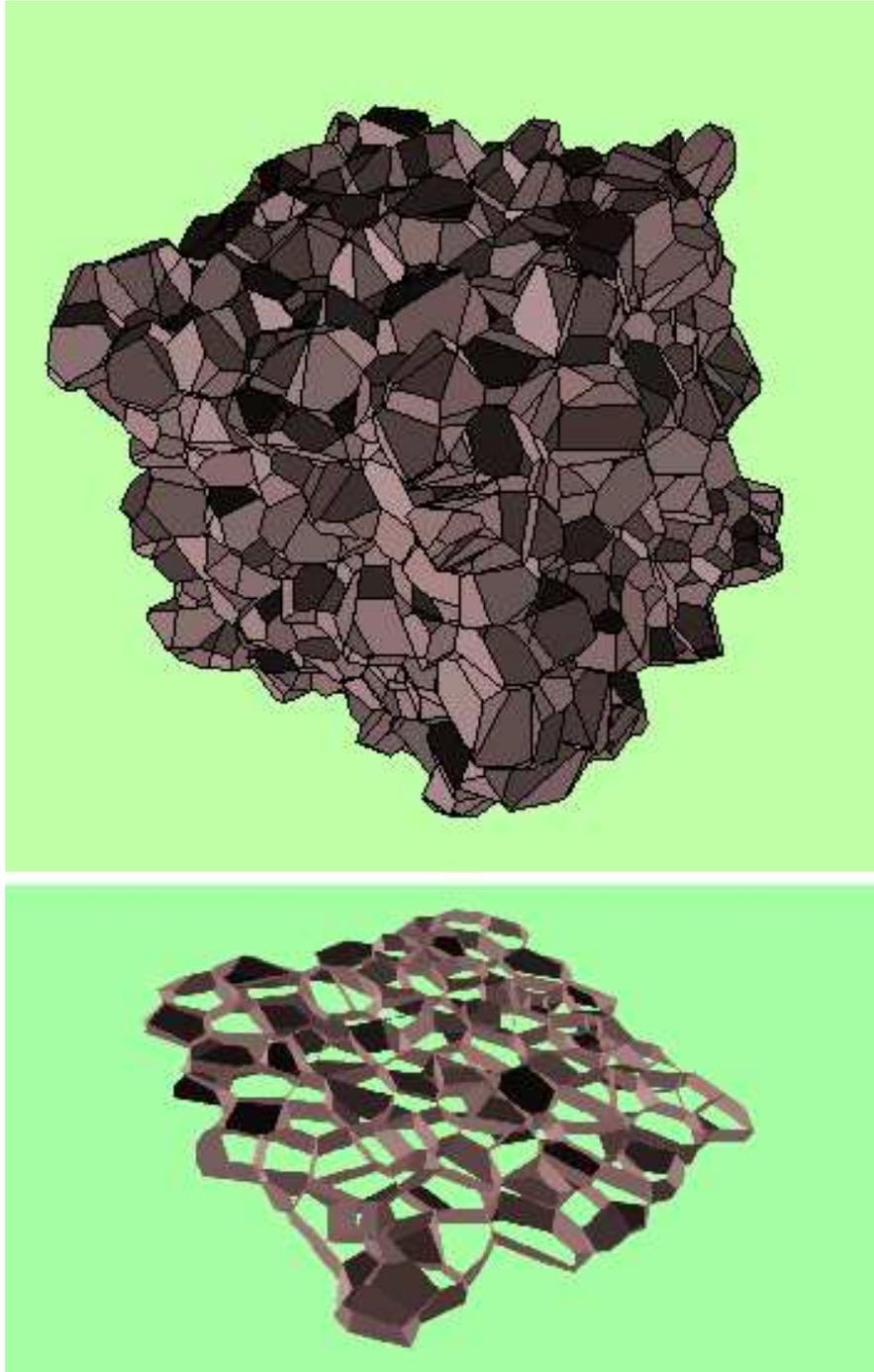}}
  \vskip -0.0truecm
\caption{A full 3-D tessellation comprising 1000 Voronoi cells/polyhedra 
generated by 1000 Poissonian distributed nuclei. Courtesy: Jacco Dankers}
\label{fig:vor3dtess}
\end{figure*} 
\subsection{Stochastic and Computational Geometry}
{\it Random spatial tessellations} are a fundamental concept in the fields of {\it Stochastic Geometry} and {\it Computational 
Geometry}. 

{\it Stochastic Geometry}, or geometric probability theory, is the subject in mathematics concerned with the problems that arise 
when we ascribe probability distributions to geometric objects such as points, lines, and planes (usually in Euclidian spaces), or 
to geometric operations such as rotations or projections \citep[see e.g.][]{stoyan1987,stoyan1995}. A formal and restricted definition 
of stochastic geometry was given by \cite{stoyan1987}, who defined the field as the branch of mathematics {\it devoted 
to the study of geometrical structures whcih can be described by random sets, in particular by point processes, in suitable 
spaces}. Since many problems in stochastic geometry are either not solvable analytically, or only in a few 
non-generic cases, we have to resort to the help of the computer to find a  (partial) answer to the problems under consideration. This makes 
it necessary to find efficient algorithms for constructing the geometrical objects involved. 

{\it Computational Geometry} is the branch of computer science that is concerned with finding the computational procedures for solving 
geometric problems, not just the geometric problems arising from stochastic geometry but geometric problems in general 
\citep{boissonnat1998,deberg2000,goodman2004}. It is concerned with the design and analysis of algorithms and software for processing geometric objects and data. Typical 
problems are the construction of spatial tessellations, like Voronoi diagrams and Delaunay meshes, the reconstruction of objects from 
finite point samples, or finding nearest neighbours in point sets. Methods from this field have many applications in applied areas like 
Computer Graphics, Computer Vision, Robotics, Computer Aided Design and Manifacturing. 

\begin{figure*}
    \begin{center}
      \mbox{\hskip -0.1truecm\includegraphics[width=12.2cm]{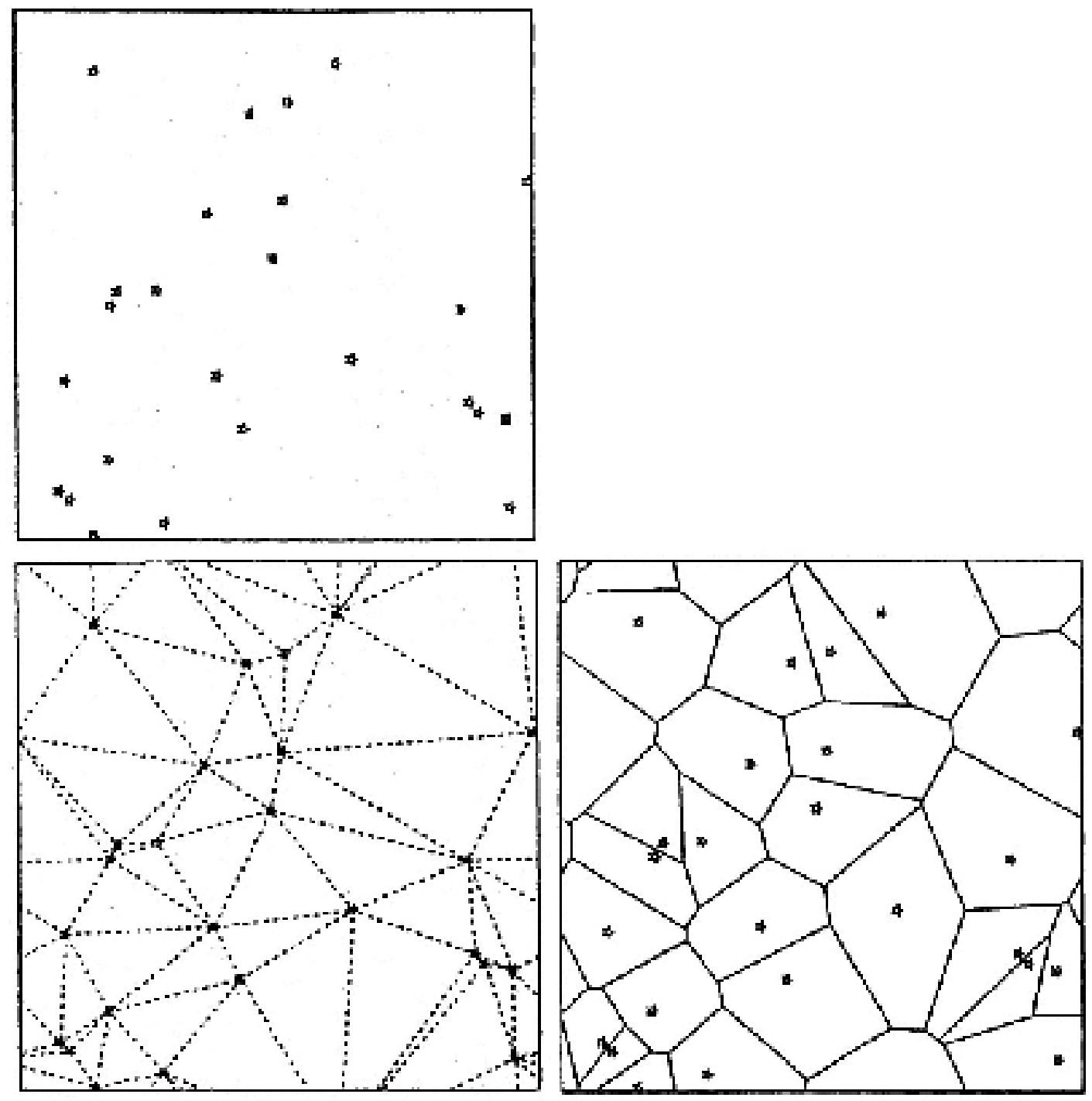}}
    \end{center}
 \vskip -0.2truecm
 \caption{An illustration of a distribution of nuclei (stars) in a square (top) and its corresponding 
Delaunay triangulation (bottom left) and Voronoi tessellation (bottom right), assuming periodic 
boundary conditions.}
  \label{fig:vordeltess}
\end{figure*} 
\subsection{Random Tessellations}
\noindent Random tessellations or mosaics occur as primary objects of various 
processes of division of space $\Re^d$ into convex cells or as secondary or 
auxiliary objects in e.g. various statistical problems.

Simply stated, a {\it tessellation} is an arrangement of polytopes (2-D: polygons, 
3-D: polyehdra) fitting together without overlapping so as to cover d-dimensional 
space $\Re^d \quad (d=1,2,\ldots$), or a subset $X \subset \Re^d$. Usually one requires
that the cells are convex and compact with disjoint interiors, although tessellations 
may also involve non-convex cells (an example are Johnson-Mehl tessellations). Posing
the additional requirement of convexity implies that all interfaces separating 
pairs of cells are planar (for 3-D), and all edges common to the boundaries of 
three or more cells are linear, implying each cell to be a (convex) polyhedron.
Formally, a {\it tessellation} in $\Re^d$ is a set ${\cal T}=\{X_i\}$ of 
$d$-dimensional sets $X_i \subset \Re^d$ called {\it cells} such that 

\begin{eqnarray}
&&{\tilde X}_i \cap {\tilde X}_j = \emptyset \qquad\hbox{for\ } i \not= j,\nonumber\\
&&\bigcup_i X_i = \Re^d,\ \\
&&\#\{X_i \in {\cal T}: X_i \cap B \not= \emptyset\} < \infty \qquad \forall \hbox{\  
bounded\ } B \subset \Re^d,\nonumber 
\end{eqnarray}
\noindent with ${\tilde X}_i$ the interior of cell $X_i$ \citep{moeller1989,moeller1994}. The first 
property implies the interiors of the cells to be disjoint, the second one that the 
cell aggregate $\{X_i\}$ is space-filling, and the third one that ${\cal T}$ is a 
countable set of cells.

\subsection{Voronoi Tessellation}
\label{sec:vortess}
\noindent The Voronoi tessellation ${\cal V}$ of a point set ${\cal P}$ is the division of space 
into mutually disjunct polyhedra, each {\it Voronoi polyhedron consisting of the part of 
space closer to the defining point than any of the other points} \citep{voronoi1908,okabe2000}.
Assume that we have a distribution of a countable set ${\cal P}$ of nuclei $\{{\bf x}_i\}$ in 
$\Re^d$. Let ${\bf x}_1,{\bf x}_2,{\bf x}_3,\ldots$ be the coordinates of the nuclei. Then 
the {\it Voronoi region} ${\cal V}_i$ of nucleus $i$ is defined by the points $\bf x$ (fig.~\ref{fig:vordeltess}), 

\bigskip
\begin{center}
\framebox[11.7truecm]{\vbox{\Large
\begin{eqnarray}
\hskip -0.5truecm
{\cal V}_i = \{{\bf x} \vert d({\bf x},{\bf x}_i) < d({\bf x},{\bf x}_j)
\quad \forall j \not= i\} \nonumber
\label{eq:voronoidef}
\end{eqnarray}}}
\begin{equation}
\end{equation}
\vskip 0.25truecm
\end{center}
\noindent where $d({\bf x},{\bf y})$ is the Euclidian distance between ${\bf x}$
and ${\bf y}$. In other words, ${\cal V}_i$ is the set of points which is nearer
to ${\bf x}_i$ than to ${\bf x}_j,\quad j \not=i$. From this basic definition, we can 
directly infer that each Voronoi region ${\cal V}_i$ is the intersection of the open 
half-spaces bounded by the perpendicular bisectors (bisecting planes in 3-D) of 
the line segments joining the nucleus $i$ and any of the the other nuclei. This  
implies a Voronoi region ${\cal V}_i$ to be a convex polyhedron (a polygon when 
in 2-D), a {\it Voronoi polyhedron}. Evidently, the concept can be extended 
to any arbitrary distance measure (Icke, priv. comm.). The relation between the 
point distribution ${\cal P}$ and its Voronoi tessellation can be clearly 
appreciated from the two-dimensional illustration in fig.~\ref{fig:vordeltess}.
\begin{table*}
\begin{center}
{\large 3-D Voronoi Tessellation Elements\\}
{(see also fig.~\ref{fig:vorelm})\ \\
\ \\}
{{
\begin{tabular}{|lll|llll|lllll|}
\hline
&&&&&&&&&&&\ \\
&&{\Large ${\cal V}_i$}&&&{\it Voronoi cell}&&&-& Polyhedron&&\\
&&&&&&&&-&defined by nucleus $i\in{\cal P}$&&\\
&&&&&&&&-&volume of space closer to $i$&&\\
&&&&&&&&&than any other nucleus $m\in{\cal P}$&&\\
&&&&&&&&&&&\ \\
&&{\Large $\Sigma_{ij}$}&&&{\it Voronoi wall (facet)}&&&-& Polygon&&\\
&&&&&&&&-&defined by nuclei $(i,j)\in{\cal P}$ &&\\
&&&&&&&&-&all points ${\bf x}$ with equal distance to $(i,j)$&&\\
&&&&&&&&&\& larger distance to any other nucleus $m \in {\cal P}$&&\\
&&&&&&&&-&constitutes part surface cells: ${\cal V}_i$, ${\cal V}_j$&&\\
&&&&&&&&&&&\ \\
&&{\Large $\Lambda_{ijk}$}&&&{\it Voronoi edge}&&&-& Line segment&&\\
&&&&&&&&-&defined by nuclei $(i,j,k)\in{\cal P}$&&\ \\
&&&&&&&&-&all points ${\bf x}$ with equal distance to $(i,j,k)$&&\ \\
&&&&&&&&&\& larger distance to any other nucleus $m \in {\cal P}$&&\\
&&&&&&&&-&constitutes part rim Voronoi cells: ${\cal V}_i$, ${\cal V}_j$, ${\cal V}_k$&&\\
&&&&&&&&&constitutes part rim Voronoi walls: $\Sigma_{ij}$, $\Sigma_{ik}$ and $\Sigma_{jk}$&&\\
&&&&&&&&&&&\ \\
&&{\Large ${\cal D}_{ijkl}$}&&&{\it Voronoi vertex}&&&-&Point&&\\
&&&&&&&&-&defined by nuclei $(i,j,k,l)\in{\cal P}$&&\ \\
&&&&&&&&-&equidistant to nuclei $(i,j,k,l)$&&\ \\
&&&&&&&&-&closer to $(i,j,k,l)$ than to any other nucleus $m\in{\cal P}$&&\ \\
&&&&&&&&-&circumcentre of (Delaunay) tetrahedron $(i,j,k,l)$&&\ \\
&&&&&&&&&&&\\
\hline
\end{tabular}}}\\
\end{center}
\label{table:vorelm}
\end{table*}
\begin{figure*}
    \begin{center}
      \mbox{\hskip -0.2truecm\includegraphics[width=11.9cm]{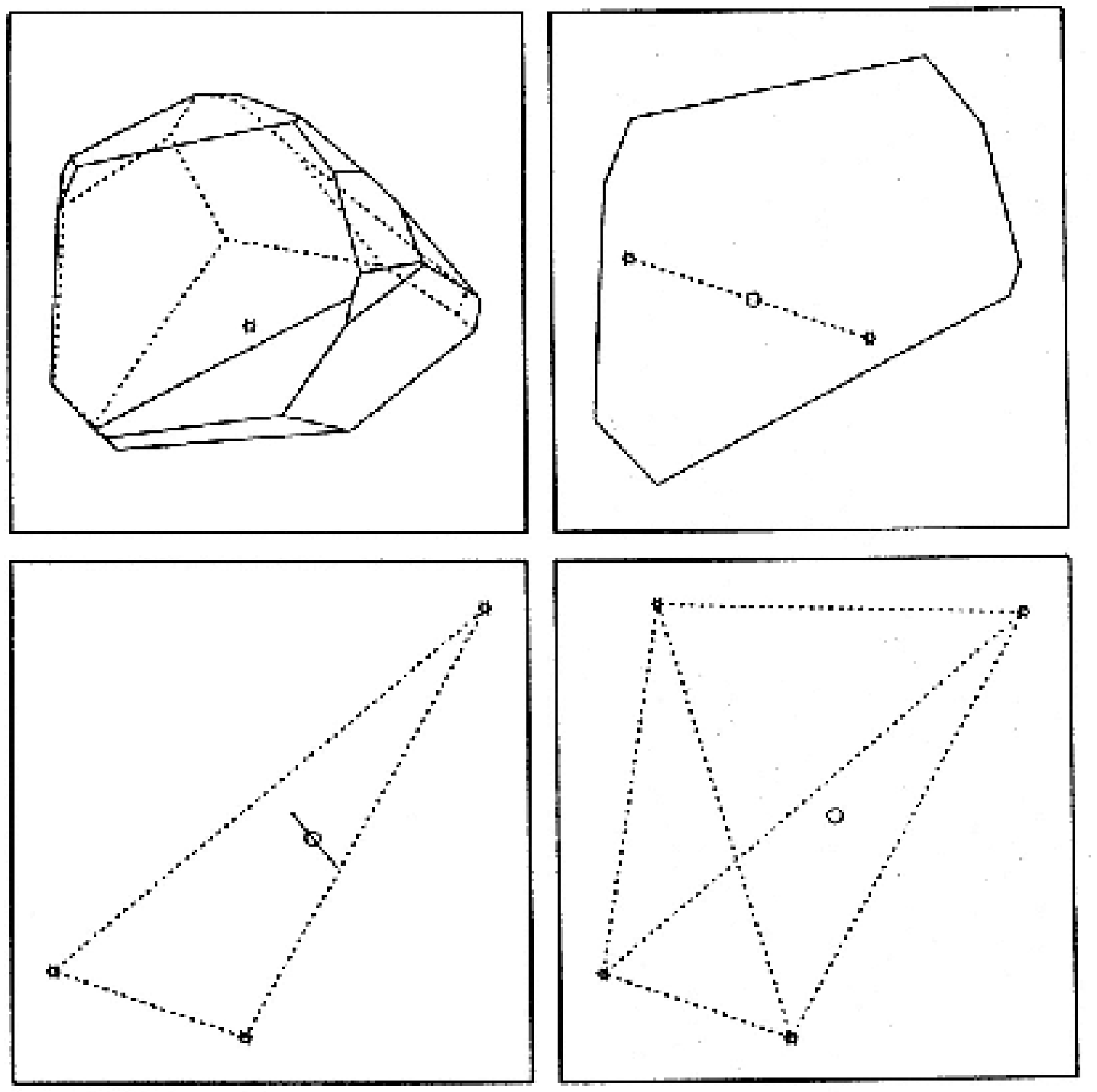}}
    \end{center}
 \caption{The four Voronoi elements of a Voronoi tessellation generated by a nucleus set {\cal P}. See 
table~\ref{table:vorelm}}
  \label{fig:vorelm}
\end{figure*} 
\noindent The complete set of Voronoi polyhedra constitute a space-filling tessellation of
mutually disjunct cells, the {\it Voronoi tessellation} ${\cal V}({\cal P})$ relative to 
${\cal P}$. A good impression of the morphology of a complete Voronoi tessellation can 
be seen in fig.~\ref{fig:vor3dtess}, a tessellation of 1000 cells generated by a Poisson 
distribution of 1000 nuclei in a cubic box. The Voronoi foam forms a packing of Voronoi 
cells, each cell being a convex polyhedron enclosed by the bisecting planes between the 
nuclei and their {\it natural neighbours}.  

\subsubsection{Voronoi elements}
Taking the three-dimensional tessellation as the archetypical representation
of structures in the physical world, the Voronoi tessellation ${\cal V}({\cal P}$ 
consists of four constituent {\it elements}: {\it Voronoi cells}, {\it Voronoi 
walls}, {\it Voronoi edges} and {\it Voronoi vertices}. Table~\ref{table:vorelm} 
provides a listing of these elements together with a description of their 
relation with respect to the nuclei of the generating point set ${\cal P}$, 
augmented by the accompanying illustration in fig.~\ref{fig:vorelm}. 

\subsubsection{Generalized Voronoi Tessellations}
The Voronoi tessellation can be generalized. \cite{miles1970} defined the
{\it generalized Voronoi tessellation ${\cal V}_n$}. The original Voronoi 
tessellation is ${\cal V}_1={\cal V}$.

This involves the extension of the definition of the Voronoi cell ${\cal V}_i$ generated 
by one nucleus $i$ to that of a {\it higher-order} Voronoi cell ${\cal V}^{k}(i_{1},\ldots,i_{k})$ 
generated by a set of $k$ nuclei $\{i_{1},\ldots,i_{k}\}\in{\cal P}$. Each $k-order$ Voronoi cell 
${\cal V}^{k}(i_{1},\ldots,i_{k})$ consists of that part of space in which the points 
${\bf x}$ have the $k$ nuclei $\{i_{1},\ldots,i_{k}\}\in{\cal P}$ as their $k$ nearest 
neighbours. In addition to \cite{miles1970,miles1972,miles1982} see \citet[][chapter 3]{okabe2000} 
for more details and references.

\subsubsection{Voronoi Uniqueness}
An inverse look at tessellations reveals the special and degenerate nature of Voronoi 
tessellations. Given a particular tessellation one might wonder whether there is a point 
process which would have the tessellation as its Voronoi tessellation. One may demonstrate 
that in general this is not true. By defining a geometric procedure to reconstruct 
the generating nucleus distribution one may straightforwardly infer that there is not 
a unique solution for an arbitrary tessellation. 

This may be inferred from the study by \cite{chiuweystoy1996}, who defined and developed a nucleus 
reconstruction procedure to demonstrate that a two-dimensional section through a three- or higher-dimensional 
Voronoi tessellation will itself not be a Voronoi tessellation. By doing so their work clarified the 
special and degenerate nature of these tessellations. 

\subsection{the Delaunay Tessellation}
\label{sec:delaunay}
Pursuing our census of Voronoi tessellation elements (table~\ref{table:vorelm}), we found that each set of 
nuclei ${i,j,k,l}$ corresponding to a Voronoi vertex ${\cal D}(i,j,k,l)$ defines a unique 
tetrahedron. This is known as {\it Delaunay tetrahedron} \citep{delaunay1934}. 

Each {\it Delaunay tetrahedron} is defined by the {\it set of four points whose circumscribing 
sphere does not contain any of the other points} in the generating set 
\citep[][triangles in 2D: see fig.~\ref{fig:vordeltess}]{delaunay1934}. For the countable set 
${\cal P}$ of points $\{{\bf x}_i\}$ in $\Re^d$, a 
Delaunay tetrahedron ${\cal D}_m$ is the simplex $T$ defined by $(1+d)$ points 
$\{{\bf x}_{i1},\ldots,{\bf x}_{i(d+1)}\}\,\in\,{\cal P}$ (the vertices of this $d$-dimensional 
tetrahedron) such that the corresponding circumscribing sphere ${\cal S}_m({\bf y}_m)$ with 
circumcenter ${\bf C}_m$ and radius $R_m$ does not contain any other point of ${\cal P}$,

\bigskip
\begin{center}
\framebox[11.7truecm]{\vbox{\Large
\begin{eqnarray}
\hskip 0.5truecm
{\cal D}_M\,=\,T({\bf x}_{i1},\ldots,{\bf x}_{i(d+1)}) 
\qquad&\hbox{with}& \ d({\bf C}_m,{\bf x}_j)\,>\,R_m \nonumber \\ 
&\forall& j \not= i1,\ldots,i(d+1)\,\nonumber
\label{eq:delaunaydef}
\end{eqnarray}}}
\begin{equation}
\end{equation}
\end{center}
\bigskip

\noindent Following this definition, the {\it Delaunay tessellation} of a point set ${\cal P}$ is 
the uniquely defined and volume-covering tessellation of mutually disjunct {\it Delaunay tetrahedra}.  

Figure~\ref{fig:vordeltess} depicts the Delaunay tessellation resulting from a given 2-D distribution 
of nuclei. On the basis of the figure we can immediately observe the intimate relation between a 
{\it Voronoi tessellation} and its {\it Delaunay tessellation}. 
The Delaunay and Voronoi tessellations are like the opposite sides of the same coin, they are 
each others {\it dual}: one may directly infer one from the other and vice versa. The combinatorial 
structure of either tessellation is completely determined from its dual.

\begin{figure*}[b]
    \begin{center}
      \mbox{\hskip -0.2truecm\includegraphics[width=12.2cm]{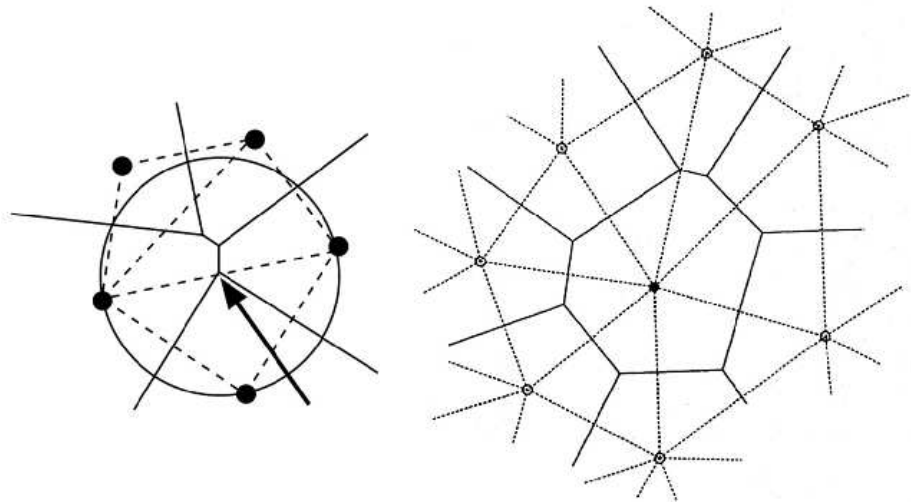}}
    \end{center}
 \caption{The {\it dual} relationship between Voronoi (solid) and Delaunay (dashed) tessellations of 
a set of nuclei (circles). Left: Zoom-in on the three Delaunay triangles corresponding to a set 
of five nuceli (black dots) and the corresponding Voronoi edges. The circle is the circumcircle 
of the lower Delaunay triangle, its centre (arrow) is a vertex of the Voronoi cell. Note that the 
circle does not contain any other nucleus in its interior ! Right: a zoom in on the Voronoi cell 
${\cal V}_i$ of nucleus $i$ (black dot). The Voronoi cell is surrounded by its related 
Delaunay triangles, and clearly delineates its {\it Natural Neighbours} (open circles).}
  \label{fig:vordelcont}
\end{figure*} 

The duality between Delaunay and Voronoi tessellations may be best appreciated on behalf of the 
following properties: 
\begin{itemize}
\item[$\bullet$] {\it Circumcentre \& Voronoi vertex}:\\
The center of the circumsphere of a Delaunay tetrahedron is a vertex of the Voronoi tessellation. 
This follows from the definition of the Voronoi tessellation, wherein the four nuclei which form 
the Delaunay tetrahedron are equidistant from the vertex.
\item[$\bullet$] {\it Contiguity condition}:\\
The circumsphere of a Delaunay tetrahedron is empty and cannot contain any nucleus in the 
set ${\cal P}$. If there would be such a nucleus it would be closer to the centre than 
the four tetrahedron defining nuclei. This would render it impossible for it being the 
vertex of all corresponding Voronoi cells.
\end{itemize}

\subsubsection{Natural Neighbours}
A pair of nuclei $i$ and $j$ whose Voronoi polyhedra ${\cal V}_i$ and ${\cal V}_j$
have a face in common is called a {\it contiguous pair} and a member of the
pair is said to be {\it contiguous} to the other member. Contiguous pairs of nuclei 
are each other's {\it natural neighbour}. 

{\it Natural neighbours} of a point $i$ are the points $j$ with whose Voronoi ${\cal V}_j$ 
its Voronoi cell ${\cal V}_i$ shares a face or, in other words the points with which 
it is connected via a Delaunay tetrahedron. This unique set of neighbouring points 
defines the neighbourhood of the point and represents the cleanest definition of the 
surroundings of a point (see fig.~\ref{fig:vordelcont}), an aspect which turns out to be 
of {\it seminal} importance for the local interpolation method(s) discussed in this 
contribution.. 

\subsection{Voronoi and Delaunay Statistics}
\label{sec:vordelstat}
\noindent In particular for practical applications the knowledge of statistical 
properties of Voronoi and Delaunay tessellations as a function of the generating 
stochastic point processes is of considerable interest. However, despite their seemingly 
simple definition it has proven remarkably difficult to obtain solid 
analytical expressions for statistical properties of Voronoi tessellations. 
Moreover, with the exception of some general tessellation properties nearly all 
analytical work on Voronoi and Delaunay tessellations has concentrated on those 
generated by homogeneous Poissonian point distributions. 

Statistical knowledge of Delaunay tessellations generated by Poissonian
nuclei in $d$- dimensional space is relatively complete. Some
important distribution functions are known, mainly due to the work of 
\cite{miles1970,miles1972,miles1974,kendall1989,moeller1989,moeller1994}. For 
Voronoi tessellations, even for Poissonian nuclei analytical results are quite 
rare. Only very few distribution functions are known, most results are 
limited to a few statistical moments: expectation values, variances and 
correlation coefficients. Most of these results stem from the pioneering works 
of \cite{meyering1953,gilbert1962,miles1970,moeller1989}. \cite{moeller1989} provides  
analytical formulae for a large number of first-order moments of $d$-dimensional 
Poisson-Voronoi tessellations, as well as of $s$-dimensional section through them, 
following the work of \cite{miles1972,miles1974,miles1982} \citep[also see][]{weygaertphd1991}. 
Numerical Monte Carlo evaluations have proven te be the main 
source for a large range of additional statistical results. For an extensive listing 
and discussion we refer the reader to \cite{okabe2000}. 

Of prime importance for the tessellation interpolation techniques  discussed in this review 
are two fundamental characteristics: (1) the number of natural neighbours of 
the generating nuclei and (2) the volume of Voronoi and Delaunay cells. 

In two dimensions each point has on average exactly $6$ natural neighbours \citep[see e.g.][]{ickewey1987}, 
irrespective of the character of the spatial point distribution. This also implies that 
each point belongs to on average $6$ Delaunay triangles. Going from two to three dimensions, 
the character of the tessellation changes fundamentally:
\begin{itemize}
\item[$\bullet$] {\it Dependence Point Process}:\\ the average number of natural neighbours is 
no longer independent of the underlying point distribution
\item[$\bullet$] {\it Integer}:\\ The average number of natural neighbours is not an integer:\\ 
for a Poisson distribution it is $\sim 13.4$ ! 
\item[$\bullet$] {\it Delaunay tetrahedra/Voronoi Vertices}:\\
For a Poisson distribution, the \# vertices per Voronoi cell $\sim 27.07$.\\ 
While in the two-dimensional case the number of vertices per Voronoi cell has same value as 
the number of {\it natural neighbours}, in three dimensions it is entirely different !\\ 
\end{itemize}
As yet it has not been possible to derive, from first principle, a closed 
analytical expression for the distribution functions of the volumes of Voronoi 
polyhedra and Delaunay cells in a Poisson Voronoi tessellation. However, the fitting formula 
suggested by \cite{kiang1966} has proven to represent a reasonably good approximation. Accordingly, 
the probability distribution of the volume of of a Voronoi polyhedron in $d$-dimensional space 
$\Re^d$ follows a gamma distribution
\begin{equation} 
   f_{\cal V}(V_{\cal V})\,{\rm d} V_{\cal V}\,=\,\frac{q}{\Gamma(q)}\,\,\left(q \frac{V_{\cal V}}{\langle V_{\cal V}\rangle}\right)^{(q-1)}\,
\exp{\left(-q\,\frac{V_{\cal V}}{\langle V_{\cal V}\rangle}\right)}\,\,{\rm d}{\left(\frac{V_{\cal V}}{\langle V_{\cal V}\rangle}\right)}\,,
\label{eq:vorvolpdf}
\end{equation}
\noindent with $V_{\cal V}$ the size of the Voronoi cell and $\langle V_{\cal V}\rangle$ the average cell size. The conjecture 
of \cite{kiang1966} is that the index has a value $q=2d$ for a tessellation in $d$-dimensional space (ie. $q=4$ for 
2-D space and $q=6$ for 3-D space). Even though other studies 
indicated slightly different values, we have found the Kiang suggestion to be quite accurate (see 
sect.~\ref{sec:samplingnoise}, also see \cite{schaapphd2007}).

While the distribution for Voronoi cell volumes involves a conjecture, for Delaunay cells ${\cal D}$ it is 
possible to derive the ergodic distribution from first principle. Its $d+1$ 
vertices completely specify the Delaunay tetrahedron ${\cal D}$. If ${\bf c}$, 
$R$ its circumradius, its vertices are the points $\{{\bf c} + R {\bf u}_i\}$. 
In this the unit vectors $\{{\bf u}_i\}$ $(i=0,\ldots,d+1)$, directed towards 
the vertices, determine the shape of the Delaunay tetrahedron. Given that 
$\Delta_d$ is the volume of the unity simplex $\{{\vec u}_0,\ldots,{\vec u}_d\}$, 
and the volume $V_{\cal D}$ of the Delaunay tetrahedron given by 
\begin{equation}
V,=\,\Delta_d \,R^d\,.
\end{equation}
\noindent \cite{miles1974,moeller1989,moeller1994} found that for a Poisson point process of 
intensity $n$ in $d$-dimensional space $\Re^d$ the distribution is specified by 
\begin{eqnarray}
f_{\cal D}({\cal D})&\,=\,&f_{\cal D}(\{{\vec u}_0,\ldots,{\vec u}_d\},R)\nonumber\\
\ \\
&\,=\,&a(n,d)\,\Delta_d\,\,R^{d^2-1}\,\exp{\left(-n \omega_d\,R^d\right)}\nonumber\,,
\label{eq:delvolpdf}
\end{eqnarray}
\noindent with $\omega_d$ the volume of the unity sphere in $d$-dimensional
space ($\omega_2=2\pi$ and $\omega_3=4\pi$) and $a(n,d)$ is a constant dependent on number density $n$ and 
dimension $d$. 
In other words, the circumradius $R$ of the Delaunay tetrahedron is ergodically independent 
of its shape, encapsulated in $\Delta_d$. From this one finds that the 
the distribution law of $n \omega_d R^d$ is the $\chi^2_{2d}/2$
distribution \citep{kendall1989}:
\begin{equation}
f(R)\,{\rm d}R\,=\,\frac{(n \omega_d R^d)^{d-1}}{(d-1)!}\,\,\exp{(-n \omega_d R^d)}\,\,
{\rm d}(n \omega_d R^d)\,.
\end{equation}
\medskip
It is of considerable importance to realize that even for a uniform density field, 
represented by a discrete point distribution of intensity $n$, neither the Voronoi 
nor the Delaunay tessellation will involve a volume distribution marked by a 
considerable spread and skewness. 

\begin{figure*}
     \begin{minipage}{\textwidth}
    \begin{center}
      \mbox{\hskip -0.2truecm\includegraphics[width=12.0 cm]{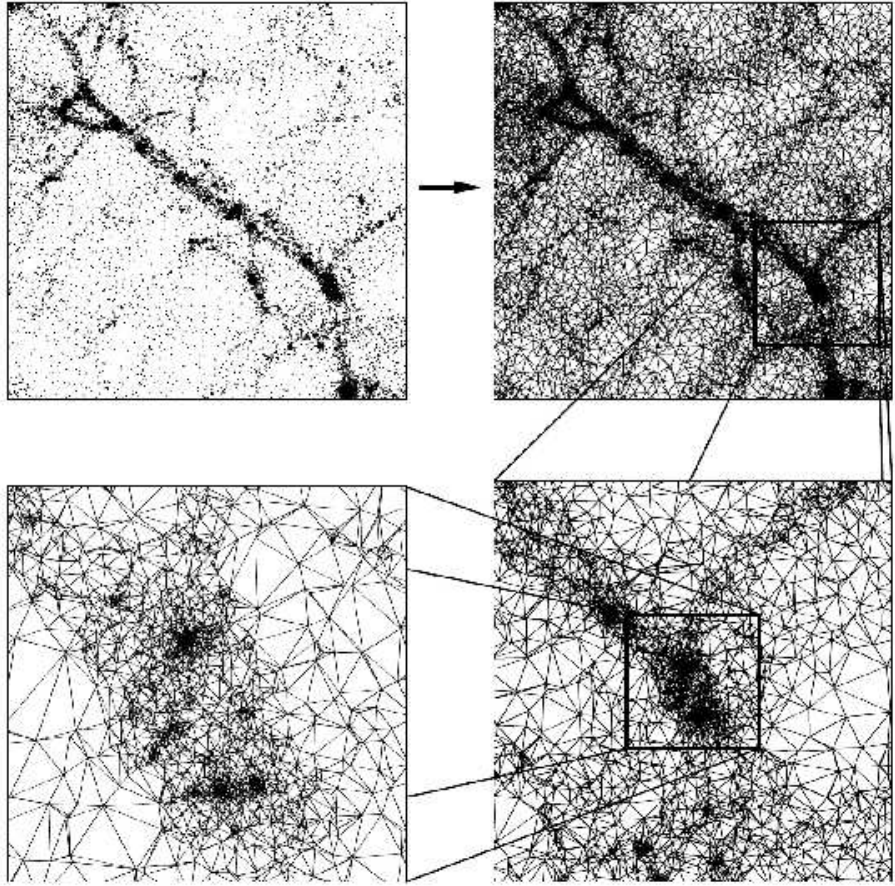}}
    \end{center}
  \end{minipage} 
 \caption{The Delaunay tessellation of a point distribution in and around a filamentary feature. The 
generated tessellations is shown at three successive zoom-ins. The frames form a testimony of the 
strong adaptivity of the Delaunay tessellations to local density and geometry of the spatial point 
distribution. }
  \label{fig:delpntadapt}
\end{figure*} 

\subsection{Spatial Adaptivity}
The distribution of the Delaunay and Voronoi cells adjusts itself to the characteristics of the point 
distribution: in sparsely sampled regions the distance between the {\it natural neighbours} is large, 
and also if the spatial distribution is anisotropic this will be reflected in their distribution. This 
may be readily appreciated from fig.~\ref{fig:delpntadapt}. At three successive spatial scales the 
Delaunay tessellations has traced the density and geometry of the local point distribution to a 
remarkable degree. On the basis of these observations it is straightforward to appreciate that the 
corresponding Delaunay tessellation forms an ideal adaptive multi-dimensional interpolation grid. 

Note that not only the size, but also the shape of the Delaunay simplices is fully determined
by the spatial point distribution. The density and geometry of the local point distribution 
will therefore dictate the resolution of spatial interpolation and reconstruction 
procedures exploiting the Delaunay tessellation. The prime representatives of such 
methods are the Natural Neighbour techniques (see sect.~\ref{sec:nn}) and the Delaunay Tessellation 
Field Estimator. 

These techniques exploit the fact that a Delaunay tetrahedron may be regarded as the optimal multidimensional 
interpolation interval. The corresponding minimal coverage characteristics of the Delaunay tessellation thus 
imply it to be optimal for defining a network of multidimensional interpolation intervals. The resulting interpolation 
kernels of Natural Neighbour and DTFE interpolation not only embody an optimal spatial resolution, but also involve a 
high level of adaptivity to the local geometry of the point distribution (see sect.~\ref{sec:nn} and sect.~\ref{sec:dtfe}). 
Within this context it is no coincidence that in many computer visualisation applications Delaunay tessellations have 
acquired the status of optimal triangulation. Moreover, the superb spatial adaptivity of the volumes of Delaunay and Voronoi 
polyhedra to the local point density may be readily translated into measures for the 
value of the local density. This forms a crucial ingredient of the DTFE formalism (see sect.~\ref{sec:dtfe}).

\subsection{Voronoi and Delaunay Tessellations: context}
The earliest significant use of Voronoi regions seems to have occurred in the work of \cite{dirichlet1850} and 
\cite{voronoi1908} in their investigations on the reducibility of positive definite quadratic forms. However, Dirichlet 
and Voronoi tessellations 
as applied to random point patterns appear to have arisen independently in various fields of science and technology 
\citep{okabe2000}.  For example, in crystallography, one simple model of crystal growth starts with a fixed
collection of sites in two- and three-dimensional space, and allows crystals to begin growing from each site, 
spreading out at a uniform rate in all directions, until all space is filled. The ``crystals'' then consist of all
points nearest to a particular site, and consequently are just the Voronoi regions for the original set of points. 
Similarly, the statistical analysis  of metereological data led to the formulation of Voronoi 
regions under the name {\it Thiessen polygons} \citep{thiessen1911}. 

Applications of Voronoi tessellations can therefore be found in fields as diverse as agriculture and forestry  
\citep{fischermiles1973}, astrophysics \citep[e.g.][]{kiang1966,ickewey1987,ebeling1993,bernwey96,molchanov1997,schaapwey2000,
cappellari2003,ritzericke2006}, ecology, zoology and botany, cell biology \citep{shapiro2006}, protein research \citep{liang1998a,liang1998b,
liang1998c}, cancer research \citep{torquato2000,schaller2005}, chemistry, crystal growth and structure, materials science 
\citep[see e.g.][incl. many references]{torquato2002}, geophysics \citep{sbm1995}, geography and geographic information 
systems \citep{boots1984,gold1997}, communication theory \citep{baccelli1999} and art and archaeology \citep{kimia2001,leymarie2003}. 
Due to the diversity of these applications it has acquired a set of alternative names, such as Dirichlet regions, Wigner-Seitz cells, and 
Thiessen figures. 

\section{Natural Neighbour Interpolation}
\label{sec:nn}
The {\it Natural Neighbour Interpolation} formalism is a generic higher-order multidimensional 
interpolation, smoothing and modelling procedure utilizing the concept of natural neighbours to obtain 
locally optimized measures of system characteristics. Its theoretical basis was developed and 
introduced by \cite{sibson1981}, while extensive treatments and elaborations of nn-interpolation may be 
found in \cite{watson1992,sukumarphd1998}. 
\begin{figure*}
    \begin{center}
      \mbox{\hskip -0.2truecm\includegraphics[width=12.2cm]{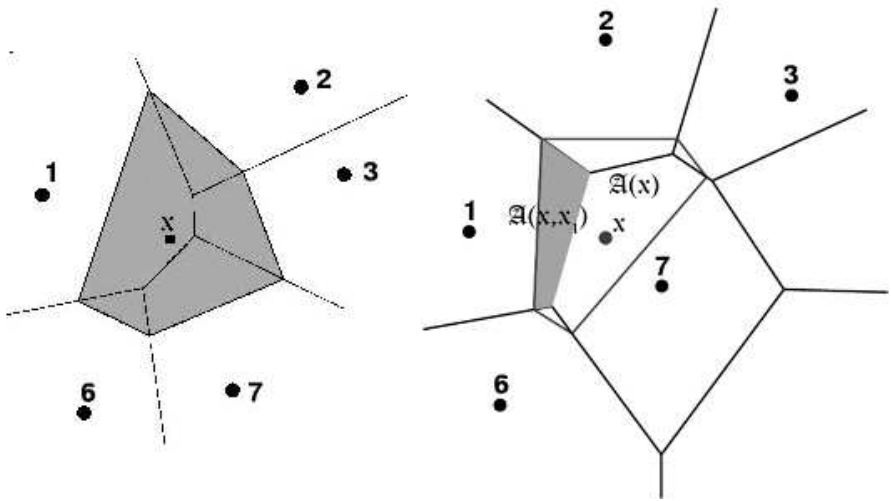}}
    \end{center}
 \caption{Natural Neighbour Interpolants. Example of NN interpolation in 
two dimensions. Left: the Voronoi cell ${\cal V}({\bf x}$ generated by a 
point ${\bf x}$. Right: the 2nd order Voronoi cell ${\cal V}_2({\bf x},{\bf x}_1)$, the region 
of space for which the points ${\bf x}$ and ${\bf x_j}$ are the closest points. 
Image courtesy M. Sambridge and N. Sukumar. Also see Sambridge, Braun \& McQueen 1995 and Sukumar 1998.}
  \label{fig:sibson1}
\end{figure*} 
Natural neighbour interpolation produces a conservative, artifice-free, result 
by finding weighted averages, at each interpolation point, of the functional values associated with 
that subset of data which are natural neighbors of each interpolation point. Unlike other schemes, 
like Shepard's interpolant \citep{shepard1968}, where distance-based weights are used, the Sibson natural neighbour 
interpolation uses area-based weights. According to the nn-interpolation scheme the 
interpolated value ${\widehat f}({\bf x})$ at a position ${\bf x}$ is given by 
\begin{equation}
{\widehat f}({\bf x})\,=\,\sum_i\,\phi_{nn,i}({\bf x})\,f_i\,,
\label{eq:nnint}
\end{equation}
\begin{figure*}
    \begin{center}
      \mbox{\hskip -0.1truecm\includegraphics[width=12.2cm]{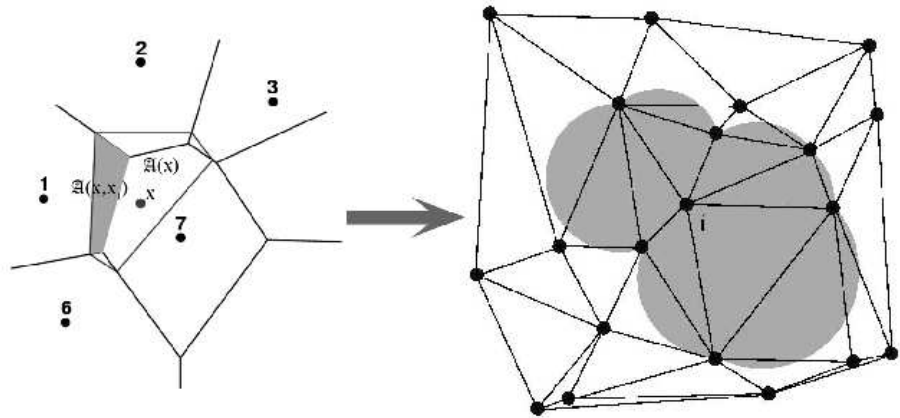}}
    \end{center}
 \caption{Natural Neighbour Interpolants. Example of NN interpolation in 
two dimensions. The basis of the nn-interpolation kernel $\phi_{nn,1}$, 
the contribution of nucleus ${\bf x}_1$ to the interpolated field values. 
Image courtesy M. Sambridge and N. Sukumar. Also see Sambridge, Braun \& McQueen 1995 and Sukumar 1998.}
  \label{fig:sibson2}
\end{figure*} 
\begin{figure*}[b]
    \begin{center}
      \mbox{\hskip -0.2truecm\includegraphics[width=12.0cm]{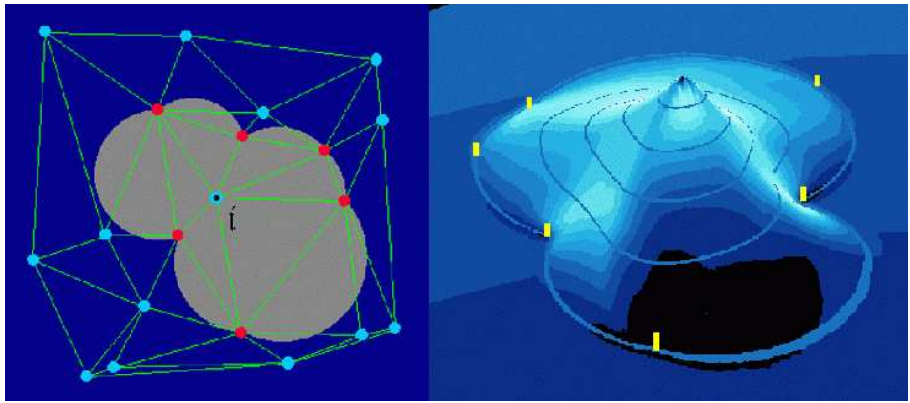}}
    \end{center}
 \caption{Natural Neighbour Interpolants. Example of NN interpolation in 
two dimensions. 3-D illustration of the nn-interpolation kernel $\phi_{nn,1}$, the contribution 
of nucleus ${\bf x}_1$ to the interpolated field values. Image courtesy M. Sambridge, also see Braun \& Sambridge 1995. 
Reproduced by permission of Nature.}
  \label{fig:sibson3}
\end{figure*} 
in which the summation is over the natural neighbours of the point ${\bf x}$, i.e. the 
sample points $j$ with whom the order-2 Voronoi cells ${\cal V}_2({\bf x},{\bf x}_j)$ are 
not empty (fig.~\ref{fig:sibson1},~\ref{fig:sibson2}). Sibson interpolation is based upon 
the interpolation kernel $\phi({\bf x},{\bf x}_j)$ to be equal to the normalized order-2 Voronoi cell, 
\begin{equation}
\phi_{nn,i}({\bf x})\,=\,{\displaystyle {{\cal A}_{2}({\bf x},{\bf x}_i)} \over 
{\displaystyle {\cal A}({\bf x})}}\,,
\label{eq:nnintint}
\end{equation}
in which ${\cal A}({\bf x})=\sum_j {\cal A}({\bf x},{\bf x}_j)$ is the volume of the 
potential Voronoi cell of point ${\bf x}$ if it had been added to the point sample ${\cal P}$ and  
the volume ${\cal A}_{2}({\bf x},{\bf x}_i)$ concerns the order-2 Voronoi cell 
${\cal V}_2({\bf x},{\bf x}_i)$, the region of space for which the points ${\bf x}$ and ${\bf x_i}$ 
are the closest points.  Notice that the interpolation kernels $\phi$ are always positive and sum to one. 
\begin{figure*}[t]
    \begin{center}
      \mbox{\hskip -0.2truecm\includegraphics[width=12.0cm]{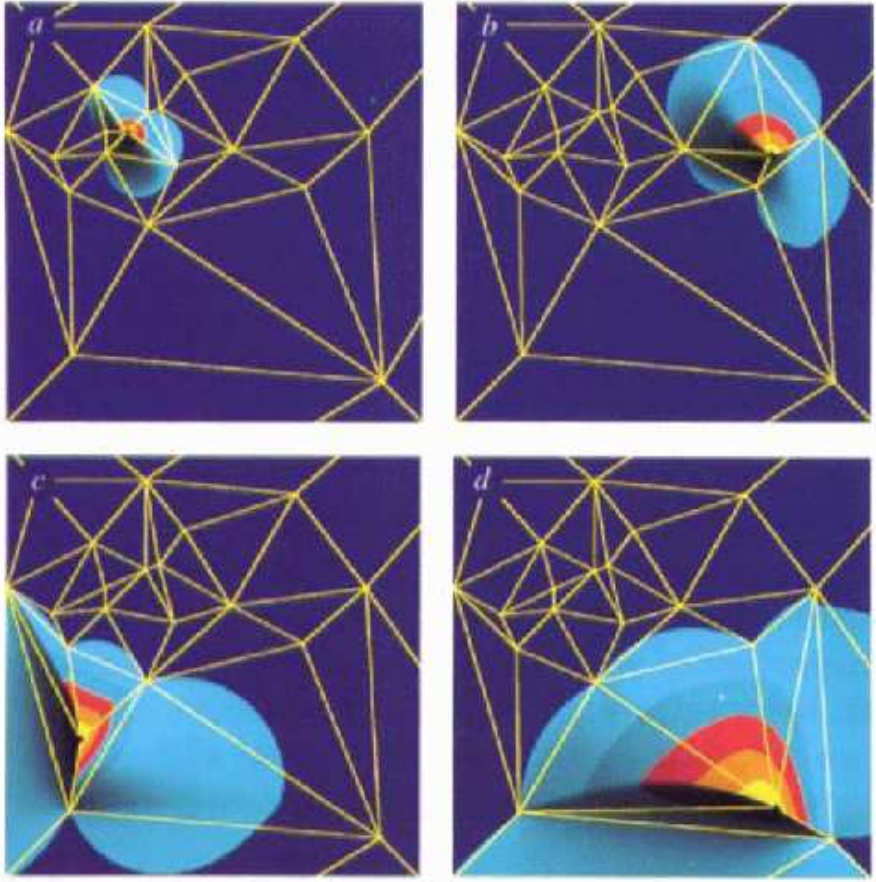}}
    \end{center}
 \caption{Natural Neighbour Kernels: illustration of the locally defined natural neighbour 
kernels $\phi_{nn,j}$ for four different nuclei $j$. Image courtesy M. Sambridge, 
also see Braun \& Sambridge 1995. Reproduced by permission of Nature.}
  \label{fig:sibson4}
\end{figure*} 

The resulting function is continuous everywhere within the convex hull of the data, and has a continuous 
slope everywhere except at the data themselves (fig.~\ref{fig:sibson3}, fig.~\ref{fig:sibson4}). Beautiful two-dimensional examples 
of nn-interpolation applications testify of its virtues (see fig.~\ref{fig:landscapint}). 

An interesting study of its performance, in comparison with other interpolation/approximation methods, concerned 
the test involving data acquired by the Mars Orbiter Laser Altimeter (MOLA), one of the instruments on the 
Mars Global Surveyor (deceased November 2006). Applying various schemes towards 
the reconstruction of a terrain near the Korolev crater, a large crater in the north polar region of Mars, 
demonstrated that nn-neighbour interpolation produced the most impressive reproduction of the original terrain 
\citep[]{abramov2004}. In comparison to the other methods -- including nearest neighbour intp., spline intp., and 
linear DTFE type interpolation -- the nn-neighbour map not only looks most natural but also proves to contain fewer 
artefacts, both in number and severity. The test reconstructions of the Jokulsa and Fjollum region of Iceland in 
fig.~\ref{fig:landscapint} provide ample proof. 

Natural neighbour interpolation may rightfully be regarded as the most general and robust method 
for multidimensional interpolation available to date. This smooth and local spatial interpolation technique 
has indeed gradually acquired recognition as a dependable, optimal, local method. For the two-dimensional 
applications it has seen highly interesting applications in geophysical fluid dynamics calculations 
\citep{braunsambridge1995,sbm1995}, and in equivalent schemes for solid mechanics problems \citep{sukumarphd1998}. 
Both applications used n-n interpolation to solve partial differential equations, showing its great potential in the field 
of computational fluid mechanics. 

While almost optimal in the quality of its reconstructions, it still involves a heavy computational effort. This 
is certainly true for three or higher dimensional spaces. This has prodded us to define a related local, adaptive 
method which can be applied to large astronomical datasets. 

\section{DTFE: the Delaunay Tessellation Field Estimator}
\label{sec:dtfe}
For three-dimensional samples with large number of points, akin to those found in large cosmological computer simulations, 
the more complex geometric operations involved in the pure nn-neighbour interpolation still represent a computationally 
challenging task. To deal with the large point samples consisting of hundreds of thousands to several millions of points 
we chose to follow a related nn-neigbhour based technique that restricts itself to pure linear interpolation. 
DTFE uses to obtain optimal {\it local} estimates of the spatial density \citep[see][sect 8.5]{okabe2000}, while 
the tetrahedra of its {\it dual} Delaunay tessellation are used as multidimensional intervals for {\it linear} interpolation of 
the field values sampled or estimated at the location of the sample points \citep[ch. 6]{okabe2000}. The DTFE technique allows 
us to follow the same geometrical and structural adaptive properties of the higher order nn-neighbour methods while allowing 
the analysis of truely large data sets. The presented tests in this review will demonstrate that DTFE indeed is able to highlight 
and analyze essential elements of the cosmic matter distribution.

\subsection{DTFE: the Point Sample}
\noindent The DTFE procedure, outlined in the flow diagram in fig.~\ref{fig:dtfescheme}, involves a sequence of steps. 
It starts with outlining the discrete point sample ${\cal P}$ in $d$-dimensional space $Re^d$,  
\begin{equation}
{\cal P}=\{{\bf x}_1,\ldots,{\bf x}_N\}\,,
\end{equation}
In the applications described in this study we restrict ourselves to Euclidian spaces, in 
particular $2-$ or $3-$dimensional Euclidian spaces. At the locations of the points in the countable 
set ${\cal P}$ the field values $\{f({\bf x}_i),i=1,\ldots,N\}$ are sampled, or can be estimated on 
the basis of the spatial point distribution itself. The prime example of the latter is when the field 
$f$ involves the density field itself. 

On the basis of the character of the sampled field $f({\bf x})$ we need to distinguish two options. 
The first option is the one defined in \cite{bernwey96}, the straightforward multidimensional linear interpolation of 
measured field values on the basis of the Delaunay tessellation. \cite{schaapwey2000} and \cite{schaapphd2007} 
extended this to the recovery of the density or intensity field. DTFE is therefore characterized by a second option, 
the ability to reconstruct the underlying density field from the discrete point process itself. 

The essential extension of DTFE wrt. \cite{bernwey96} is that it allows the option of using the {\it point sample 
process} ${\cal P}$ itself as a measure for the value of the {\it density} at its position. The latter 
poses some stringent requirements on the sampling process. It is crucial for the discrete spatial point 
distribution to constitute a fair sample of the underlying continuous density field. In other words, 
the discrete point sample ${\cal P}$ needs to constitute a general Poisson point process of the density 
field. 

Such stringent requirements on the spatial point process ${\cal P}$ are not necessary when the sampled field 
has a more generic character. As long as the sample points are spread throughout most of the sample volume 
the interpolation procedure will yield a properly representative field reconstruction. It was this situation 
with respect to cosmic velocity fields which lead to the original definition of the Delaunay spatial 
interpolation procedure forming the basis of DTFE \citep{bernwey96,bernwey97}. 

\subsection{DTFE: Linear Interpolation}
At the core of the DTFE procedure is the use of the Delaunay tessellation of the discrete point 
sample process (see sect.~\ref{sec:delaunay}) as an adaptive spatial linear interpolation grid. 
Once the Delaunay tessellation of ${\cal P}$ is 
determined, it is used as a {\it multidimensional linear interpolation grid} for a 
field $f({\bf r})$. That is, each Delaunay tetrahedron is supposed to be a region with a constant 
field gradient, $\nabla f$. 

The linear interpolation scheme of DTFE exploits the same spatially adaptive characteristics of the Delaunay tessellation 
generated by the point sample ${\cal P}$ as that of regular natural neighbour 
schemes (see sect.~\ref{sec:nn}, eqn.~\ref{eq:nnint}). For DTFE the interpolation kernel $\phi_{dt,i}({\bf x})$ 
is that of regular linear interpolation within the Delaunay tetrahedron in which ${\bf x}$ is located,
\begin{equation}
{\widehat f}_{dt}({\bf x})\,=\sum_i\,\phi_{dt,i}({\bf x})\,f_i,
\label{eq:dtfeint}
\end{equation}
in which the sum is over the four sample points defining the Delaunay tetrahedron. Note that not only the size, but also 
the shape of the Delaunay simplices is fully determined by the spatial point distribution. As a result the resolution of 
the DTFE procedure depends on both the density and geometry of the local point distribution. Not only does the 
DTFE kernel embody an optimal spatial resolution, it also involves a high level of adaptivity to the local geometry 
of the point distribution (see sect.~\ref{sec:dtfekernel}).

Also note that for both the nn-interpolation as well as for the linear DTFE interpolation, 
the interpolation kernels $\phi_i$ are unity at sample point location ${\bf x}_i$ and 
equal to zero at the location of the other sample points $j$ (e.g. see fig.~\ref{fig:sibson3}),
\begin{eqnarray}
\phi_i({\bf x}_j)\,=\,
{\begin{cases}
1\hskip 2.truecm {\rm if}\,i=j\,,\\
0\hskip 2.truecm {\rm if}\,i\ne j\,,
\end{cases}}\,
\end{eqnarray}
where ${\bf x}_j$ is the location of sample point $j$.

In practice, it is convenient to replace eqn.~\ref{eq:dtfeint} with its equivalent expression 
in terms of the (linear) gradient ${\widehat {\nabla f}}\bigr|_m$ inside the Delaunay simplex $m$,
\begin{equation}
{\widehat f}({\bf x})\,=\,{\widehat f}({\bf x}_{i})\,+\,{\widehat {\nabla f}} \bigl|_m \,\cdot\,({\bf x}-{\bf x}_{i}) \,.
\label{eq:fieldval}
\end{equation}
\noindent The value of ${\widehat {\nabla f}}\bigr|_m$ can be easily and uniquely determined from the $(1+D)$ field values $f_j$ at 
the sample points constituting the vertices of a Delaunay simplex. Given the location ${\bf r}=(x,y,z)$ of the four 
points forming the Delaunay tetrahedra's vertices, ${\bf r}_0$, ${\bf r}_1$, ${\bf r}_2$ and ${\bf r}_3$, 
and the value of the sampled field at each of these locations, $f_0$, $f_1$, $f_2$ and $f_3$ and defining the 
quantities 
\begin{eqnarray}
\Delta x_n &\,=\,& x_n-x_0\,;\nonumber\\
\Delta y_n &\,=\,& y_n -y_0\,;\qquad \hbox{for}\ n=1,2,3\\ 
\Delta z_n &\,=\,&z_n -z_0\nonumber
\end{eqnarray}
\noindent as well as $\Delta f_n\,\equiv\,f_n-f_0\,(n=1,2,3)$ the gradient $\nabla f$ follows from the inversion
\begin{eqnarray}
\nabla f\,=\,
\begin{pmatrix}
{\displaystyle \partial f \over \displaystyle \partial x}\\
\ \\
{\displaystyle \partial f \over \displaystyle \partial y}\\
\ \\
{\displaystyle \partial f \over \displaystyle \partial z}
\end{pmatrix}
\,=\,{\bf A}^{-1}\,
\begin{pmatrix}
\Delta f_{1} \\ \ \\ \Delta f_{2} \\ \ \\ \Delta f_{3} \\ 
\end{pmatrix}\,;\qquad
{\bf A}\,=\,
\begin{pmatrix}
\Delta x_1&\Delta y_1&\Delta z_1\\
\ \\
\Delta x_2&\Delta y_2&\Delta z_2\\
\ \\
\Delta x_3&\Delta y_3&\Delta z_3
\end{pmatrix}
\label{eq:fieldgrad}
\end{eqnarray}
\noindent Once the value of $\nabla_f$ has been determined for each Delaunay tetrahedron in the tessellation, it is 
straightforward to determine the DTFE field value ${\widehat f}({\bf x})$ for any location ${\bf x}$ by means 
of straightforward linear interpolation within the Delaunay tetrahedron in which ${\bf x}$ is located 
(eqn.~\ref{eq:fieldval}).

The one remaining complication is to locate the Delaunay tetrahedron ${\cal D}_m$ in which a particular 
point ${\bf x}$ is located. This is not as trivial as one might naively think. It not necessarily concerns 
a tetrahedron of which the nearest nucleus is a vertex. Fortunately, a very efficient method, the {\it 
walking triangle algorithm}~\citep{lawson1977,sloan1987} has been developed. Details of the method may be 
found in \cite{sbm1995,schaapphd2007}. 

\subsection{DTFE: Extra Features}
While DTFE in essence is a first order version of {\it Natural Neighbour Interpolation} procedure, following the same adaptive 
multidimensional interpolation characteristics of the Delaunay grid as the higher-order nn-neighbour techniques, it also 
incorporates significant extensions and additional aspects. In particular, DTFE involves two extra and unique features 
which are of crucial importance for the intended cosmological context and application:
\begin{itemize}
\item[$\bullet$] {\it Volume-weighted}:\\
The interpolated quantities are {\it volume-weighted}, instead of the implicit {\it mass-weighted} averages 
yielded by conventional grid interpolations.
\item[$\bullet$] {\it Density estimates}:\\
The spatial adaptivity of the Delaunay/Voronoi tessellation to the underlying point distribution is used 
to estimate the local density. 
\end{itemize}
\begin{figure*} 
  \vskip 0.5truecm
  \mbox{\hskip -1.5truecm\includegraphics[width=22.0cm,angle=90.0]{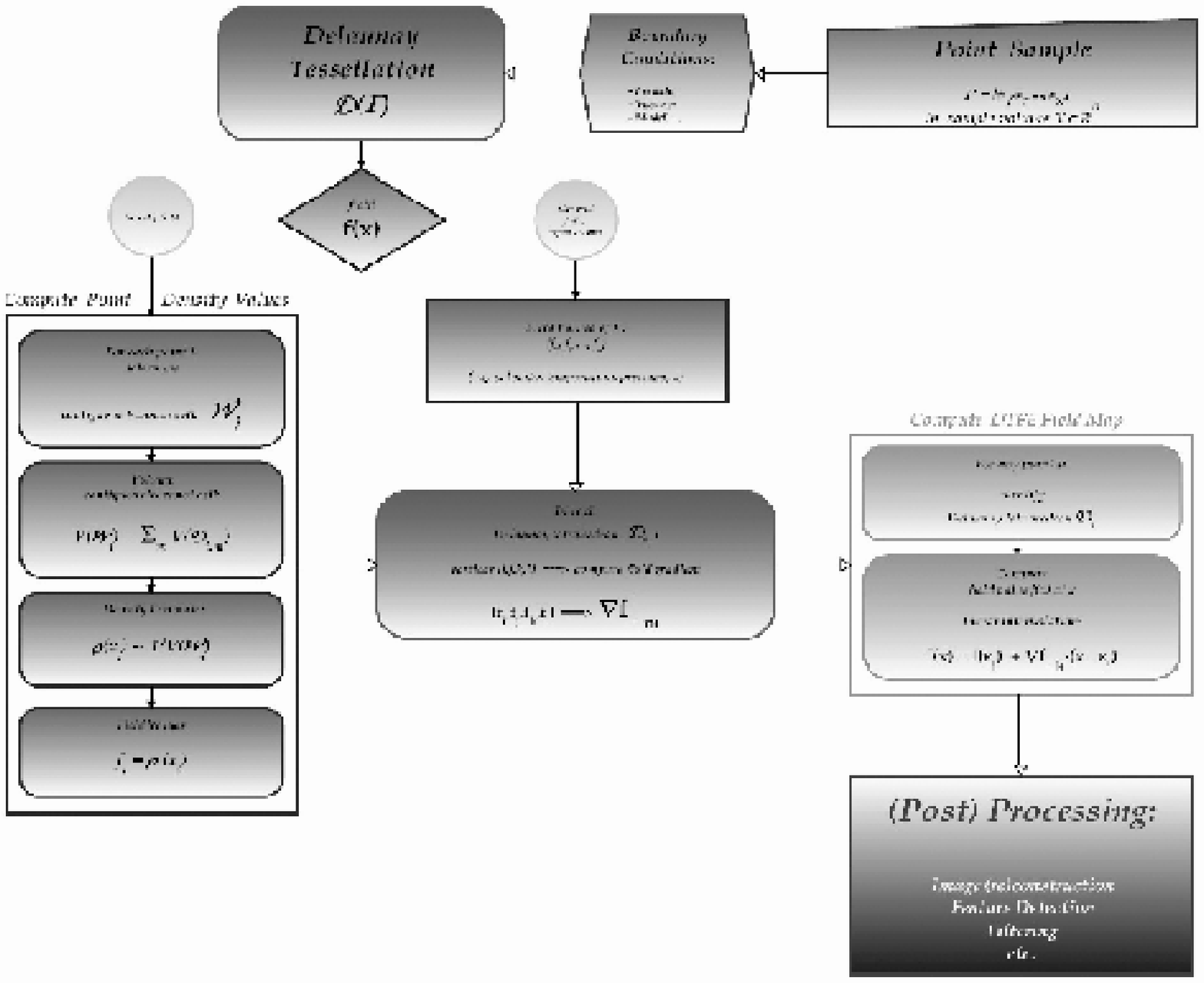}}
  \caption{Flow diagram illustrating the essential ingredients of the DTFE procedure.}
\label{fig:dtfescheme} 
\end{figure*} 
\begin{figure*}
  \centering
    \includegraphics[width=11.5cm]{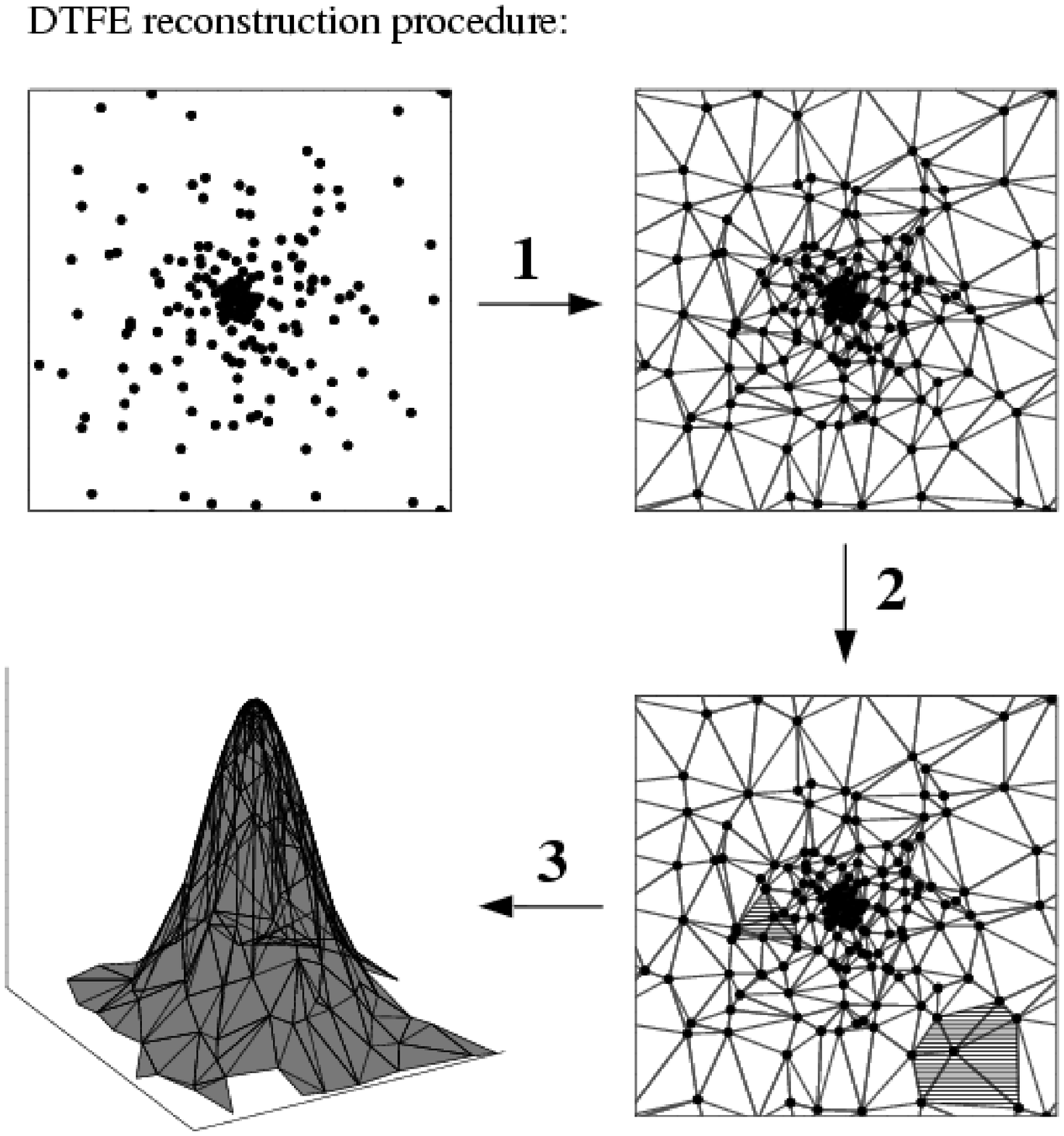}
    \caption{Summary: overview of the essential steps of the DTFE reconstruction
    procedure. Given a point distribution (top left), one has to
    construct its corresponding Delaunay tessellation (top right),
    estimate the density at the position of the sampling points by
    taking the inverse of the area of their corresponding contiguous
    Voronoi cells (bottom right) and finally to assume that the
    density varies linearly within each Delaunay triangle, resulting
    in a volume-covering continuous density field (bottom left).}
\label{fig:dtfepanel} 
\end{figure*} 
\section{DTFE: the field reconstruction procedure}
\label{sec:dtfe_proc}
The complete DTFE reconstruction procedure, essential steps of which are illustrated 
in Fig.~\ref{fig:dtfepanel}, can be summarized in terms of the flow diagram 
in Fig.~\ref{fig:dtfescheme} and consists of the following sequence of steps. 
\begin{enumerate}
\item[$\bullet$] {\bf Point sample}\\
Defining the spatial distribution of the point sample:
\begin{enumerate}
\item[+] {\it Density field}:\\ point sample needs 
to be a general Poisson process of the (supposed) underlying density field, i.e. 
it needs to be an unbiased sample of the underlying density field.
\item[+] {\it General (non-density) field}:\\
no stringent requirements upon the stochastic representativeness of the sampling 
process will be necessary except that the sample volume is adequately covered.
\end{enumerate}
\medskip
\item[$\bullet$] {\bf Boundary Conditions}\\
An important issue, wrt the subsequent Delaunay tessellation computation and the self-consistency 
of the DTFE density and velocity field reconstructions, is that of the assumed 
boundary conditions. These will determine the Delaunay and Voronoi cells that overlap 
the boundary of the sample volume. Dependent upon the sample at hand, a variety of 
options exists:\\
\begin{enumerate}
\item[+] {\it Vacuum boundary conditions:}\\ outside the sample volume there are no points. This will lead 
to infinitely extended (contiguous) Voronoi cells surrouding sample points near the boundary. Evidently, 
these cells cannot be used for DTFE density field estimates and field interpolations: the volume of the 
DTFE reconstruction is smaller than that of the sample volume. \\
\item[+] {\it Periodic boundary conditions:}\\ the point sample is supposed to be repeated periodically 
in boundary boxes, defining a toroidal topology for the sample volume. The resulting Delaunay and 
Voronoi tessellations are also periodic, their total volume exactly equal to the sampel volume. While specific 
periodic tessellation algorithms do exist~\citep{weygaert1994}, this is not yet true for most available 
routines in standard libraries. For the analysis of N-body 
simulations this is the most straightforward and preferrable choice.\\
\item[+] {\it Buffer conditions:}\\ the sample volume box is surrounded by a bufferzone filled with 
a synthetic point sample. The density of the synthetic buffer point sample should be related to the 
density in the nearby sample volume. The depth of the bufferzone depends on the density of the synthetic point 
sample, it should be sufficiently wide for any Delaunay or Voronoi cell related to a sample point not 
to exceed the bufferzone. A clear condition for a sufficiently deep bufferzone has been specified by 
\cite{neyrinck2005}.\\
\ \\
When involving velocity field analysis, the velocities of the buffer points should also follow the velocity 
distribution of the sample and be in accordance with the continuity equation. Relevant examples of possible 
choices are:\\
\itemitem{-} {\it internal:} the analyzed sample is a subsample embedded within a large sample volume, 
a sufficient number of these sample points outside the analysis volume is taken along in the DTFE 
reconstruction.
\itemitem{-} {\it random cloning technique:} akin to the technique described by \cite{yahil1991}.
\itemitem{-} {\it constrained random field:} realizations employing the existing correlations in the 
field \citep[][]{edbert1987,hofmrib1991,weyedb1996,zaroubi1995}.
\end{enumerate}
\medskip
\item[$\bullet$] {\bf Delaunay Tessellation}\\
Construction of the Delaunay tessellation from the point sample (see fig.~\ref{fig:delpntadapt}). 
While we still use our own Voronoi-Delaunay code~\citep{weygaert1994}, at present there is a score 
of efficient library routines available. Particularly noteworthy is the \cgal initiative, 
a large library of computational geometry routines\footnote{\cgal is a \texttt{C++} library of 
algorithms and data structures for Computational Geometry, see \url{www.cgal.org}.}\\
\medskip
\item[$\bullet$] {\bf Field values point sample}\\
Dependent on whether it concerns the densities at the sample points or a 
measured field value there are two options:
\begin{enumerate}
\item[+] {\it General (non-density) field}:\\ (Sampled) value of field at sample point. \\
\item[+] {\it Density field}:
The density values at the sampled points are determined from the corresponding Voronoi tessellations.
The estimate of the density at each sample point is the normalized inverse of the volume of its {\it contiguous} 
Voronoi cell ${\cal W}_i$ of each point $i$. The {\it contiguous Voronoi cell} of a point $i$ is the union of 
all Delaunay tetrahedra of which point $i$ forms one of the four vertices (see fig.~\ref{fig:voronoicont} for an 
illustration). We recognize two applicable situations:\\
\itemitem{-} {\it uniform sampling process}: the point sample is an unbiased sample of the underlying density field. Typical 
example is that of $N$-body simulation particles. For $D$-dimensional space the density estimate is, 

\begin{equation}
{\widehat \rho}({\bf x}_i)\,=\,(1+D)\,\frac{w_i}{V({\cal W}_i)} \,.
\label{eq:densvor}
\end{equation}
\noindent with $w_i$ the weight of sample point $i$, usually we assume the same 
``mass'' for each point. \\
\itemitem{-} {\it systematic non-uniform sampling process}: sampling density according to specified 
selection process quantified by an a priori known selection function $\psi({\bf x})$, varying as 
function of sky position $(\alpha,\delta)$ as well as depth/redshift. For $D$-dimensional space the 
density estimate is , 
\begin{equation}
{\widehat \rho}({\bf x}_i)\,=\,(1+D)\,\frac{w_i}{\psi({\bf x}_i)\,V({\cal W}_i)} \,.
\label{eq:densvornu}
\end{equation}
\end{enumerate}
\medskip
\item[$\bullet$] {\bf Field Gradient}\\
 Calculation of the field gradient estimate $\widehat{\nabla f}|_m$ 
 in each $D$-dimensional Delaunay simplex $m$ ($D=3$: tetrahedron; $D=2$: triangle) 
by solving the set of linear equations for the field values at the positions 
of the $(D+1)$ tetrahedron vertices,\\
\begin{eqnarray}
\widehat{\nabla f}|_m \ \ \Longleftarrow\ \ 
{\begin{cases}
f_0 \ \ \ \ f_1 \ \ \ \ f_2 \ \ \ \ f_3 \\
\ \\
{\bf r}_0 \ \ \ \ {\bf r}_1 \ \ \ \ {\bf r}_2 \ \ \ \ {\bf r}_3 \\
\end{cases}}\,
\label{eq:dtfegrad}
\end{eqnarray}
Evidently, linear interpolation for a field $f$ is only meaningful when the field 
does not fluctuate strongly. Particularly relevant for velocity field reconstructions 
is that there should be no orbit crossing flows within the volume of the Delaunay 
cell which would involve multiple velocity values at any one location. In other words, 
DTFE velocity field analysis is only significant for {\it laminar} flows. 

Note that in the case of the sampled field being the velocity field ${\bf v}$ we may 
not only infer the velocity gradient in a Delaunay tetrahedron, but also the directly 
related quantities such as the {\it velocity divergence}, {\it shear} and {\it vorticity}. 
\medskip
\begin{figure}
  \centering
    \mbox{\hskip -0.1truecm\includegraphics[width=11.8cm]{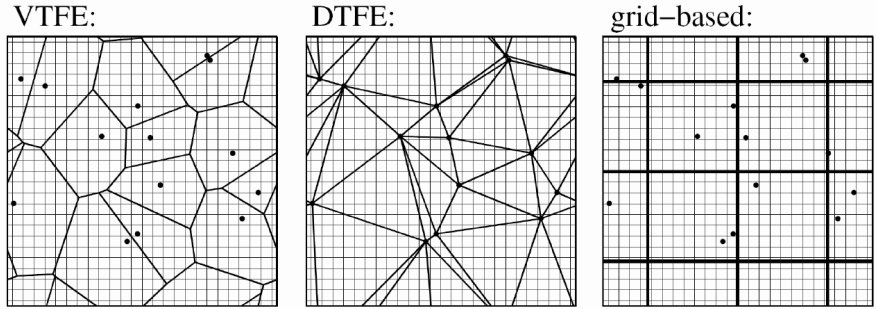}}
    \caption{Two-dimensional 
    display-grids in the VTFE, DTFE and grid-based reconstruction
    methods. The grid is overlaid on top of the basic underlying
    structure used to reconstruct the density field. SPH-like methods
    are not shown, because of the inherent difficulty in visualizing
    their underlying structure, which does not consist of a
    subdivision of space in distinct non-overlapping structural
    elements, but of circles of different radius at each position in
    space.}
\label{fig:dtfevtfegrid} 
\end{figure} 
\item[$\bullet$] {\bf Interpolation}.\\
The final basic step of the DTFE procedure is the field interpolation. The processing 
and postprocessing steps involve numerous interpolation calculations, for each of the 
involved locations ${\bf x}$.
\medskip
Given a location ${\bf x}$, the Delaunay tetrahedron $m$ in which it is embedded is 
determined. On the basis of the field gradient $\widehat{\nabla f}|_m$ the field value 
is computed by (linear) interpolation (see eq.~\ref{eq:fieldval}), 
\begin{equation}
{\widehat f}({\bf x})\,=\,{\widehat f}({\bf x}_{i})\,+\,{\widehat {\nabla f}} \bigl|_m \,\cdot\,({\bf x}-{\bf x}_{i}) \,.
\end{equation}
In principle, higher-order interpolation procedures are also possible. Two relevant high-order procedures are:
\itemitem{-} Spline Interpolation
\itemitem{-} Natural Neighbour Interpolation (see eqn.~\ref{eq:nnint} and eqn.~\ref{eq:nnintint}).\\
Implementation of natural neighbour interpolation, on the basis of \cgal routines, is presently in progress. 
\bigskip
\begin{figure}
  \vskip -0.25truecm
  \centering
    \mbox{\hskip -0.5truecm\includegraphics[width=12.5cm]{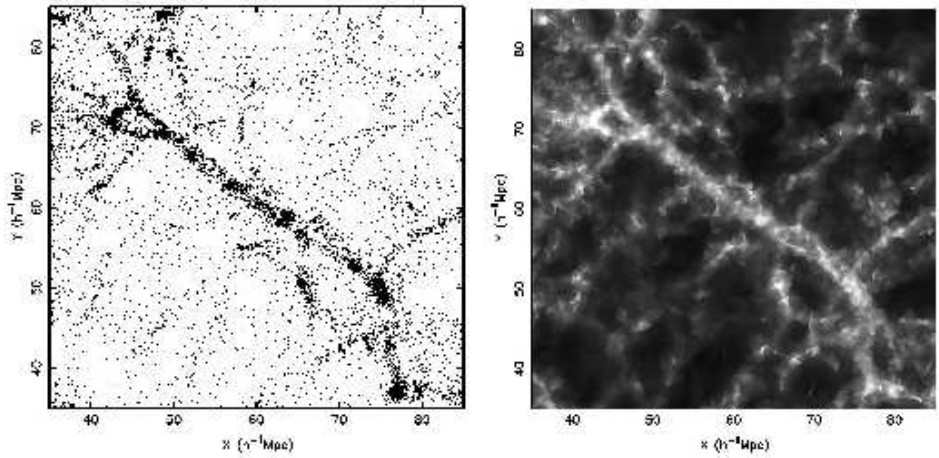}}
    \caption{Image of a characteristic filamentary region in the
 outcome of a cosmological $N$-body simulation. Left-hand
 frame: particle distribution in a thin slice through the simulation
 box. Right-hand frame: two-dimensional slice through the three-dimensional DTFE density
 field reconstruction. From \citep{schaapphd2007}.}
\label{fig:filpartdtfe} 
\end{figure} 
\item[$\bullet$] {\bf Processing}.\\
Though basically of the same character for practical purposes we make a distinction between 
straightforward processing steps concerning the production of images and simple smoothing 
filtering operations on the one hand, and more complex postprocessing on the other hand. 
The latter are treated in the next item. Basic to the processing steps is the determination 
of field values following the interpolation procedure(s) outlined above.\\ 
Straightforward ``first line'' field operations are {\it ``Image reconstruction''} and, 
subsequently, {\it ``Smoothing/Filtering''}.\\
\begin{enumerate}
\item[+] {\it Image reconstruction}.\\ For a set of {\it image points}, usually grid points, 
determine the {\it image value}: formally the average field value within the corresponding gridcell.
In practice a few different strategies may be followed, dictated by accuracy requirements. These 
are:\\
\itemitem{-} {\it Formal geometric approach}: integrate over the field values within 
each gridcell. This implies the calculation of the intersection of the relevant 
Delaunay tetrahedra and integration of the (linearly) running field values within the 
intersectio. Subsequently the integrands of each Delaunay intersection are added and 
averaged over the gridcell volume.
  \itemitem{-} {\it Monte Carlo approach}: approximate the integral by taking the 
average over a number of (interpolated) field values probed at randomly distributed 
locations within the gridcell around an {\it image point}. Finally average over the obtained field values 
within a gridcell. 
   \itemitem{-} {\it Singular interpolation approach}: a reasonable and usually satisfactory alternative to the 
formal geometric or Monte Carlo approach is the shortcut to limit the field value calculation to that at 
the (grid) location of the {\it image point}. This offers a reasonable approximation for gridcells which 
are smaller or comparable to that of intersecting Delaunay cells, on the condition the field gradient within 
the cell(s) is not too large.

\item[+] {\it Smoothing} and {\it Filtering}:
\itemitem{-} Linear filtering of the field ${\widehat f}$: convolution of the field 
${\widehat f}$ with a filter function $W_s({\bf x},{\bf y})$, usually user-specified, 
   \begin{equation}
     f_s({\bf x})\,=\,\int\,{\widehat f}({\bf x'})\, W_s({\bf x'},{\bf y})\,d{\bf x'}     
   \end{equation}
\itemitem{-} Median (natural neighbour) filtering: the DTFE density field is adaptively 
smoothed on the bais of the median value of densities within the {\it
contiguous Voronoi cell}, the region defined by the a point and its
{\it natural neighbours} \citep[see][]{platen2007}.
\itemitem{-} (Nonlinear) diffusion filtering: filtering of (sharply defined) features in 
images by solving appropriately defined diffusion equations \citep[see e.g][]{mitra2000}.
\end{enumerate}
\medskip
\begin{figure} 
  \vskip -0.25truecm
  \centering
  \mbox{\hskip -0.15truecm\includegraphics[width=12.0cm]{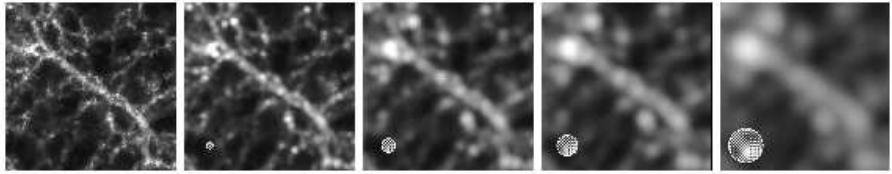}}
    \caption{Filtering of density fields. The left-hand frame
depicts the original DTFE density field. The subsequent frames show
filtered DTFE density fields. The FWHM of the Gaussian filter is indicated by the shaded circle 
in the lower left-hand corner of these frames. From \citep{schaapphd2007}.}
    \vskip -0.5truecm
\label{fig:filfilt} 
\end{figure} 
\medskip
\item[$\bullet$] {\bf Post-processing}.\\
The real potential of DTFE fields may be found in sophisticated applications, 
tuned towards uncovering characteristics of the reconstructed fields. 
An important aspect of this involves the analysis of structures in the 
density field. Some notable examples are:\\
  \begin{enumerate}
    \item[+] Advanced filtering operations. Potentially interesting 
applications are those based on the use of wavelets \citep{martinez2005}.
    \item[+] Cluster, Filament and Wall detection by means of the 
{\it Multiscale Morphology Filter} \citep{aragonphd2007,aragonmmf2007}.
    \item[+] Void identification on the basis of the {\it cosmic watershed} 
algorithm~\citep{platen2007}.
    \item[+] Halo detection in N-body simulations \citep{neyrinck2005}.
    \item[+] The computation of 2-D surface densities for the study of 
gravitational lensing \citep{bradac2004}.
  \end{enumerate} 
\medskip
In addition, DTFE enables the simultaneous and combined analysis of density 
fields and other relevant physical fields. As it allows the simultaneous 
determination of {\it density} and {\it velocity} fields, it can serve as 
the basis for studies of the dynamics of structure formation in the cosmos. 
Its ability to detect substructure as well as reproduce the morphology 
of cosmic features and objects implies DTFE to be suited for assessing their
dynamics without having to invoke artificial filters.\\
\begin{enumerate}
  \item[+] DTFE as basis for the study of the full {\it phase-space} structure of structures 
and objects. The phase-space structure dark haloes in cosmological 
structure formation scenarios has been studied by \cite{arad2004}. 
\end{enumerate}
\end{enumerate}

\section{DTFE: densities and velocities}
\begin{figure} 
  \centering
    \vskip -0.5truecm
    \mbox{\hskip -0.25truecm\includegraphics[height=12.4cm,angle=270.0]{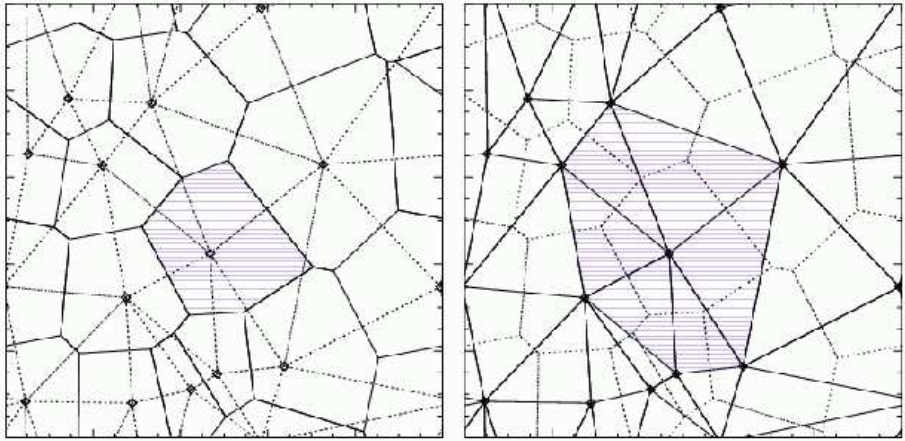}}
    \vskip 0.0truecm
    \caption{Illustration of a contiguous Voronoi cell. A contiguous Voronoi cell 
is the union of all Delaunay tetrahedra of which a point $i$ is one of the vertices.}
    \vskip -0.5truecm
\label{fig:voronoicont} 
\end{figure} 
\subsection{DTFE density estimates}
The DTFE procedure extends the concept of interpolation of field values sampled 
at the point sample ${\cal P}$ to the estimate of the density ${\widehat \rho}({\bf x})$ 
from the spatial point distribution itself. This is only feasible if the 
spatial distribution of the discrete point sample forms a fair and 
unbiased reflection of the underlying density field. 

It is commonly known that an optimal estimate for the spatial density at the location of a point ${\bf x}_i$ in a 
discrete point sample ${\cal P}$ is given by the inverse of the volume of the corresponding 
Voronoi cell \citep[see][section 8.5, for references]{okabe2000}. Tessellation-based methods for estimating the density have 
been introduced by \cite{brown1965} and \cite{ord1978}. In astronomy, \cite{ebeling1993} were the first to use tessellation-based density 
estimators for the specific purpose of devising source detection algorithms. This work has recently been applied to cluster detection 
algorithms by \citep{ramella2001,kim2002,marinoni2002,lopes2004}. Along the same lines, \cite{ascalbin2005} suggested that the use of 
a multidimensional binary tree might offer a computationally more efficient alternative. However, these studies have been restricted to 
raw estimates of the local sampling density at the position of the sampling points and have not yet included the more elaborate interpolation 
machinery of the DTFE and Natural Neighbour Interpolation methods. 

\begin{figure}
  \centering
    \mbox{\hskip -0.25truecm\includegraphics[width=12.5cm]{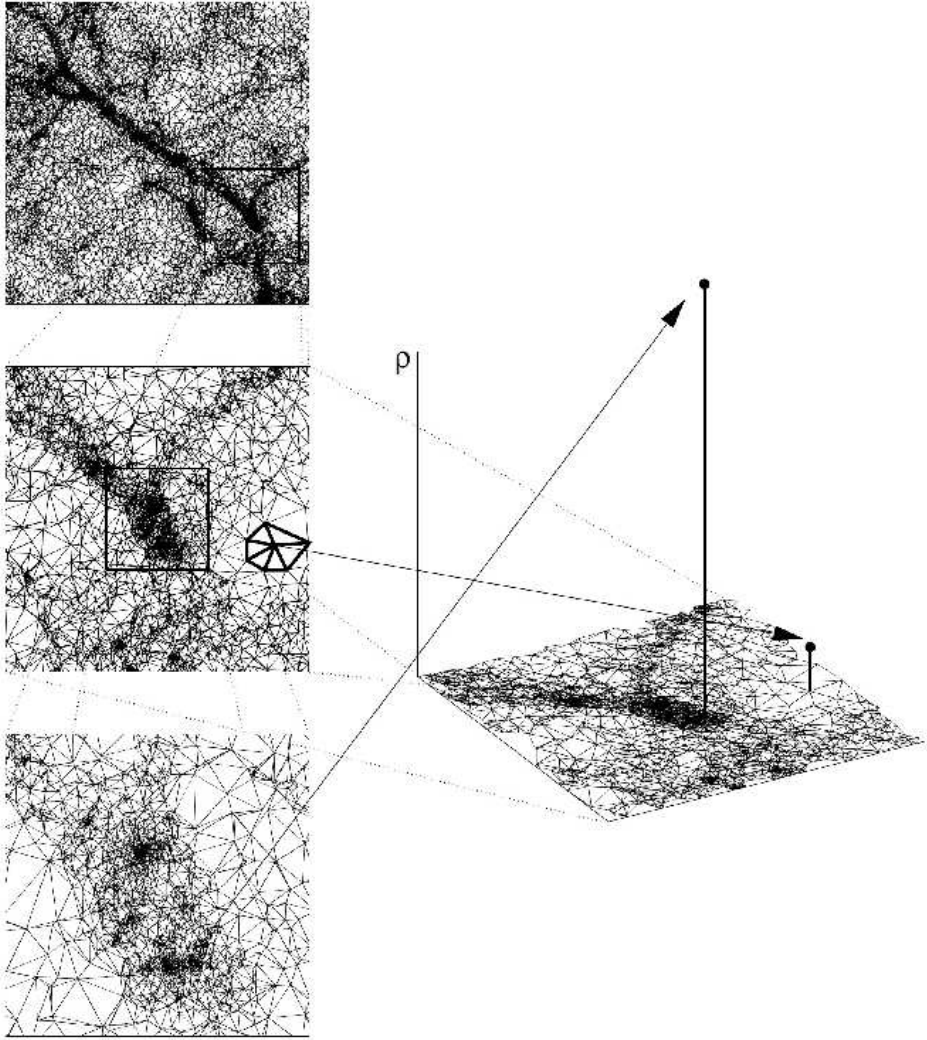}}
    \caption{Relation between density and volume contiguous Voronoi cells. Two example 
points embedded within a filamentary structure (see fig.~\ref{fig:delpntadapt}).}
    \vskip -0.5truecm
\label{fig:voronoicont} 
\end{figure} 
\subsubsection{Density definition}
\noindent The density field reconstruction of the DTFE procedure consists of two steps, 
the zeroth-order estimate ${\widehat \rho}_0$ of the density values at 
the location of the points in ${\cal P}$ and the subsequent linear 
interpolation of those zeroth-order density estimates over the corresponding Delaunay 
grid throughout the sample volume. This yields the DTFE density field estimate 
${\widehat \rho}({\bf x})$. 

It is straightforward to infer (see next sect.~\ref{sec:masscons}) that if the the zeroth-order 
estimate of the density values would be the inverse of the regular Voronoi volume the condition 
of mass conservation would not be met. Instead, the DTFE procedure employs a slightly modified 
yet related zeroth-order density estimate, the normalized inverse of the volume $V({\cal W}_i)$ 
of the {\it contiguous Voronoi cell} ${\cal W}_i$ of each point $i$. For $D$-dimensional 
space this is 
\begin{equation}
{\widehat \rho}({\bf x}_i)\,=\,(1+D)\,\frac{m_i}{V({\cal W}_i)} \,.
\label{eq:densdtfe}
\end{equation}
The {\it contiguous Voronoi cell} of a point $i$ is the union of all Delaunay tetrahedra 
of which point $i$ forms one of the four vertices (see fig.~\ref{fig:voronoicont}). 

\subsubsection{Mass Conservation}
\label{sec:masscons}
An {\it essential} requirement for the cosmological purposes of our 
interpolation scheme is that the estimated DTFE density field ${\widehat \rho}({\bf x})$ 
should guarantee {\it mass conservation}: the total mass corresponding to the density 
field should be equal to the mass represented by the sample points. Indeed, 
this is an absolutely crucial condition for many applications of a 
physical nature. Since the mass $M$ 
is given by the integral of the density field $\rho({\bf x})$ over space, this 
translates into the integral requirement 
\begin{eqnarray}
{\widehat M}&\,=\,&\int {\widehat \rho}({\bf x})\,d{\bf x}\nonumber\\
&\,=\,&\sum_{i=1}^{N} m_i\,=\,M\,=\,{\rm cst.},
\label{eq:integratemass}
\end{eqnarray}
with $m_i=m$ is the mass per sample point: the interpolation procedure 
should conserve the mass $M$. 

The integral eq.~\ref{eq:integratemass} is equal to the volume below the linearly varying 
${\widehat \rho}$-surface in $({\bf x},{\widehat \rho})$-space. In this space each Delaunay   
tetrahedron $m$ is the base ``hyper-plane'' of a polyhedron ${\cal D}^+_m$ (here we 
use the term ``tetrahedron'' for any multidimensional Delaunay simplex). The 
total mass corresponding to the density field may therefore be written as the sum of the 
volumes of these polyhedra, 
\begin{equation} 
  M~=~\sum_{m=1}^{N_T} V({\cal D}^*_m) \,,
\end{equation} 
\noindent with $N_T$ being the total number of Delaunay tetrahedra 
in the tessellation and $V({\cal D}^*_m)$ the volume of polyhedron 
${\cal D}^+_m$. This volume may be written as the average density at the 
vertices of Delaunay tetrahedron ${\cal D}_m$ times its 
volume $V({\cal D}_m)$
\begin{equation} 
  V({\cal D}^*_m)~=~\frac{1}{D+1}\, 
  \left({\widehat \rho}_{m1}+{\widehat \rho}_{m2}+\ldots+
{\widehat \rho}_{m(D+1)}\right)\, V({\cal D}_m) \, .
\end{equation} 
The points $\{m1,m2,\ldots,m(D+1)\}$ are the nuclei which are vertices of the 
Delaunay tetrahedron ${\cal D}^*_m$. The total mass $M$ contained in the 
density field is the sum over all Delaunay tetrahedra within the 
sample volume:
\begin{equation}  
  M~=~\frac{1}{D+1}\,\sum_{m=1}^{N_T}\,
\left({\widehat \rho}_{m1}+{\widehat \rho}_{m2}+\ldots+
{\widehat \rho}_{m(D+1)}\right)\, V({\cal D}_m) \, .
\label{eq:massordered}
\end{equation} 
A simple reordering of this sum yields 
\begin{equation}  
  M~=~\frac{1}{D+1}\,\,\sum_{i=1}^N{\widehat \rho}_i\,\sum_{m=1}^{N_{D,i}} 
  V({\cal D}_{m,i}) \,, 
\label{eq:reorderedsum}
\end{equation} 
in which ${\cal D}_{m,i}$ is one of the $N_{D,i}$ Delaunay tetrahedra of which 
nuclues $i$ is a vertex. The complete set ${\cal D}_{m,i}$ constitutes teh 
{\it contiguous Voronoi cell} ${\cal W}_i$ of nucleus $i$. Mass $M$ may therefore 
be written as 
\begin{equation}  
  M~=~\frac{1}{D+1}\,\,\sum_{i=1}^N{\widehat \rho}_i\,V(\mathcal{W}_i) \,.
\label{eq: massform}
\end{equation}
This equation immediately implies that $M$ is equal to the 
total mass of the sampling points (eq.~\ref{eq:integratemass}) on 
the condition that the density at the location of the sampling points 
is 
\begin{equation} 
\label{3eq: contigvordens} 
    {\widehat \rho}({\bf x}_i)~=~(D+1)\,\frac{m_i}{V(\mathcal{W}_i)} \, .
\end{equation} 
This shows that the DTFE density estimate (eq.~\ref{eq:densdtfe}), 
proportional to the inverse of contiguous Voronoi cell volume, 
indeed guarantees mass conservation. The corresponding normalization 
constant ($1+D$) stands for the number of times each Delaunay 
tetrahedron is used in defining the DTFE density field, equal 
to the number of (1+D) sample points constituting its vertices. 

\subsubsection{Non-uniform sampling}
Astronomical situations often involve a non-uniform sampling process. Often the non-uniform 
selection may be quantified by an a priori known selection function $\psi({\bf x})$. This situation 
is typical for galaxy surveys: $\psi({\bf x})$ may encapsulate differences in sampling density as 
function of 
\begin{itemize}
\item[$\bullet$] Sky position $(\alpha,\delta)$\\
In practice, galaxy (redshift) surveys are hardly ever perfectly uniform. Various 
kinds of factors -- survey definition, observing conditions, instrumental 
aspects, incidental facts  -- will result in a non-uniform coverage of the 
sky. These may be encapsulated in a {\it survey mask} $\psi(\alpha,\delta)$. 
\item[$\bullet$] Distance/redshift: $\psi_z(r)$\\
Magnitude- or flux-limited surveys induce a systematic redshift selection. 
At higher redshifts, magnitude-limited surveys can only probe galaxies whose luminosity 
exceeds an ever increasing value $L(z)$. The corresponding radial selection function 
$\psi_z$ is given by 
\begin{equation} 
   \psi_z(z)~=~\frac{\int_{L(z)}^\infty \Phi(L) {\rm d}L}{\int_{L_0}^\infty \Phi(L) {\rm d}L} \, ,
\label{eq: selfnc}
\end{equation}
where $\Phi(L)$ is the galaxy luminosity function.
\end{itemize}

\noindent Both selection effects will lead to an underestimate of density value. To correct 
for these variations in sky completeness and/or redshift/depth completeness the 
estimated density at a sample point/galaxy $i$ is weighted by the inverse 
of the selection function,
\begin{equation}
\psi({\bf x}_i)\,=\,\psi_s(\alpha_i,\delta_i)\,\psi_z(r_i)\,,
\end{equation}
\noindent so that we obtain,
\begin{equation}
{\widehat \rho}({\bf x}_i)\,=\,(1+D)\,\frac{m_i}{\psi({\bf x}_i)\,V({\cal W}_i)} \,.
\label{eq:densvornu}
\end{equation}.
While it is straightforward to correct the density estimates accordingly, 
a complication occurs for locations where the completeness is very low or 
even equal to zero.  In general, regions with redshift completeness zero should 
be excluded from the correction procedure, even though for specific cosmological 
contexts it is feasible to incorporate field reconstruction procedures utilizing 
field correlations. A {\it constrained random field} approach \citep{edbert1987,
hofmrib1991,zaroubi1995} uses the autocorrelation function of the presumed 
density field to infer a realization of the corresponding Gaussian field. Recently, 
\citep{aragonphd2007} developed a DTFE based technique which manages to 
reconstruct a structured density field pattern within selection gaps 
whose size does not exceed the density field's coherence scale.

\subsection{DTFE Velocity Fields}

\subsubsection{DTFE density and velocity field gradients}
The value of the density and velocity field gradient in each Delaunay tetrahedron is directly and 
uniquely determined from the location ${\bf r}=(x,y,z)$ of the four points forming the Delaunay 
tetrahedra's vertices, ${\bf r}_0$, ${\bf r}_1$, ${\bf r}_2$ and ${\bf r}_3$, 
and the value of the estimated density and sampled velocities at each of these locations, 
$({\widehat \rho}_0,{\bf v}_0)$, $({\widehat \rho}_1,{\bf v}_1)$, $({\widehat \rho}_2,{\bf v}_2)$ and 
$({\widehat \rho}_3,{\bf v}_3)$,
\begin{eqnarray}
\begin{matrix}
\widehat{\nabla \rho}|_m\\ 
\ \\
\widehat{\nabla {\bf v}}|_m
\end{matrix} 
\ \ \Longleftarrow\ \ 
{\begin{cases}
{\widehat \rho}_0 \ \ \ \ {\widehat \rho}_1 \ \ \ \ {\widehat \rho}_2 \ \ \ \ {\widehat \rho}_3 \\
{\bf v}_0 \ \ \ \ {\bf v}_1 \ \ \ \ {\bf v}_2 \ \ \ \ {\bf v}_3 \\
\ \\
{\bf r}_0 \ \ \ \ {\bf r}_1 \ \ \ \ {\bf r}_2 \ \ \ \ {\bf r}_3 \\
\end{cases}}\,
\label{eq:dtfegrad}
\end{eqnarray}

\noindent The four vertices of the Delaunay tetrahedron are both necessary and sufficient for computing the 
entire $3 \times 3$ velocity gradient tensor $\partial v_i/\partial x_j$. Evidently, the same 
holds for the density gradient $\partial \rho/\partial x_j$. We define the matrix ${\bf A}$ 
is defined on the basis of the vertex distances $(\Delta x_n,\Delta y_n, \Delta z_n)$ (n=1,2,3),   
\begin{eqnarray}
\begin{matrix}
\Delta x_n \,=\,x_n-x_0\\
\Delta y_n \,=\,y_n -y_0\\
\Delta z_n \,=\,z_n -z_0\\
\end{matrix}
\quad \Longrightarrow\quad
{\bf A}\,=\,
\begin{pmatrix}
\Delta x_1&\Delta y_1&\Delta z_1\\
\ \\
\Delta x_2&\Delta y_2&\Delta z_2\\
\ \\
\Delta x_3&\Delta y_3&\Delta z_3
\end{pmatrix}
\end{eqnarray}
\noindent Similarly defining 
$\Delta {\bf v}_n\,\equiv\,{\bf v}_n-{\bf v}_0\,(n=1,2,3)$ and 
$\Delta {\rho}_n\,\equiv\,{\rho}_n-{\rho}_0\,(n=1,2,3)$ it is 
straightforward to compute directly and simultaneously the 
density field gradient $\nabla \rho|_m$ and the velocity field gradient 
$\nabla {\bf v}|_m\,=\,\partial v_i/\partial x_j\,$ in Delaunay tetrahedron 
$m$ via the inversion,
\begin{eqnarray}
\begin{pmatrix}
{\displaystyle \partial \rho \over \displaystyle \partial x}\\
\ \\
{\displaystyle \partial \rho \over \displaystyle \partial y}\\
\ \\
{\displaystyle \partial \rho \over \displaystyle \partial z}
\end{pmatrix}
\,=\,{\bf A}^{-1}\,
\begin{pmatrix}
\Delta \rho_{1}\\
\ \\
\Delta \rho_{2}\\
\ \\
\Delta \rho_{3}\\
\end{pmatrix}\,;\qquad\qquad
\begin{pmatrix}
{\displaystyle \partial v_x \over \displaystyle \partial x} & \ \ {\displaystyle \partial v_y \over \displaystyle \partial x} & \ \  
{\displaystyle \partial v_z \over \displaystyle \partial x}\\
\ \\
{\displaystyle \partial v_x \over \displaystyle \partial y} & \ \ {\displaystyle \partial v_y \over \displaystyle \partial y} & \ \ 
{\displaystyle \partial v_z \over \displaystyle \partial y} \\
\ \\
{\displaystyle \partial v_x \over \displaystyle \partial z} & \ \ {\displaystyle \partial v_y \over \displaystyle \partial z} & \ \ 
{\displaystyle \partial v_z \over \displaystyle \partial z}
\end{pmatrix}
&\,=\,&{\bf A}^{-1}\,
\begin{pmatrix}
\Delta v_{1x} & \ \ \Delta v_{1y} & \ \ \Delta v_{1z} \\ \ \\ 
\Delta v_{2x} & \ \ \Delta v_{2y} & \ \ \Delta v_{2z} \\ \ \\ 
\Delta v_{3x} & \ \ \Delta v_{3y} & \ \ \Delta v_{3z} \\ 
\end{pmatrix}\,.
\label{eq:velgrad}
\end{eqnarray}

\subsubsection{Velocity divergence, shear and vorticity}
From the nine velocity gradient components $\partial v_i / \partial x_j$ we can directly 
determine the three velocity deformation modes, the velocity divergence $\nabla \cdot {\bf v}$, the 
shear $\sigma_{ij}$ and the vorticity ${\bf \omega}$, 
\begin{eqnarray}
\nabla \cdot {\bf v}&\,=\,&
\left({\displaystyle \partial v_x \over \displaystyle \partial x} +
{\displaystyle \partial v_y \over \displaystyle \partial y} +
{\displaystyle \partial v_z \over \displaystyle \partial z}\right)\,,\nonumber\\
\ \nonumber\\
\sigma_{ij}&\,=\,&
{1 \over 2}\left\{ 
{\displaystyle \partial v_i \over \displaystyle \partial x_j} +
{\displaystyle \partial v_j \over \displaystyle \partial x_i}
\right\}\,-\, {1 \over 3} \,(\nabla\cdot{\bf v})\,\delta_{ij} \,,\\
\ \nonumber\\
\omega_{ij}&\,=\,&
{1 \over 2}\left\{
{\displaystyle \partial v_i \over \displaystyle \partial x_j} -
{\displaystyle \partial v_j \over \displaystyle \partial x_i}
\right\}\,.\nonumber
\label{eq:vgradcomp}
\end{eqnarray}
\noindent where ${\bf \omega}=\nabla \times {\bf v}=\epsilon^{kij} \omega_{ij}$ (and $\epsilon^{kij}$ is the completely 
antisymmetric tensor). In the theory of gravitational instability, there will be no vorticity contribution as 
long as there has not been shell crossing (ie. in the linear and mildly nonlinear regime). 

\subsection{DTFE: velocity field limitations}
\label{sec:vellimit}
DTFE interpolation of velocity field values is only feasible in regions devoid of 
multistream flows. As soon as there are multiple flows -- notably in high-density cluster 
concentrations or in the highest density realms of the filamentary and planar caustics
in the cosmic web -- the method breaks down and cannot be apllied.

In the study presented here this is particularly so in high-density clusters. The 
complication can be circumvented by filtering the velocities over a sufficiently 
large region, imposing an additional resolution constraint on the DTFE velocity field. 
Implicitly this has actually already been accomplished in the linearization procedure 
of the velocity fields preceding the DTFE processing. 
The linearization of the input velocities involves a kernel size of 
$\sqrt{5} \hmpc$, so that the resolution of the velocity field is set 
to a lower limit of $\sqrt{5} \hmpc$. This is sufficient to assure the viability 
of the DTFE velocity field reconstructions.

\subsection{DTFE: mass-weighted vs. volume-weighted}
\label{sec:massvolweight}
A major and essential aspect of DTFE field estimates is that it concerns {\it volume-weighted} 
field averages 
\begin{equation}
{\widehat f}_{vol}({\bf x})\,\equiv\,{\displaystyle \int\,d{\bf y}\, f({\bf y})\,W({\bf x}-{\bf y})
\over \displaystyle \int\,d{\bf y}\,\,W({\bf x}-{\bf y})}
\label{eq:fvol}
\end{equation}
instead of the more common {\it mass-weighted} field averages. 
\begin{equation}
{\widehat f}_{mass}({\bf x})\,\equiv\,{\displaystyle \int\,d{\bf y}\, f({\bf y})\,\rho({\bf y})\,W({\bf x}-{\bf y})
\over \displaystyle \int\,d{\bf y}\,\rho({\bf y})\,W({\bf x}-{\bf y})}
\label{eq:fmass}
\end{equation}
\noindent where $W({\bf x},{\bf y})$ is the adopted filter function defining
the weight of a mass element in a way that is dependent on its position $y$ with respect to 
the position ${\bf x}$. 

Analytical calculations of physical systems in an advanced stage of evolution do quite often 
concern a perturbation analysis. In a cosmological context we may think of the nonlinear evolution 
of density and velocity field perturbations. In order to avoid unnecessary mathematical 
complications most results concern {\it volume-weighted} quantities. However, when seeking 
to relate these to the results of observational results or numerical calculations, 
involving a discrete sample of measurement points, nearly all conventional techniques 
implicitly involve {\it mass-weighted} averages. This may lead to considerable confusion, even 
to wrongly motivated conclusions. 

Conventional schemes for translating a discrete sample of field values $f_i$ into a continuous 
field ${\hat f}({\bf x})$ are usually based upon a suitably weighted sum over the discretely sampled field 
values, involving the kernel weight functions $W({\bf x},{\bf y})$. It is straightforward to 
convert this discrete sum into an integral over Dirac delta functions, 
\begin{eqnarray}
{\hat f}({\bf x})&\,=\,&{\displaystyle \sum\nolimits_{i=1}^N\,{\tilde f}_i\ W({\bf x}-{\bf x}_i) \over
\displaystyle \sum\nolimits_{i=1}^N\,W({\bf x}-{\bf x}_i)}\nonumber\\
&\,=\,& {\displaystyle \int\,d{\bf y}\,f({\bf y})\,W({\bf x}-{\bf y})\ 
\sum\nolimits_{i=1}^N\,\delta_D({\bf y}-{\bf x}_i) \over \displaystyle 
\int\,d{\bf y}\,W({\bf x}-{\bf y})\ \sum\nolimits_{i=1}^N\,\delta_D({\bf y}-{\bf x}_i)}\\
&\,=\,&{\displaystyle \int\,d{\bf y}\, f({\bf y})\,\rho({\bf y})\,W({\bf x}-{\bf y})
\over \displaystyle \int\,d{\bf y}\,\rho({\bf y})\,W({\bf x}-{\bf y})}\,.\nonumber
\label{eq:fintmass}
\end{eqnarray}
\noindent Evidently, a weighted sum implicitly yields a mass-weighted field average. Notice that 
this is true for rigid grid-based averages, but also for spatially adaptive SPH like 
evaluations. 

A reasonable approximation of volume-averaged quantities may be obtained by volume averaging 
over quantities that were mass-filtered with, in comparison, a very small scale for the 
mass-weighting filter function. This prodded \cite{juszk1995} to suggest a (partial) 
remedy in the form of a two-step scheme for velocity field analysis. First, the field values 
are interpolated onto a grid according to eqn.~\ref{eq:fintmass}. Subsequently, the resulting 
grid of field values is used to determine volume-averages according to eqn.~\ref{eq:fvol}.  
We can then make the interesting observation that the asymptotic limit of this, using 
a filter with an infinitely small filter length, yields
\begin{eqnarray}
f_{mass}({\bf x}_0)\,=\,{\displaystyle \sum_i w_i f({\bf x}_i)
\over \displaystyle \sum_i w_i}\,=\,{\displaystyle f({\bf x_1})\,+\,
\sum_{i=2}^N {w_i \over w_1} f({\bf x}_i) \over \displaystyle 1 + \sum_{i=2}^N {w_i \over w_1}}
\quad \longrightarrow \quad f({\bf x}_1)\,
\end{eqnarray}
\noindent where we have ordered the locations $i$ by increasing distance to ${\bf x}_0$ and thus by decreasing 
value of $w_i$. 

The interesting conclusion is that the resulting estimate of the volume-weighted average 
is in fact the field value at the location of the closest sample point ${\bf x}_1$, $f({\bf x}_1)$.
This means we should divide up space into regions consisting of that part of space closer to a 
particular sample point than to any of the other sample points and take the field value 
of that sample point as the value of the field in that region. This is nothing but 
dividing up space according to the Voronoi tessellation of the set of sample points ${\cal P}$.  
This observation formed the rationale behind the introduction and definition of 
Voronoi and Delaunay tessellation interpolation methods for velocity fields by \cite{bernwey96}. 

While the resulting {\it Voronoi Tessellation Field Estimator} would involve a discontinuous 
field, the step towards a multidimensional linear interpolation procedure would guarantee 
the continuity of field values. The subsequent logical step, invoking the dual Delaunay tessellation 
as equally adaptive spatial linear interpolation grid, leads to the definition of the 
DTFE interpolation scheme. 

\subsection{DTFE Density and Velocity Maps: non-uniform resolution}
For a proper quantitative analysis and assessment of DTFE density and velocity field  
reconstructions it is of the utmost importance to take into account the inhomogeneous 
spatial resolution of raw DTFE maps. 

Cosmic density and velocity fields, as well as possible other fields of relevance, are composed of contributions 
from a range of scales. By implicitly filtering out small-scale contributions 
in regions with a lower sampling density DTFE will include a smaller range of spatial scales contributing 
to a field reconstruction. Even while selection function corrections are taken into account, 
the DTFE density and velocity fields will therefore involve lower amplitudes. 
DTFE velocity fields are less affected than DTFE density fields \citep{weyrom2007}, a manifestation of 
the fact that the cosmic velocity field is dominated by larger scale modes than the density field. 

In addition, it will also lead to a diminished ``morphological resolution''. The sampling density 
may decrease to such a level lower than required to resolve the geometry or morphology of 
weblike features. Once this level has been reached DTFE will no longer be suited for an analysis 
of the Cosmic Web.

\begin{figure} 
  \centering
    \mbox{\hskip 0.0truecm\includegraphics[height=19.0cm]{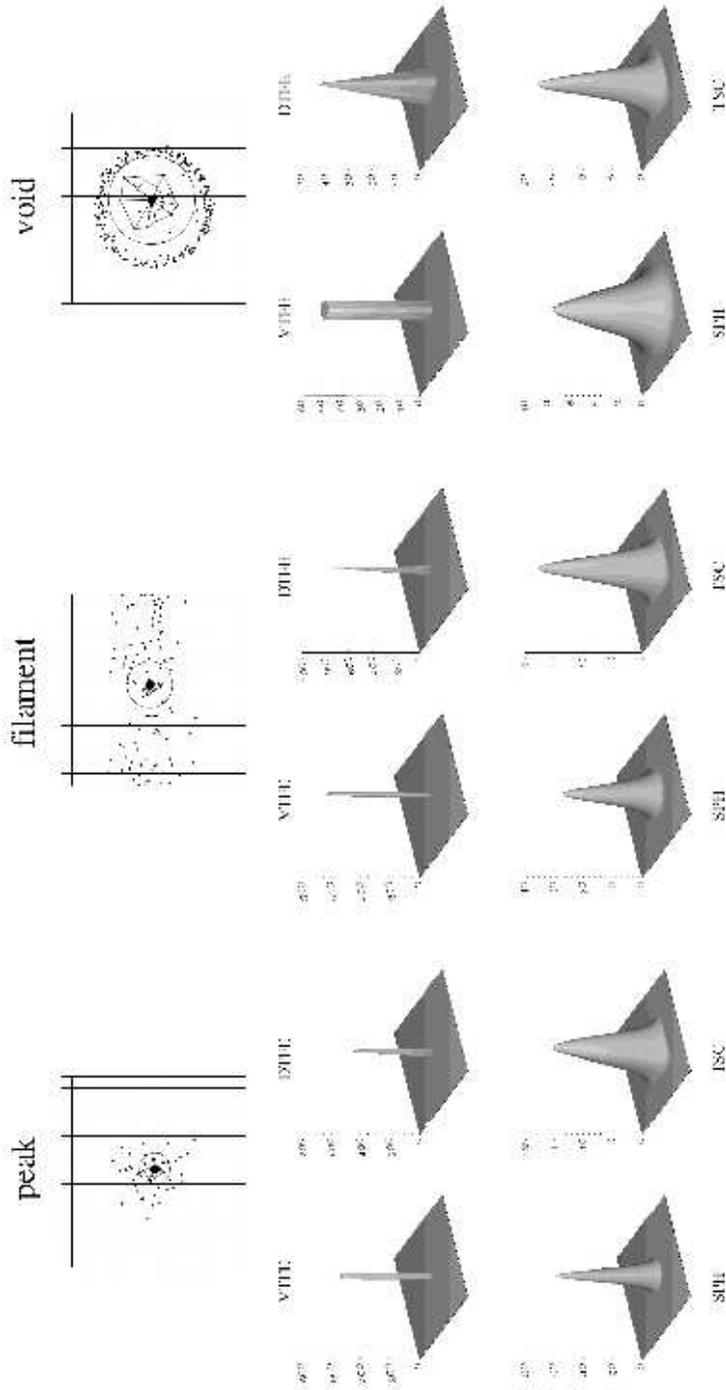}}
    \caption{Typical interpolation kernels for points embedded within three different morphological 
environments: a point in a high-density peak (bottom), a point in a filament (centre) and a 
point in a low-density field environment (top). The three kernels depicted concern three different 
environments: a point embedded within a Gaussian density peak (lefthand), a point within an 
elongated filamentary concentration of points (centre) and a point in a low-density environment 
(righthand).}
    \vskip -0.5truecm
\label{fig:dtfekernel} 
\end{figure} 
\section{DTFE: Technical Issues}
\subsection{DTFE Kernel and Effective Resolution}
\label{sec:dtfekernel}
DTFE distributes the mass $m_i$ corresponding to each sampling point $i$ over its 
corresponding contiguous Voronoi cell. A particular sampling point $i$ will therefore 
contribute solely to the mass of those Delaunay simplices of which it is a vertex. 
This restricted spatial smoothing is the essence of the strict locality of the DTFE 
reconstruction scheme. 

The expression for the interpolation kernel of DTFE provides substantial 
insight into its local spatial resolution. Here we concentrate on the 
density field kernel, the shape and scale of the interpolation kernels 
of other fields is comparable. 

In essence, a density reconstruction scheme distributes the mass 
$m_i$ of a sampling particle $i$ over a region of space according 
to a distributing function $\mathcal{F}_i({\bf x})$, 
\begin{equation} 
  \rho({\bf x})~=~\sum_{i=1}^N m_i \mathcal{F}_i({\bf x}) \, ,
  \label{eq:kerndef} 
\end{equation} 
with $\int {\rm d}{\bf x}\,\mathcal{F}_i({\bf x})\,=\,1\ \ \forall\, i$. 
Generically, the shape and scale of the {\it interpolation 
kernel} $\mathcal{F}_i({\bf x})$ may be different for each sample point. 
The four examples of nn-neighbour kernels in fig.~\ref{fig:sibson4} are a 
clear illustration of this. 

For the linear interpolation of DTFE we find that (see eqn.~\ref{eq:dtfeint})  
for a position ${\bf x}$ within a Delaunay tetrahedron $m$ defined by 
$(1+D)$ sample points $\{{\bf x}_{m0}, {\bf x}_{m1}, \dots, {\bf x}_{mD}\}$ the 
interpolation indices $\phi_{dt,i}$ are given by 
\begin{eqnarray}
\phi_{dt,i}({\bf x})\,=\,
{\begin{cases}
1\,+\,t_1\,+\,t_2\,+\,\dots\,+\,t_D\,\hfill\hfill\qquad i\in\{m0,m1,\ldots,mD\}\\
\ \\ 
0\hfill\hfill\qquad i \notin \{m0,m1,\ldots,mD\}\\
\end{cases}}\,
\end{eqnarray}
\noindent In this, for $i\in\{m0,m1,\ldots,mD\}$, the $D$ parameters $\{t_1,\ldots,t_D\}$ 
are to be solved from 
\begin{equation}
  {\bf x}\,=\,{\bf x}_{i}\,+\,t_1({\bf x}_{m1}-{\bf x}_i)\,+\,
t_2({\bf x}_{m2}-{\bf x}_i)\,+\,\ldots\,+\,t_D({\bf x}_{mD}-{\bf x}_i)\,,
\end{equation}
\noindent On the basis of eqns. (\ref{eq:kerndef}) and (\ref{eq:dtfeint}) the 
expression for the DTFE kernel is easily derived:
\begin{eqnarray} 
  \mathcal{F}_{{\rm DTFE,}i}({\bf x})\,=\,
\begin{cases}
{\displaystyle (D+1)\over \displaystyle V(\mathcal{W}_i)}\,\phi_{dt,i}({\bf x})\hfill\hfill\qquad {\rm if}\ \ \ {\bf x} 
\in \mathcal{W}_i\\
\ \\
0 \hfill\hfill\qquad {\rm if} \ \ \ {\bf x} \notin \mathcal{W}_i
\end{cases}  
\end{eqnarray} 
\noindent in which $\mathcal{W}_i$ is the contiguous Voronoi cell of sampling
point $i$. In this respect we should realize that in two dimensions the contiguous 
Voronoi cell is defined by on average 7 sample points: the point itself and its 
{\it natural neighbours}. In three dimensions this number is, on average, 13.04. 
Even with respect to spatially adaptive smoothing such as embodied by the kernels 
used in Smooth Particle Hydrodynamics, defined a certain number of nearest neighbours 
(usually in the order of 40), the DTFE kernel is indeed highly localized. 

A set of DTFE smoothing kernels is depicted in fig.~\ref{fig:dtfekernel}. Their local and 
adaptive nature may be best appreciated from the 
comparison with the corresponding kernels for regular (rigid) gridbased TSC interpolation, 
a scale adaptive SPH smoothing kernel (based on 40 nearest neighbours) and a zeroth-order 
Voronoi (VTFE) kernel (where the density at a point is set equal to the inverse of the volume 
of the corresponding Voronoi cell). The three kernels depicted concern three different 
environments: a point embedded within a Gaussian density peak (lefthand), a point within an 
elongated filamentary concentration of points (centre) and a point in a low-density environment 
(righthand). The figure clearly illustrates the adaptivity in scale and geometry of the 
DTFE kernel. 

\begin{figure} 
  \centering
    \mbox{\hskip -0.1truecm\includegraphics[height=11.8cm,angle=90.0]{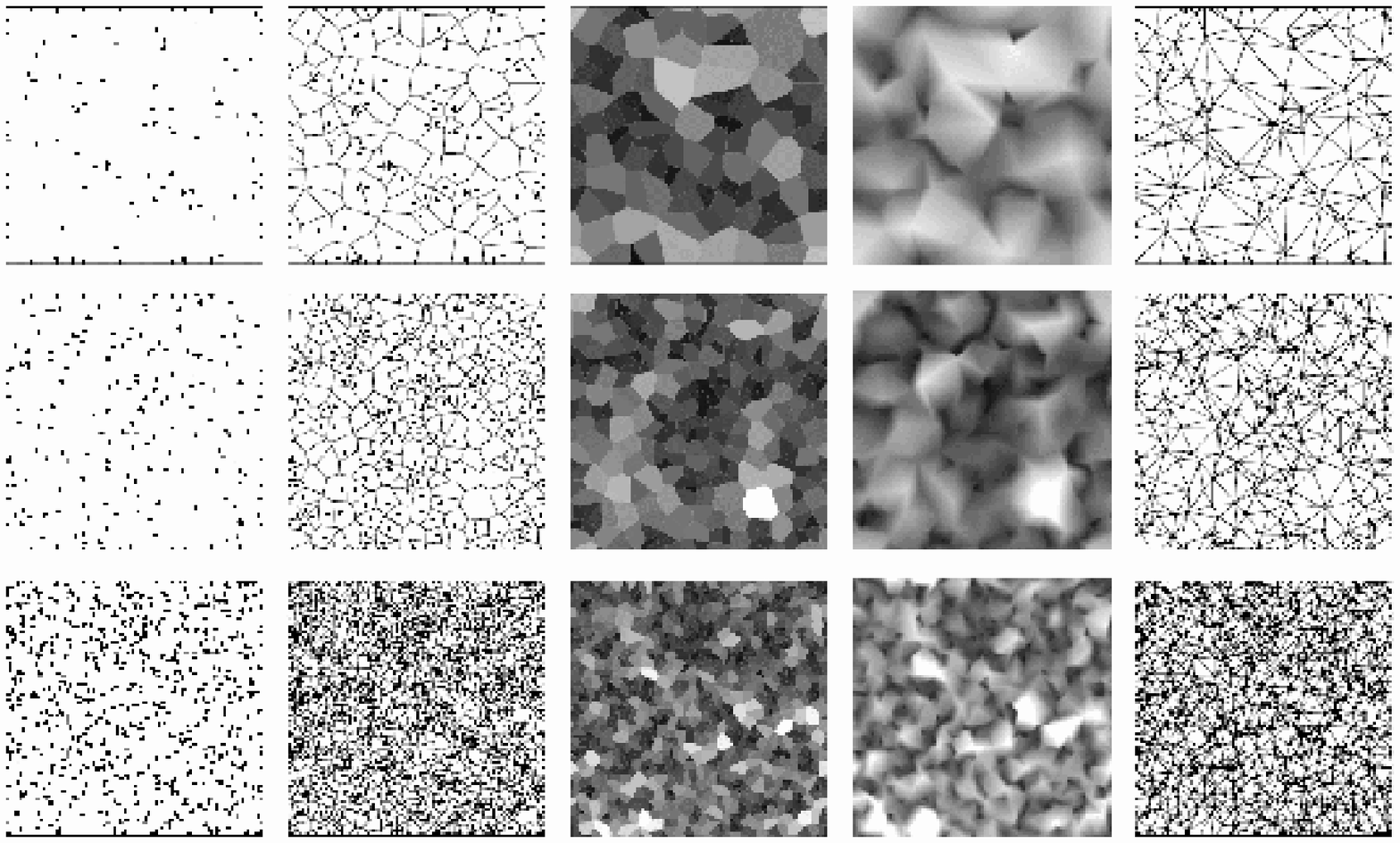}}
    \vskip -0.5truecm
\label{fig:vordelsample} 
\end{figure} 
\begin{figure}
\caption{Poisson sampling noise in uniform fields. The rows illustrate three
  Poisson point samplings of a uniform field with increasing sampling
  density (from top to bottom consisting of 100, 250 and 1000
  points). From left to right the point distribution, the
  corresponding Voronoi tessellation, the zeroth-order VTFE
  reconstructed density field, the first-order DTFE reconstructed
  density field and the corresponding Delaunay tessellation are
  shown. From Schaap 2007.}
\end{figure}
\subsection{Noise and Sampling characteristics}
\label{sec:samplingnoise}
In order to appreciate the workings of DTFE one needs to take into 
account the noise characteristics of the method. Tessellation based 
reconstructions do start from a discrete random sampling of a field. 
This will induce noise and errors. Discrete sampling noise will 
propagate even more strongly into DTFE density reconstructions as 
the discrete point sample itself will be the source for the 
density value estimates. 

In order to understand how sampling noise affects the reconstructed 
DTFE fields it is imperative to see how intrinsically uniform 
fields are affected by the discrete sampling. Even though a uniform 
Poisson process represents a discrete reflection of a uniform density 
field, the stochastic nature of the Poisson sample will induce 
a non-uniform distribution of Voronoi and Delaunay volumes (see 
sect.~\ref{sec:vordelstat}). Because the volumes of Voronoi and/or 
Delaunay cells are at the basis of the DTFE density estimates their 
non-uniform distribution propagates immediately into fluctuations 
in the reconstructed density field. 

This is illustrated in fig.~\ref{fig:vordelsample}, in which three 
uniform Poisson point samples, each of a different sampling density $n$, 
are shown together with the corresponding Voronoi and Delaunay tessellations. 
In addition, we have shown the first-order DTFE reconstructed density 
fields, along with the zeroth-order VTFE density field (where the density 
at a point is set equal to the inverse of the volume of the corresponding 
Voronoi cell). The variation in both the Delaunay and Voronoi cells, as well as 
in the implied VTFE and DTFE density field reconstructions, provide a reasonable 
impression of the fluctuations going along with these interpolation schemes. 

\begin{figure} 
  \centering
    \mbox{\hskip -0.6truecm\includegraphics[width=12.25cm]{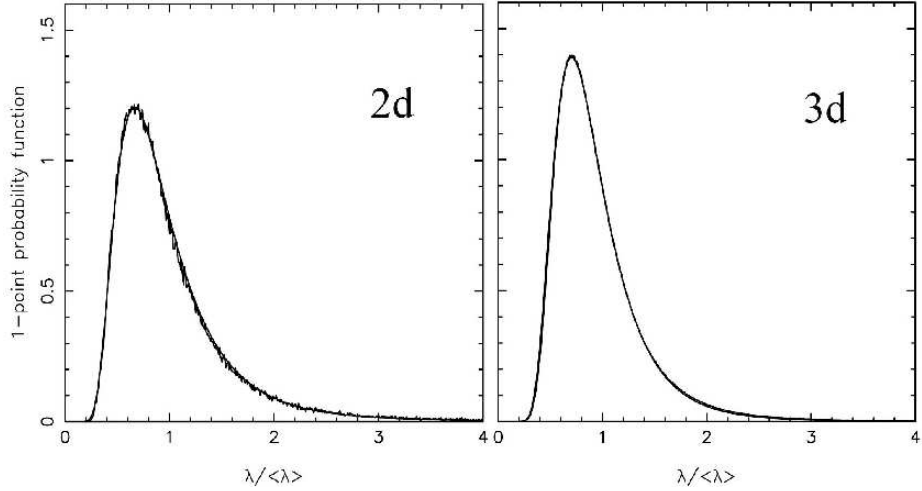}}
    \vskip -0.0truecm
    \caption{One-point distribution functions of the DTFE density field for a
Poisson point process of $10\,000$ points (for 2-d space, lefthand) and 
$100\,000$ points (for 3-d space, righthand). Superimposed are the analytical 
approximations given by eqn.~\ref{eq:dtfedenpdf}}.
\label{fig:vordelpdf} 
\end{figure}

Following the Kiang suggestion \citep{kiang1966} for the Gamma type 
volume distribution for the volumes of Voronoi cells (eqn.~\ref{eq:vorvolpdf}), we may find an 
analytical expression for the density distribution for the zeroth-order VTFE field:
\begin{eqnarray}
f({\tilde \rho})\,=\,
\begin{cases}
{\displaystyle 128 \over \displaystyle 3}\,\tilde{\rho}^{-6} \,\exp{\left(-\frac{\displaystyle 4}{\displaystyle \tilde{\rho}}\right)}\ \ \qquad 2D\\
\ \\
{\displaystyle 1944 \over \displaystyle 5}\,\tilde{\rho}^{-8} \,\exp{\left(-\frac{\displaystyle 6}{\displaystyle \tilde{\rho}}\right)}\qquad 3D
\end{cases}
\label{eq:dtfedenpdf}
\end{eqnarray} 
\noindent The normalized density value ${\tilde \rho}$ is the density estimate $\rho$ in units of the 
average density $\langle \rho \rangle$. While in principle this only concerns the zeroth-order density 
estimate, it turns out that these expression also provide an excellent description of the one-point 
distribution function of the first-order DTFE density field, both in two and three dimensions. 
This may be appreciated from Fig.~\ref{fig:vordelpdf}, showing the pdfs for DTFE density field 
realizations for a Poisson random field of $10,000$ points (2-D) and $100,000$ point (3-D). 
The superimposed analytical expressions (eqn.~\ref{eq:dtfedenpdf}) represent excellent fits. 

The 2-D and 3-D distributions are clearly non-Gaussian, involving a tail 
extending towards high density values. The positive value of the skewness 
($2\sqrt{3}$ for 2D and $\sqrt{5}$ for 3D) confirms the presence of this 
tail. Interestingly, the distribution falls off considerably more rapid for $d=3$ 
than for $d=2$. Also we see that the distributions are more narrow than in the 
case of a regular Gaussian: the variance is $1/3$ for $d=2$ and $1/5$ for d=3, 
confirmed byt the strongly positive value of the curtosis. The larger value for $d=2$ 
indicates that it is more strongly peaked than the distribution for $d=3$. 

On the basis of the above one may also evaluate the significance of DTFE density 
field reconstructions, even including that for intrinsically inhomogeneous fields. 
For details we refer to \cite{schaapphd2007}. 

\subsection{Computational Cost}.
The computational cost of DTFE is not overriding. Most computational 
effort is directed towards the computation of the Delaunay tessellation 
of a point set of $N$ particles. The required CPU time is in the 
order of $O(N\,{\rm log}\,N)$, comparable to the cost of generating
the neighbour list in SPH. The different computational steps of the 
DTFE procedure, along with their scaling as a function of 
number of sample points $N$, are:
\begin{enumerate}
\item Construction of the Delaunay tessellation: ${\mathcal O}(N \,
  {\rm log}\,N)$;
  \item Construction of the adjacency matrix: ${\mathcal O}(N)$;
  \item Calculation of the density values at each location: ${\mathcal O}(N)$;
  \item Calculation of the density gradient inside each Delaunay
    triangle: ${\mathcal O}(N)$;
  \item Interpolation of the density to an image grid: ${\mathcal O}(ngrid\hspace*{.05cm}^2 \cdot N^{1/2})$.
\end{enumerate}
Step2, involving the calculation of the adjacency matrix necessary for 
the {\it walking triangle} algorithm used for Delaunay tetrahedron 
identification, may be incorporated in the construction of the Delaunay 
tessellation itself and therefore omitted. The last step, the interpolation of 
the density to an image grid, is part of the post-processing operation and 
could be replaced by any other desired operation.  It mainly depends on 
the number of gridcells per dimension.

\section{DTFE: Hierarchical and Anisotropic Patterns}
\noindent To demonstrate the ability of the Delaunay Tessellation Field 
Estimator to quantitatively trace key characteristics of the cosmic web we 
investigate in some detail two aspects, its ability to trace the hierarchical 
matter distribution and its ability to reproduce the shape of anisotropic -- 
filamentary and planar -- weblike patterns. 

\subsection{Hierarchy and Dynamic Range}
\label{sec:dtfescaling}
\noindent The fractal Soneira-Peebles model \citep{soneirapeebles1977} has been used to assess the 
level to which DTFE is able to trace a density field over the full range 
of scales represented in a point sample. The Soneira-Peebles model is an   
analytic self-similar spatial point distribution which was defined for the purpose of 
modelling the galaxy distribution, such that its statistical properties would be tuned 
to reality. An important property of the Soneira-Peebles model is that it is one
of the few nonlinear models of the galaxy distribution whose statistical properties 
can be fully and analytically evaluated. This concerns its power-law two-point 
correlation function, correlation dimension and its Hausdorff dimension. 
\begin{figure} 
  \centering
    \mbox{\hskip -0.6truecm\includegraphics[height=19.0cm]{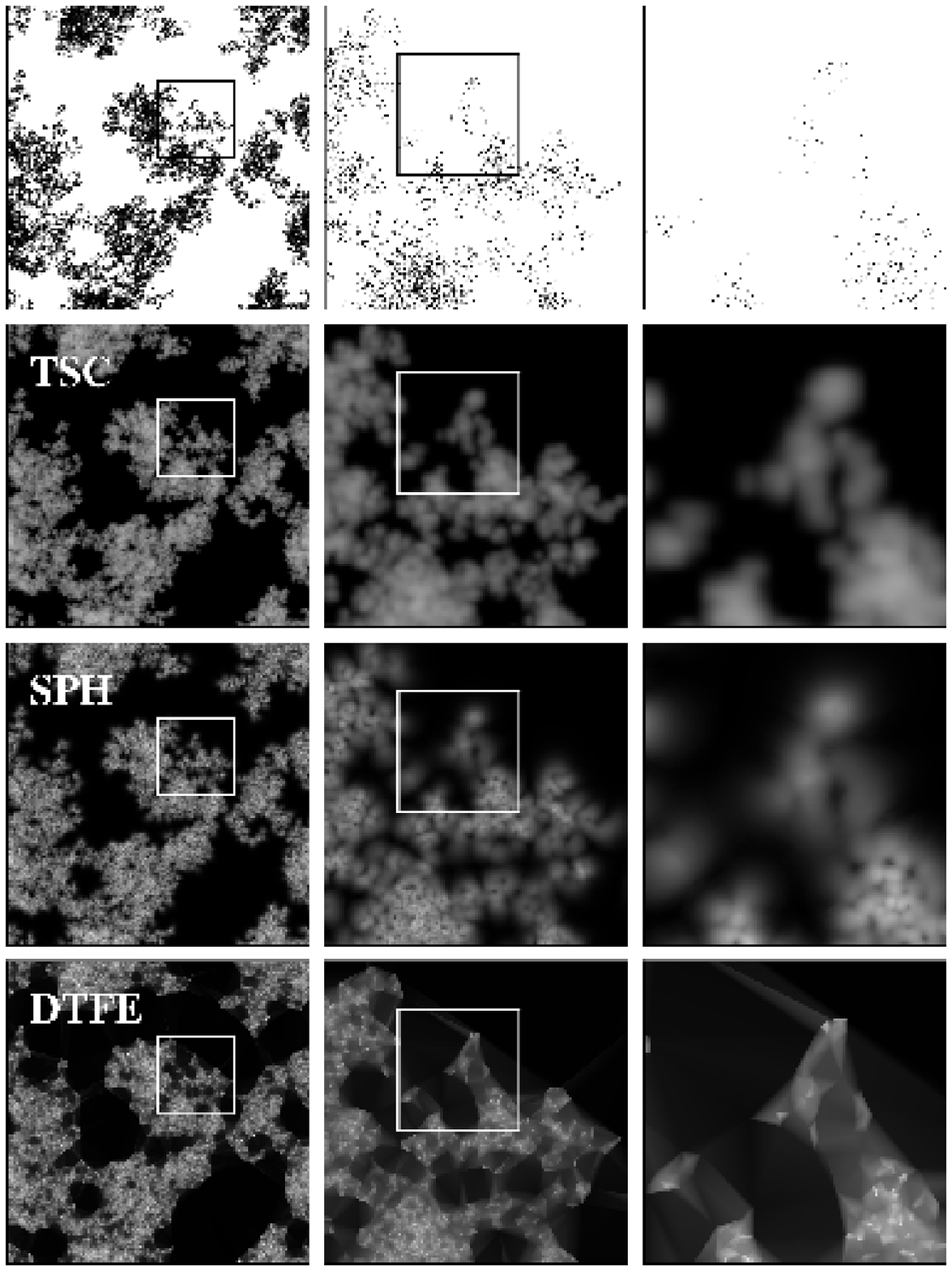}}
    \vskip -0.0truecm
\label{fig:eta4panel} 
\end{figure} 
\begin{figure*}
\caption{Singular Soneira-Peebles model with $\eta=4$, $\lambda=1.9$ and $L=8$. 
    Top row: full point distribution (left-hand frame) and zoom-ins focusing on a
    particular structure (central and right-hand frames). Rows 2~to~4:
    corresponding density field reconstructions produced using the
    TSC, SPH and DTFE methods.}.
\end{figure*}
The Soneira-Peebles model is specified by three parameters. The starting 
point of the model is a level-$0$ sphere of radius $R$. At each level-$m$ 
a number of $\eta$ subspheres are placed randomly within their parent level-$m$ 
sphere: the level-$(m+1)$ spheres have a radius $R/\lambda$ where $\lambda>1$, 
the size ratio between parent sphere and subsphere. This process is repeated 
for $L$ successive levels, yielding $\eta^L$ level-$L$ spheres of radius 
$R/\lambda^L$. At the center of each of these spheres a point is placed, 
producing a point sample of $\eta^L$ points. While this produces a pure 
{\it singular} Soneira-Peebles model, usually a set of these is 
superimposed to produce a somewhat more realistically looking model of the 
galaxy distribution, an {\it extended} Soneira-Peebles model. 

An impression of the self-similar point sets produced by the Soneira-Peebles 
process may be obtained from fig.~\ref{fig:eta4panel}. The top row contains 
a series of three point distributions, zoomed in at three consecutive levels on 
a singular Soneira-Peebles model realization with $(\eta=4,\lambda=1.90,L=8)$. 
The next three columns depict the density field renderings produced 
by three different interpolation schemes, a regular rigid grid-based TSC 
scheme, a spatially adaptive SPH scheme and finally the DTFE reconstruction. 
The figure clearly shows the supreme resolution of the latter. By comparison, 
even the SPH reconstructions appear to fade for the highest resolution frame. 
\begin{figure} 
  \centering
    \mbox{\hskip -0.0truecm\includegraphics[width=19.00cm,angle=90.0]{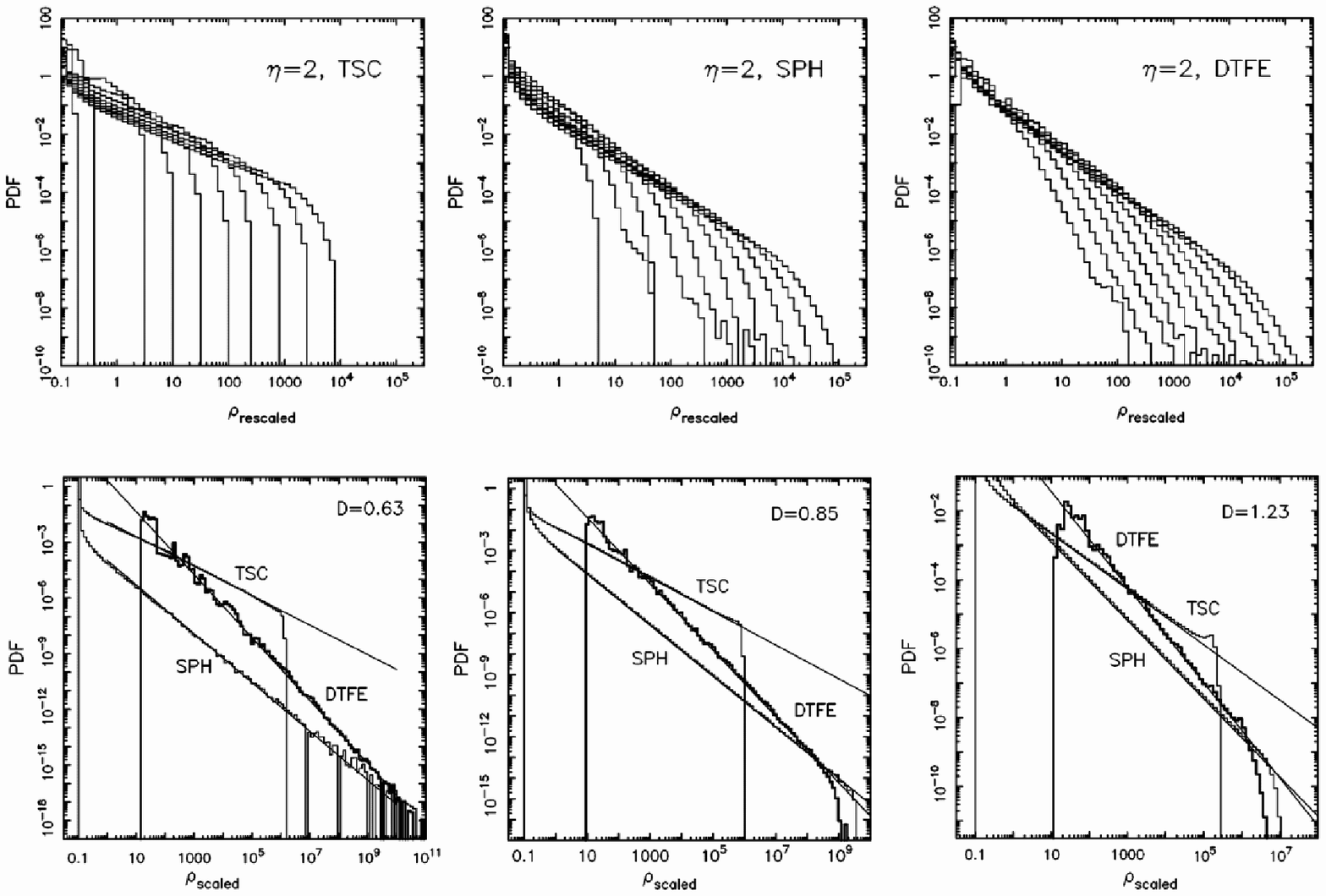}}
    \vskip -0.0truecm
\end{figure} 
\label{fig:dtfescaling}
\begin{figure}
\caption{Top row: Average PDFs of the density field in circles of different level (see text for a description) for 
the TSC, SPH and DTFE density field reconstructions. Model with $\eta=2$, $\lambda=1.75$ and $L=14$.
Bottom row: scaled PDFs of Soneira-Peebles density field reconstructions. Each frame corresponds to a Soneira-Peebles 
realization of a different fractal dimension, denoted in the upper right-hand corner. In each frame the TSC, SPH and DTFE 
reconstructed PDFs are shown.}
\label{fig:sonpeebscale} 
\end{figure}
\subsubsection{Self-similarity}
\noindent One important manifestations of the self-similarity of the defined 
Soneira-Peebles distribution is reflected in the power-law two-point correlation 
function. For $M$ dimensions 
it is given by 
\begin{eqnarray}
\xi(r)&\,\sim\,&r^{-\gamma}\,,\nonumber\\
\ \\
\gamma&\,=\,&M-\left(\frac{{\rm log} \, \eta}{{\rm log} \,
  \lambda}\right)\,\,\,{\rm for}\,\,\,\frac{R}{\lambda^{L-1}}<r<R \nonumber\,.
\label{eq:corr2ptsonpeebles}
\end{eqnarray}
The parameters $\eta$ and $\lambda$ may be adjusted such that they 
yield the desired value for the correlation slope $\gamma$. 

To probe the self-similarity we look at the one-point distribution 
of the density field, both for the point distribution as well as the 
TSC, SPH and DTFE density field reconstructions. Mathematically, the 
condition of self-similarity implies that the PDF corresponding to a 
density field $\rho({\bf x})$ inside an $n$-level circle of radius 
$R/\lambda^n$ should be equal to the PDF inside the reference circle 
of radius $R$, after the density field in the $n$-level circle has been scaled according to
\begin{equation} 
  \rho({\bf x})~\rightarrow~\rho_n({\bf x})~=~\rho({\bf x})/\lambda^{Mn} \, ,
\label{eq:scaling}
\end{equation}
\noindent in which $M$ is the dimension of space. Yet another 
multiplication factor of $\lambda^{M n}$ has to be included to 
properly normalize the PDF (per density unit). In total this results 
in an increase by a factor $\lambda^{2M n}$. Self-similar distributions 
would look exactly the same at different levels of resolution once 
scaled accordingly. This self-similarity finds its expression in a 
{\it universal} power-law behaviour of the PDFs at different levels. 

We have tested the self-similar scaling of the pdfs for a range of 
Soneira-Peebles models, each with a different fractal dimensions \citep{schaapwey2007b}. For  
a Soneira-Peebles model with parameters $(\eta=2,\lambda=1.75,L=14)$
we show the resulting scaled PDFs for the TSC, SPH and DTFE density 
field reconstructions in the top row of Fig.~\ref{fig:sonpeebscale} .
The self-similar scaling of the TSC density field reconstructions is hardly convincing. On the 
other hand, while the SPH reconstruction performs considerably better, it 
is only the DTFE rendering which within the density range probed by 
each level displays an almost perfect power-law scaling ! Also notice that 
the DTFE field manages to probe the density field over a considerably larger 
density range than e.g. the SPH density field. 
\begin{table*}
\begin{center}
\begin{tabular}{|c|c|c|c|c|}
\hline 
\hline
&&&&\\
$\hskip 0.9cm D\hskip 0.9cm $ &\hskip 0.55cm  $\alpha$(theory) {\hskip 0.55cm} &\hskip 0.55cm  $\alpha({\rm TSC})$ {\hskip 0.55cm} & \hskip 0.55cm $\alpha({\rm SPH})${\hskip 0.55cm}  &\hskip 0.55cm  $\alpha({\rm DTFE})$ {\hskip 0.55cm} \\
&&&&\\
\hline 
&&&&\\
 $0.63$ & $-1.69$ & $-0.81$ & $-1.32$ & $-1.70$ \\
 $0.86$ & $-1.57$ & $-0.82$ & $-1.24$ & $-1.60$ \\
 $1.23$ & $-1.39$ & $-0.79$ & $-1.13$ & $-1.38$ \\
&&&&\\
\hline
\hline
\end{tabular}
\end{center}
\caption{\small Slopes of the power-law region of the PDF of a Soneira-Peebles density field as reconstructed by the TSC, SPH and DTFE procedures. 
The theoretical value (Eqn.~\ref{eq:slope}) is also listed. Values are listed for three different Soneira-Peebles realizations, each 
with a different fractal dimension $D$.}
\label{table:slopes}
\end{table*}
Subsequently we determined the slope $\alpha$ of the {\it ``universal''} power-law 
PDF and compared it with the theoretical predictions. The set of three frames in the bottom row of 
Fig.~\ref{fig:sonpeebscale} show the resulting distributions with 
the fitted power-laws. Going from left to right, the frames in this figure correspond 
to Soneira-Peebles realizations with fractal dimensions of $D=0.63$, $D=0.85$ and
$D=1.23$. The slope $\alpha$ of the PDF can be found 
by comparing the PDF at two different levels, 
\begin{eqnarray}
\alpha &\,=\,& \frac{\log {\rm PDF} (\rho_1) ~-~ \log {\rm PDF} (\rho_0)}{\log \rho_1 ~-~ \log \rho_0}\nonumber\\
\,\,\\
&\,=\,&\frac{\log (\lambda^{2M n}/\eta^n)}{\log (1/\lambda^{M n})}\,=\,\frac{D}{M}~-~2 \nonumber\,,
\label{eq:slope}
\end{eqnarray}
\noindent in which $D$ is the fractal dimension of the singular Soneira-Peebles model.
When turning to table~\ref{table:slopes} we not only find that the values of $\alpha$ 
derived from the TSC, SPH and DTFE fields do differ significantly, a fact which has 
been clear borne out by fig.~\ref{fig:sonpeebscale}, but also that the DTFE density field 
PDFs do reproduce to an impressively accurate level the analytically expected power-law 
slope of the model itself \citep{schaapwey2007b}. It is hardly possible to find a more convincing argument for 
the ability of DTFE to deal with hierarchical density distributions !

\subsection{Shapes and Patterns}
DTFE's ability to trace anisotropic weblike patterns is tested on the basis of a class of heuristic models of cellular 
matter distributions, {\it Voronoi Clustering Models} \citep{schaapwey2007c,weygaert2007a}. Voronoi models use the frame of 
a Voronoi tessellation as the skeleton of the galaxy distribution, identifying the structural 
frame around which matter will gradually assemble during the emergence of cosmic structure. The interior of Voronoi 
{\it cells} correspond to voids and the Voronoi {\it planes} with sheets of galaxies. The {\it edges} delineating the rim of each wall 
are identified with the filaments in the galaxy distribution. What is usually denoted as a flattened ``supercluster'' 
will comprise an assembly of various connecting walls in the Voronoi foam, as elongated ``superclusters'' of 
``filaments'' will usually consist of a few coupled edges. The most outstanding structural elements are the 
{\it vertices}, corresponding to the very dense compact nodes within the cosmic web, the rich 
clusters of galaxies. 

The Voronoi clustering models offer flexible templates for cellular patterns, and they are easy to tune towards a particular 
spatial cellular morphology. To investigate the shape performance of DTFE we use pure {\it Voronoi Element Models}, tailor-made 
heuristic ``galaxy'' distributions either and exclusively in and around 1) the Voronoi walls, 2) the Voronoi edges, 3) the 
Voronoi vertices. Starting from a random initial distribution, all model galaxies are projected onto the relevant  wall, edge 
or vertex of the Voronoi cell in which they are located. 
\begin{figure*}
 \centering
 \vskip 0.0truecm 
 \subfigure
     {\mbox{\hskip -1.1truecm{\includegraphics[width=14.0cm]{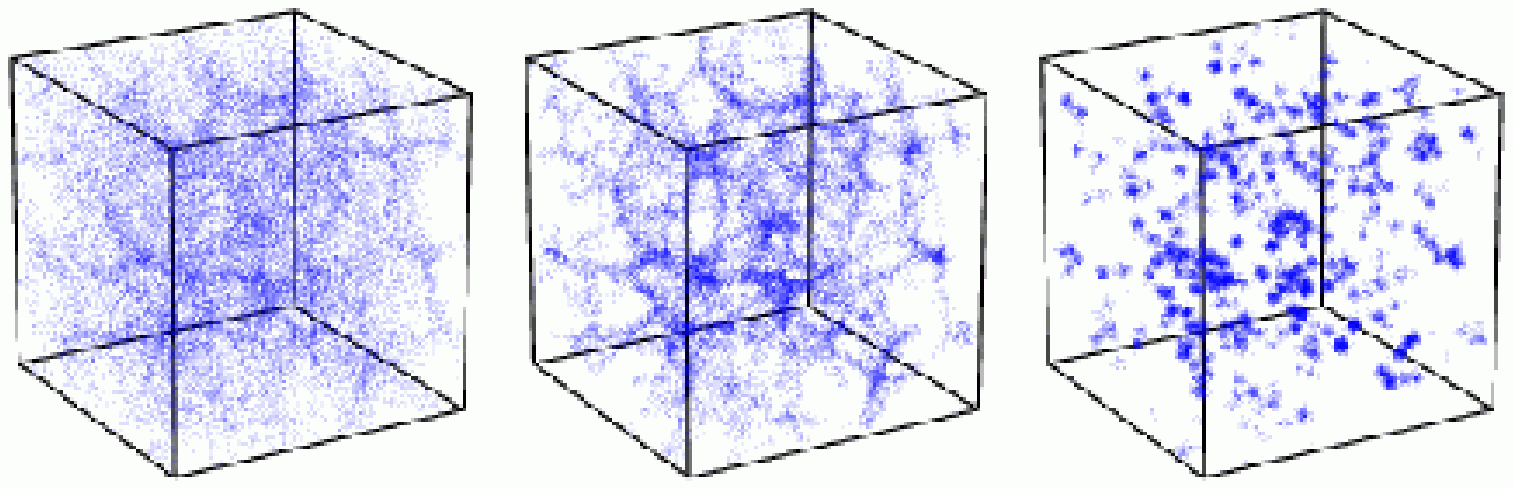}}}}
\vskip -0.5cm  
\subfigure
     {\mbox{\hskip -0.3truecm{\includegraphics[width=12.5cm]{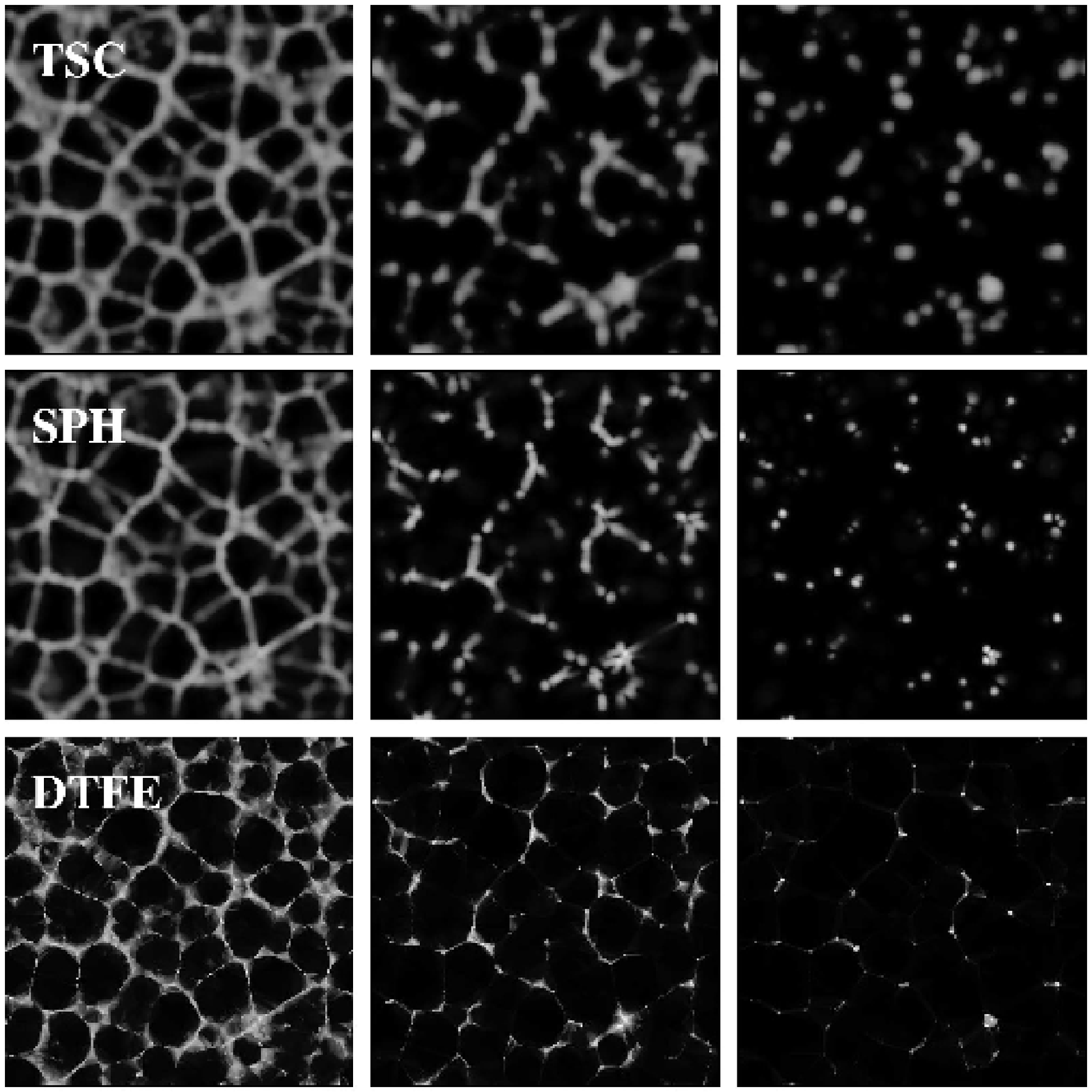}}}}
 \caption{Three Voronoi element clustering models. Top row: box with periodic boundary conditions, boxsize 100h$^{-1}$Mpc.
Left: Wall Model; Centre: Filament Model; Right: Cluster Model. Second to fourth row: Corresponding density Reconstructions of 
the three Voronoi element clustering models. Second: TSC, Third: SPH, Fourth: DTFE.} 
     \label{fig:vorgaldist}
\end{figure*}

The three boxes in the top row of fig.~\ref{fig:vorgaldist} depict a realization for a 3-D {\it Voronoi wall model}, a 
{\it Voronoi filament model} and a {\it Voronoi cluster model}. Planar sections through the TSC, SPH and DTFE density field 
reconstructions of each of these models are shown in three consecutive rows, by means of greyscale maps. The distinctly 
planar geometry of the {\it Voronoi wall model} and the one-dimensional geometry if the {\it Voronoi filament model} is 
clearly recognizable from the sharply defined features in the DTFE density field reconstruction. While the 
SPH reconstructions outline the same distinct patterns, in general the structural features have a more puffy appearance.  
The gridbased TSC method is highly insensitive to the intricate weblike Voronoi features, often the effective rigid gridscale 
TSC filter is not able to render and detect them. 

The DTFE reconstruction of the {\it Voronoi cluster models} (fig.~\ref{fig:vorgaldist}, lower righthand) does depict 
some of the artefacts induced by the DTFE technique. DTFE is succesfull in identifying nearly every cluster, even the 
poor clusters which SPH cannot find. The compact dense cluster also present a challenge in that they reveal 
low-density triangular wings in the regions between the clusters.  Even though these wings include only a minor fraction 
of the matter distribution, they are an artefact which should be accounted for. Evidently, the SPH reconstruction of 
individual clusters as spherical blobs is visually more appealing. 

\subsubsection{Voronoi Filament Model}
\begin{figure}[t]
  \centering
    \mbox{\hskip -0.7truecm\includegraphics[width=13.25cm]{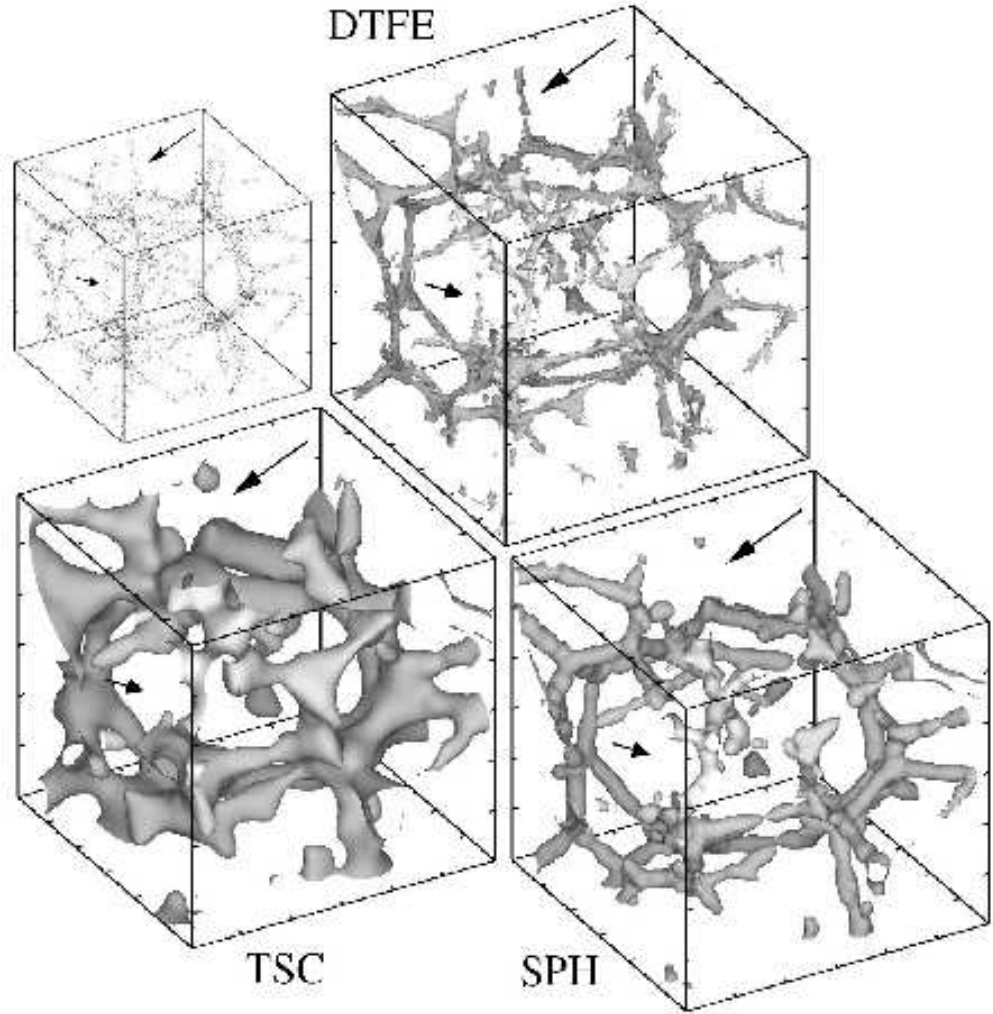}}
    \vskip -0.0truecm
    \caption{Three-dimensional visualization of the Voronoi filament model
     and the corresponding TSC, SPH and DTFE density field
     reconstructions. The density contours have been chosen such that
     $65\%$ of the mass is enclosed. The arrows indicate two
     structures which are visible in both the galaxy distribution and
     the DTFE reconstruction, but not in the TSC and SPH
     reconstructions.}.
     \vskip -0.25truecm
\label{fig:vorfil3d} 
\end{figure}

The best testbench for the ability of the different methods to recover the patterns and morphology of the fully 
three-dimensional density field is that of the contrast rich {\it Voronoi filament models}. In Fig.~\ref{fig:vorfil3d} the central 
part of a sample box of the {\it Voronoi filament model} realization is shown. Isodensity contour levels are chosen such 
that $65\%$ of the mass is enclosed within regions of density equal to or higher the corresponding density value.
The resulting TSC, SPH and DTFE density fields are depicted in the lower lefthand (TSC), lower righthand (SPH) and top frame (DTFE). 
The galaxy distribution in the upper lefthand frame does contain all galaxies within the region. Evidently, the galaxies have 
distributed themselves over a large range of densities and thus occupy a larger fraction of space than that outlined by the density 
contours.  
\begin{figure}[t]
  \centering
    \mbox{\hskip -0.8truecm\includegraphics[width=12.60cm]{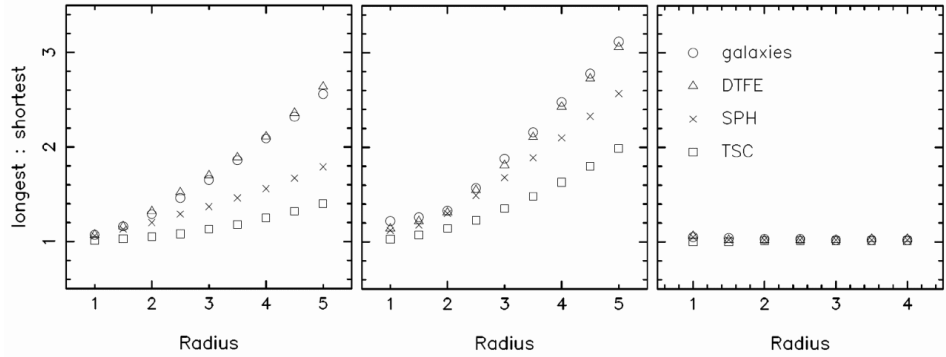}}
    \vskip -0.0truecm
    \caption{Anisotropy measurements for the Voronoi models. Plotted is the longest-to-shortest axis ratio of the
       intertia tensor inside a series of concentric spheres centered on a characteristic structure as a function of 
       the radius of the sphere. The radius is given in units of the standard
       deviation ($\sigma$) of the corresponding Gaussian density
       profiles. The lefthand frame corresponds to the Voronoi wall
       model, the central frame to the Voronoi filament model and the
       righthand frame to the Voronoi cluster model. In each frame
       the results are shown for the TSC, SPH and DTFE
       reconstructions, as well as for the galaxy distribution. The
       meaning of the symbols is depicted in the right-hand frame.}.
     \vskip -0.25truecm
\label{fig:vorshape} 
\end{figure} 
The appearances of the TSC, SPH and DTFE patterns do differ substantially. Part of this is due to a different effective scale of 
the filter kernel. The $65\%$ mass contour corresponds to a density contour $\rho=0.55$ in the TSC field, $\rho=1.4$ in the SPH 
reconstruction and $\rho=2.0$ in the DTFE reconstruction ($\rho$ in units of the average density). The fine filamentary maze seen 
in the galaxy distribution is hardly reflected in the TSC grid based reconstruction even though the global structure, the almost 
ringlike arrangement of filaments, is clearly recognizable. The SPH density field fares considerably better, as it outlines 
the basic configuration of the filamentary web. Nonetheless, the SPH rendered filaments have a width determined by the 
SPH kernel scale, resulting in a pattern of tubes. Bridging substantial density gradients is problematic for SPH reconstructions, 
partly due to the lack of directional information. 

It is the DTFE reconstruction (top frame fig.~\ref{fig:vorfil3d}) which yields the most outstanding 
reproduction of the filamentary weblike character of the galaxy distribution. A detailed comparison between 
the galaxy distribution and the density surfaces show that it manages to trace the most minute details 
in the cosmic web. Note that the density contours do enclose only $65\%$ of the mass, and thus relates to a smaller volume 
than suggested by the features in the galaxy distribution itself. The success of the DTFE method is underlined by identifying 
a few features in the galaxy distribution which were identified by DTFE but not by SPH and TSC. The arrows in 
Fig.~\ref{fig:vorfil3d} point at two tenuous filamentary features visible in the galaxy distribution as well as in 
the DTFE field, yet entirely absent from the TSC and SPH fields. In comparison to the inflated contours of the 
SPH and TSC reconstructions, the structure outlined by the DTFE density field has a more intricate, 
even somewhat tenuous, appearance marked by a substantial richness in structural detail and contrast.
Some artefacts of the DTFE method are also visible: in particular near intersections of filaments we tend 
to find triangular features which can not be identified with similar structures in the galaxy distribution. 
Nearby filaments are connected by relatively small tetrahedra, translating into high density features of 
such shape. 

\subsubsection{Shape and Morphology Analysis}
An important measure of the local density distribution concerns the shape of the density contour levels. 
Various representative features in the three Voronoi element models were identified, followed by a 
study of their shape over a range of spatial scales. The axis ratios of the local density distribution, 
within a radius $R$, was computed from the eigenvalues of the mass intertia tensor. The results of the shape 
analysis are shown in Fig.~\ref{fig:vorshape}. From left to right, the three frames present the axis ratio 
of the longest over the smallest axis, $a_1/a_3$, for walls, filaments and clusters, as a function of the 
scale $R$ over which the shape was measured. The open circles represent the shape of the particle distribution, 
the triangles the shape found in the equivalent DTFE density field, while crosses and squares stand for the 
findings of SPH and TSC. 

In the case of the Voronoi cluster models all three density field reconstructions agree with the 
sphericity of the particle distributions. In the central and righthand frame of Fig.~\ref{fig:vorshape} we 
see that the intrinsic shapes of the walls and filaments become more pronounced as the radius $R$ increases. 
The uniform increase of the axis ratio $a_1/a_3$ with $R$ is a reflection of the influence of the 
intrinsic width of the walls and filaments on the measured shape. For small radii the mass 
distribution around the center of one of these features is largely confined to the interior of 
the wall or filament and thus near-isotropic. As the radius $R$ increases in value, the intrinsic 
shape of these features comes to the fore, resulting in a revealing function of shape 
as function of $R$. 

The findings of our analysis are remarkably strong and unequivocal: over the complete range 
of radii we find a striking agreement between DTFE and the corresponding particle distribution. 
SPH reveals systematic and substantial differences in that they are tend to be more 
spherical than the particle distribution, in particular for the strongly anisotropic 
distributions of the walls and filaments. In turn, the SPH shapes are substantially 
better than those obtained from the TSC reconstructions. The rigidity 
of the gridbased TSC density field reconstructions renders them the worst descriptions of the 
anisotropy of the local matter distribution. These results show that DTFE is indeed capable of 
an impressively accurate description of the shape of walls, filaments and clusters. It 
provides a striking confirmation of the earlier discussed visual impressions. 

\section{DTFE: Velocity Field Analysis}
\noindent De facto the DTFE method has been first defined in the context of a description and analysis of
cosmic flow fields which have been sampled by a set of discretely and sparsely sampled galaxy peculiar 
velocities. \cite{bernwey96} demonstrated the method's superior performance with respect to conventional interpolation 
procedures. In particular, they also proved that the obtained field estimates involved those of the proper 
{\it volume-weighted} quantities, instead of the usually implicit {\it mass-weighted} quantities (see 
sect.~\ref{sec:massvolweight}). This corrected a few fundamental biases in estimates of higher order 
velocity field moments. 

\begin{figure*}
\begin{center}
\vskip -0.0truecm
\mbox{\hskip -0.6truecm\includegraphics[width=12.1cm]{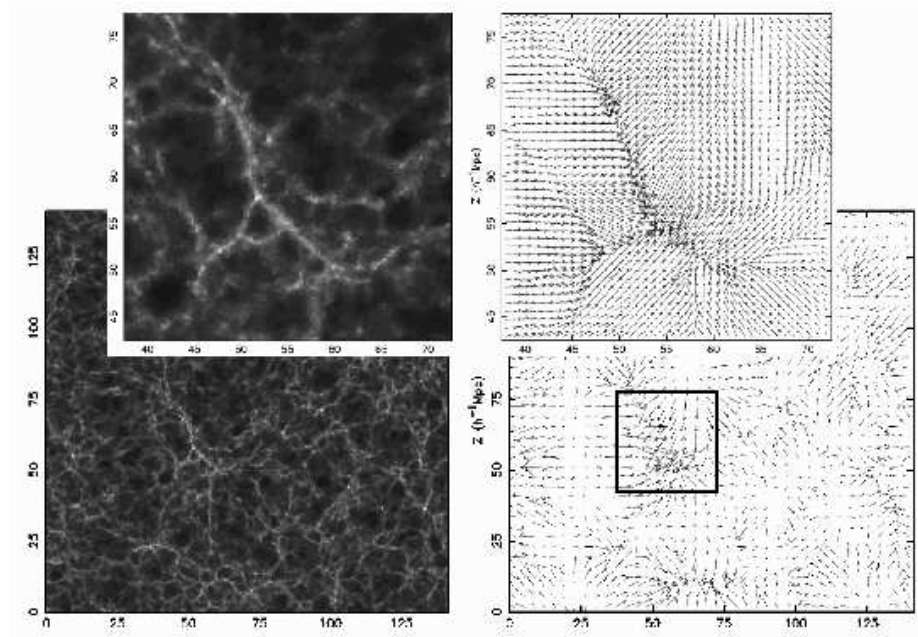}}
\vskip 0.0truecm
\caption{The density and velocity field of the LCDM GIF N-body simulation, computed by DTFE. The 
bottom two frames show the density (bottom left) and velocity field (bottom right) in a central 
slice through the complete simulation box. The density map is a grayscale map. The velocity 
field is shown by means of velocity vectors: the vectors are the velocities component within the 
slice, their size proportional to the velocity's amplitude. By means of DTFE the images in both top frames zoom in 
on the density structure (left) and flow field (bottom) around a filamentary structure (whose 
location in the box is marked by the solid square in the bottom righthand panel). The shear flow along 
the filaments is meticulously resolved. Note that DTFE does not need extra operations to zoom in, 
one DTFE calculation suffices to extract all necessary information. From Schaap 2007.}
\label{fig:giffil}
\end{center}
\end{figure*}

The potential of the DTFE formalism may be fully exploited within the context of the analysis 
of $N$-body simulations and the galaxy distribution in galaxy redshift surveys. Not only it allows 
a study of the patterns in the nonlinear matter distribution but also a study of the 
related velocity flows. Because DTFE manages to follow both the density distribution and the 
corresponding velocity distribution into nonlinear features it opens up the window towards 
a study of the dynamics of the formation of the cosmic web and its corresponding elements. Evidently, 
such an analysis of the dynamics is limited to regions and scales without multistream flows 
(see sect.~\ref{sec:vellimit}). 

A study of a set of GIF $\Lambda$CDM simulations has opened up a detailed view of the 
dynamics in and around filaments and other components of the cosmic web by allowing 
an assessment of the detailed density and velocity field in and around them 
(see fig.~\ref{fig:giffil}). DTFE density and velocity fields may be depicted at any 
arbitrary resolution without involving any extra calculation: zoom-ins represent 
themselves a real magnification of the reconstructed fields. This is in stark contrast to 
conventional reconstructions in which the resolution is arbitrarily set by the users and 
whose properties are dependent on the adopted resolution.

\begin{figure*}
\begin{center}
\mbox{\hskip -0.6truecm\includegraphics[width=13.1cm]{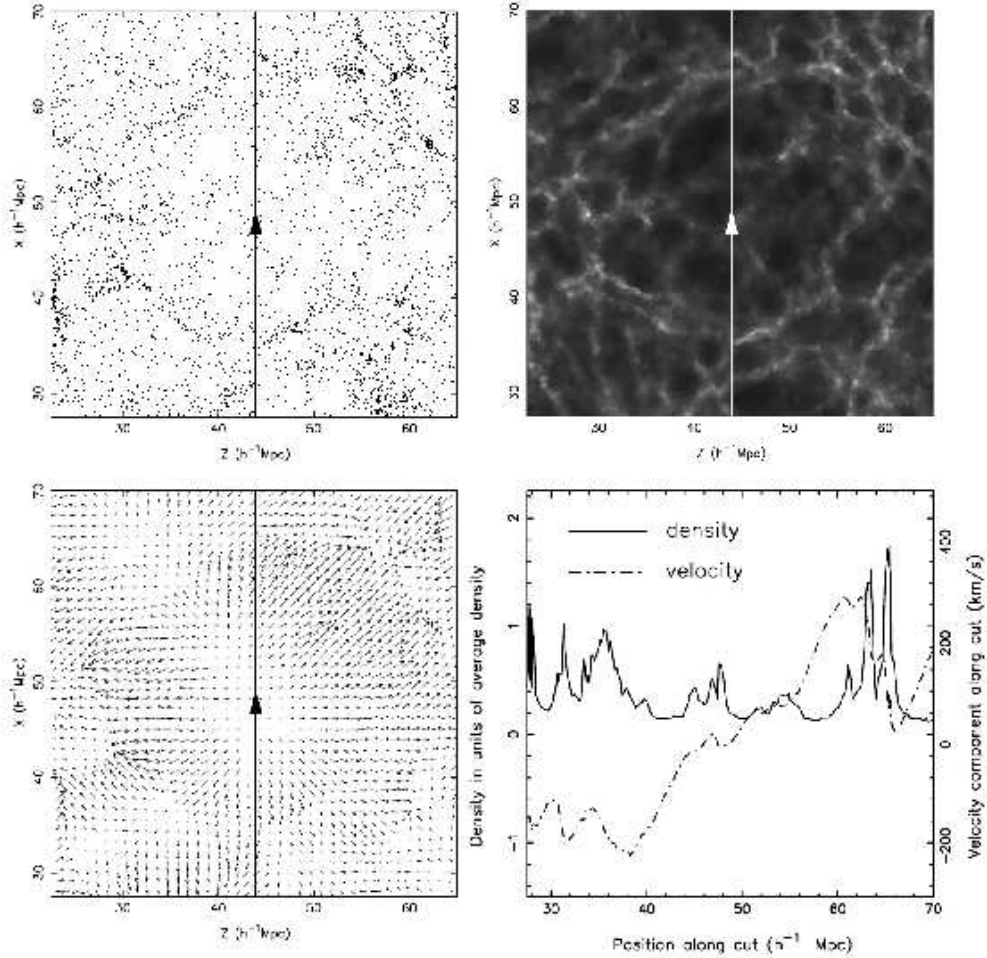}}
\vskip 0.0truecm
\caption{The density and velocity field around a void in the GIF LCDM simulation. The 
top righthand panel shows the N-body simulation particle distribution within a slice 
through the simulation box, centered on the void. The top righthand panel shows the 
grayscale map of the DTFE density field reconstruction in and around the void, the 
corresponding velocity vector plot is shown in the bottom lefthand panel. Notice the 
detailed view of the velocity field: within the almost spherical global outflow 
of the void features can be recognized that can be identified with the diluted 
substructure within the void. Along the solid line in these panels we determined 
the linear DTFE density and velocity profile (bottom righthand frame). We can recognize 
the global ``bucket'' shaped density profile of the void, be it marked by substantial 
density enhancements. The velocity field reflects the density profile in detail, 
dominated by a global super-Hubble outflow. From Schaap 2007.}
\label{fig:gifvoid}
\end{center}
\end{figure*}
\subsection{A case study: Void Outflow}
\label{sec:voidflow}
In Fig.~\ref{fig:gifvoid} a typical void-like region is shown, together with the resulting DTFE 
density and velocity field reconstructions. It concerns a voidlike region selected from a $\Lambda$CDM 
GIF simulation \citep{kauffmann1999}. Figure~\ref{fig:gifdtfeweb} shows a major part of the 
(DTFE) density field of the entire GIF simulation which contains the void. 
It concerns a $256^3$ particles GIF $N$-body simulation, encompassing 
a $\Lambda$CDM ($\Omega_m=0.3,\Omega_{\Lambda}=0.7,H_0=70\, {\rm km/s/Mpc}$) 
density field within a (periodic) cubic box with length $141h^{-1} 
{\rm Mpc}$ and produced by means of an adaptive ${\rm P^3M}$ $N$-body code. 

The top left frame shows the particle distribution in and around the void 
within this $42.5\hmpc$ wide and $1\hmpc$ thick slice through the 
simulation box. The corresponding density field (top righthand frame) 
and velocity vector field (bottom lefthand frame) are a nice illustration of 
the ability of DTFE to translate the inhomogeneous particle distribution into 
highly resolved continuous and volume-filling fields. 

The DTFE procedure  clearly manages to render the void as a slowly varying
region of low density. Notice the clear distinction between the empty
(dark) interior regions of the void and its edges. In the interior of
the void several smaller {\it subvoids} can be distinguished, with 
boundaries consisting of low density filamentary or planar structures.
Such a hierarchy of voids, with large voids containing the traces of 
the smaller ones from which it formed earlier through merging, 
has been described by theories of void evolution \citep{regoes1991,dubinski1993,
weykamp1993,shethwey2004}. 

\bigskip
The velocity vector field in and around the void represents a nice 
illustration of the qualities of DTFE to recover the general velocity 
flow. It also demonstrates its ability identify detailed features within 
the velocity field. The flow in and around the void is dominated by the 
outflow of matter from the void, culminating 
into the void's own expansion near the outer edge. The comparison 
with the top two frames demonstrates the strong relation with 
features in the particle distribution and the DTFE density field. Not only 
is it slightly elongated along the direction of the void's shape, but it is also 
sensitive to some prominent internal features of the void. Towards the ``SE'' 
direction the flow appears to slow down near a ridge, while near the 
centre the DTFE reconstruction identifies two expansion centres. 

The general characteristics of the void expansion are most evident 
by following the density and velocity profile along a one-dimensional 
section through the void. The bottom-left frame of fig.~\ref{fig:gifvoid} shows 
these profiles for the linear section along the solid line indicated in the other three 
frames. The first impression is that of the {\it bucket-like} shape of the void, be it 
interspersed by a rather pronounced density enhancement near its centre. 
This profile shape does confirm to the general trend of low-density regions 
to develop a near uniform interior density surrounded by sharply defined 
boundaries. Because initially emptier inner regions expand faster than the 
denser outer layers the matter distribution gets evened out while the 
inner layers catch up with the outer ones. The figure forms a telling 
confirmation of DTFE being able to recover this theoretically expected 
profile by interpolating its density estimates across the particle diluted 
void (see sect.~\ref{sec:voidflow}).

\bigskip
The velocity profile strongly follows the density structure of the void. The linear 
velocity increase is a manifestation of its general expansion. The near constant velocity 
divergence within the void conforms to the {\it super-Hubble flow} expected for the near 
uniform interior density distribution (see sect.~\ref{sec:voidflow}). Because voids are emptier 
than the rest of the universe they will expand faster than the rest of the universe with a net 
velocity divergence equal to
\begin{eqnarray}
   \theta&\,=\,&{\displaystyle \nabla\cdot{\bf v} \over \displaystyle H}\,=\,3 (\alpha-1)\,,\qquad\alpha=H_{\rm void}/H\,,
\end{eqnarray}
\noindent where $\alpha$ is defined to be the ratio of the super-Hubble expansion rate of the 
void and the Hubble expansion of the universe.

Evidently, the highest expansion ratio is that for voids which are completely empty, 
ie. $\Delta_{{\rm void}}=-1$. The expansion ratio $\alpha$ for such voids may be 
inferred from Birkhoff's theorem, treating these voids as empty FRW universes 
whose expansion time is equal to the cosmic time. For a matter-dominated Universe 
with zero cosmological constant, the maximum expansion rate that a void may 
achieve is given by  
\begin{equation}
\theta_{\rm max}\,=\,1.5\Omega_m^{0.6}\,,
\label{eq:voidmax}
\end{equation}
\noindent with $\Omega_m$ the cosmological mass density parameter. For empty voids in a Universe with a 
cosmological constant a similar expression holds, be it that the value of $\alpha$ will have to be numerically 
calculated from the corresponding equation. In general the dependence on $\Lambda$ is only weak. Generic voids 
will not be entirely empty, their density deficit $|\Delta_{\rm void}|\approx 0.8-0.9$ (cf. eg. the linear density 
profile in fig.~\ref{fig:gifvoid}). The expansion rate $\theta_{void}$ for 
such a void follows from numerical evaluation of the expression
\begin{eqnarray}
\theta_{\rm void}&\,=\,&\frac{3}{2}\,\frac{\Omega_{m}^{0.6}-\Omega_{m,{\rm void}}^{0.6}}{1+\frac{1}{2}\Omega_{m,{\rm void}}^{0.6}}\,;
\qquad \Omega_{m, {\rm void}}\,=\,\frac{\Omega_m ({\Delta}_{\rm void} +1)}{(1+\frac{1}{3}\theta)^{2}} \, 
\end{eqnarray}
\noindent 
\noindent in which $\Omega_{m,{\rm void}}$ is the effective cosmic density parameter inside the 
void. 

\begin{figure}[t]
  \centering
    \mbox{\hskip -1.1truecm\includegraphics[width=13.00cm]{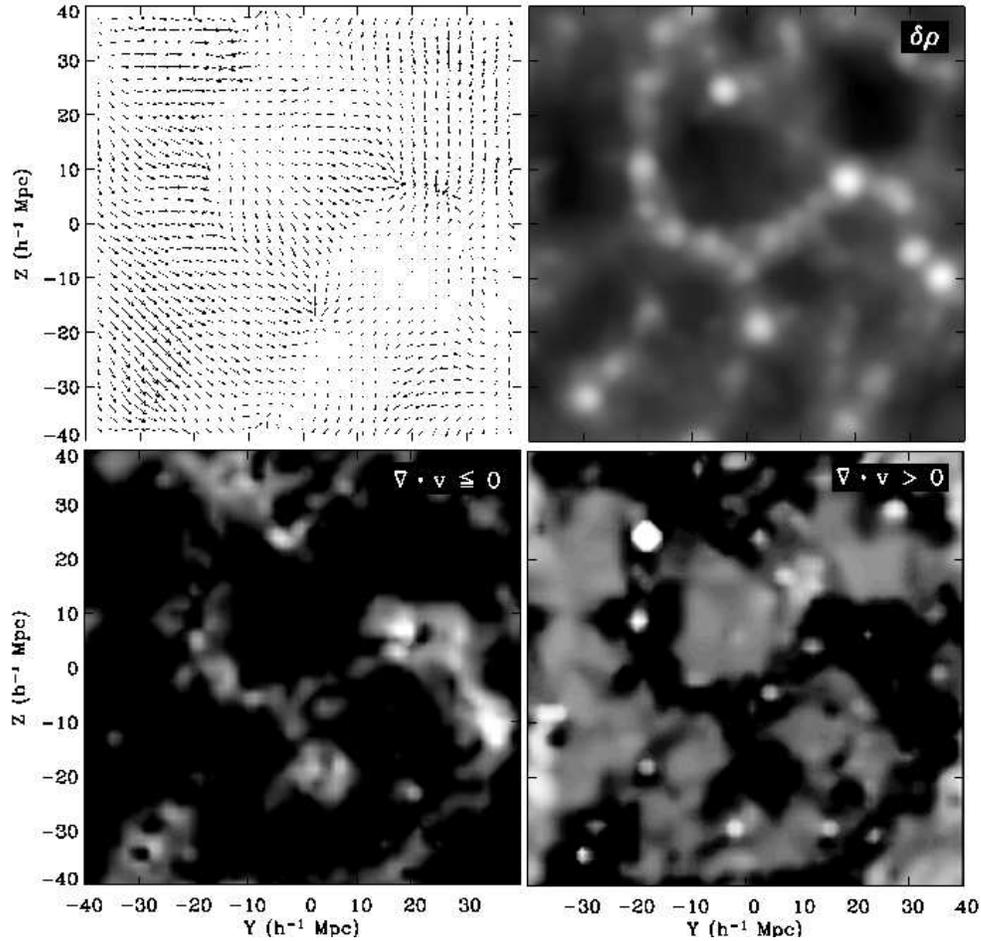}}
    \vskip -0.0truecm
    \caption{The DTFE reconstructed velocity and velocity divergence field of a 
(low-resolution) SCDM N-body simulation, compared with the corresponding DTFE density field 
(top righthand frame). The velocity divergence field is split in two parts. The negative 
velocity divergence regions, those marked by inflow, are shown in the bottom lefthand 
frame. They mark both the high-density cluster regions as well as the more moderate 
filamentary regions. The positive divergence outflow regions in the bottom righthand 
panel not only assume a larger region of space but also have a more roundish morphology. 
They center on the large (expanding) void regions in the matter distribution. The figure 
represents a nice illustration of DTFE's ability to succesfully render the non-Gaussian 
nonlinear velocity field. From Romano-D\'{\i}az 2004.}
\label{fig:dtfeveldensdivv} 
\end{figure} 
\subsection{Velocity Field Evolution}
\label{sec:velstat}
On a mildly nonlinear scale the velocity field contains important information on 
the cosmic structure formation process. Density and velocity perturbations are supposed 
to grow gravitationally out of an initial field of Gaussian density and velocity perturbations. 
Once the fluctuations start to acquire values in the order of unity or higher the growth rapidly 
becomes more complex. The larger gravitational acceleration exerted by the more massive structures 
in the density field induce the infall of proportionally large amounts of matter and hence to an 
increase of the growth rate, while the opposite occurs in and around the underdense regions. 

The overdensities collapse into compact massive objects whose density excess may achieve values 
exceeding unity by many orders of magnitude. Meanwhile the underdensities expand and occupy 
a growing fraction of space while evolving into empty troughs with a density deficit tending towards 
a limiting value of -1.0. The migration flows which are induced by the evolving matter distribution 
are evidently expected to strongly reflect the structure formation process. 

In practice a sufficiently accurate analysis of the nonlinear cosmic velocities is far from trivial. 
The existing estimates of the peculiar velocities of galaxies are still ridden by major uncertainties 
and errors. This is aggravated by the fact that the sampling of the velocity field is discrete, rather 
poor and diluted and highly inhomogeneous. While the conventional approach involves a smoothing 
of the velocity field over scales larger than $10\hmpc$ in order to attain a map of the linear 
velocity field, we have argued in section~\ref{sec:massvolweight} that implicitly this usually 
yields a {\it mass-weighted} velocity field. DTFE represents a major improvement on this. Not only 
does it offer an interpolation which guarantees an optimal resolution, thereby opening the 
window onto the nonlinear regime, it also guarantees a {\it volume-weighted} flow field (see 
sect.~\ref{eq:fvol}). 

\subsubsection{Density and Velocity Divergence}
The implied link between the matter distribution and the induced migration 
flows is most strongly expressed through the relation between the density field $\delta({\bf x}$ and 
the velocity divergence field. The bottom frames of fig.~\ref{fig:dtfeveldensdivv} contain greyscale 
maps of the DTFE normalized velocity divergence estimate $\widehat \theta$ (with $H_0$ the Hubble 
constant),
\begin{equation}
{\widehat \theta}\,\equiv\,{\displaystyle \widehat{\nabla \cdot {\bf v}} \over \displaystyle H_0}\,=
{\displaystyle 1 \over \displaystyle H_0}\,\left({\widehat{\frac{\partial v_x}{\partial x}}} +
  {\widehat{\frac{\partial v_y}{\partial y}}} + {\widehat{\frac{\partial v_z}{\partial z}}}\right) \,,
  \label{eq:dtfe_div_dtfe}
\end{equation}
\noindent for an N-body simulation. For illustrative purposes we have depicted the regions of 
negative and positive velocity divergence separately. The comparison between these maps and the 
corresponding density field (upper righthand frame) and velocity field (upper lefthand) provides 
a direct impression of their intimate relationship. The two bottom frames clearly delineate the 
expanding and contracting modes of the velocity field. The regions with a negative divergence are 
contracting, matter is falling in along one or more directions. The inflow region delineates 
elongated regions coinciding with filaments and peaks in the density field. The highest inflow 
rates are seen around the most massive matter concentrations. Meanwhile the expanding regions 
may be seen to coincide with the surrounding large and deep underdense voids, clearly occupying 
a larger fraction of the cosmic volume. 

The strong relation between the density contrast and velocity divergence is a manifestation of the  
continuity equation. In the linear regime this is a strict linear one-to-one relation, 
\begin{equation}
{\displaystyle \nabla \cdot {\bf v} ({\bf x,t}) \over \displaystyle H}\,=\,- a(t) f(\Omega_m,\Omega_{\Lambda}) 
\,\,\delta({\bf x},t)\,,
\label{eq:divvdenlin}
\end{equation}
\noindent linking the density perturbation field $\delta$ to the peculiar velocity field ${\bf v}$ via the 
factor $f(\Omega_m)$ \citep[see][]{peebles80}. There remains a 1-1 relation between velocity 
divergence and density into the mildly nonlinear regime (see eqn.~\ref{eq:divvdennlin}). This explains 
why the map of the velocity divergence in fig.~\ref{fig:dtfeveldensdivv} is an almost near perfect negative image 
of the density map. 

Even in the quasi-linear and mildly nonlinear regime the one-to-one correspondance between velocity divergence and density 
remains intact, be it that it involves higher order terms \citep[see][for an extensive review]{bernardeau2002}. Within the 
context of Eulerian perturbation theory \citet[][(B)]{bernardeau1992} derived an accurate 2nd order approximation form 
the relation between the divergence and the density perturbation $\delta({\bf x})$. \citet[][(N)]{nusser1991} 
derived a similar quasi-linear approximation within the context of the Lagrangian Zel'dovich approximation. According 
to these approximate nonlinear relations, 
\begin{eqnarray}
{\displaystyle 1 \over \displaystyle H}\,\nabla \cdot {\bf v} ({\bf x})\,=\,
\begin{cases}
{\displaystyle 3 \over \displaystyle 2} f(\Omega_m) \left[1-(1+\delta({\bf x}))^{2/3}\right]\,\qquad\hfill\hfill \hbox{\rm (B)}\\
\ \\
-f(\Omega_m)\,{\displaystyle \delta({\bf x}) \over \displaystyle 1 + 0.18\,\delta({\bf x})}\,\qquad\hfill\hfill \hbox{\rm (N)}\\
\end{cases}
\label{eq:divvdennlin}
\end{eqnarray}
\noindent for a Universe with Hubble parameter $H(t)$ and matter density parameter $\Omega_m$. The studies by 
\cite{bernwey96,weyrom2007,schaapphd2007} have shown that the DTFE velocity field reconstructions are indeed able 
to reproduce these quasi-linear relations. 

\subsubsection{Velocity Field Statistics}
\noindent The asymmetric nonlinear evolution of the cosmic velocity and density field manifests itself in an 
increasing shift of the statistical distribution of density and velocity perturbations away from the initial Gaussian 
probability distribution function. The analytical framework of Eulerian perturbation theory provides a reasonably accurate 
description for the early nonlinear evolution \citep[see][for a review]{bernardeau2002}. 
As for the velocity field, robust results on the evolution and distribution of the velocity divergence, 
$\nabla \cdot {\bf v}$, were derived in a series of papers by Bernardeau \citep[e.g.][]{bernardeau1995}. 
The complete probability distribution function (pdf) of the velocity divergence may be evaluated via the summation 
of the series of cumulants of the non-Gaussian distribution function. The velocity divergence pdf is strongly 
sensitive to the underlying cosmological parameters, in particular the cosmological density parameter $\Omega_m$. 
It represents a considerable advantage to the more regular analysis of the cosmic velocity flows on large linear scales. 
While the latter yields constraints on a combined function of $\Omega_m$ and the bias $b$ between the galaxy and 
matter distribution, access to nonlinear scales would break this degeneracy. 

\begin{figure} 
  \vskip 0.0truecm
  \centering
    \mbox{\hskip -1.0truecm\includegraphics[width=14.00cm]{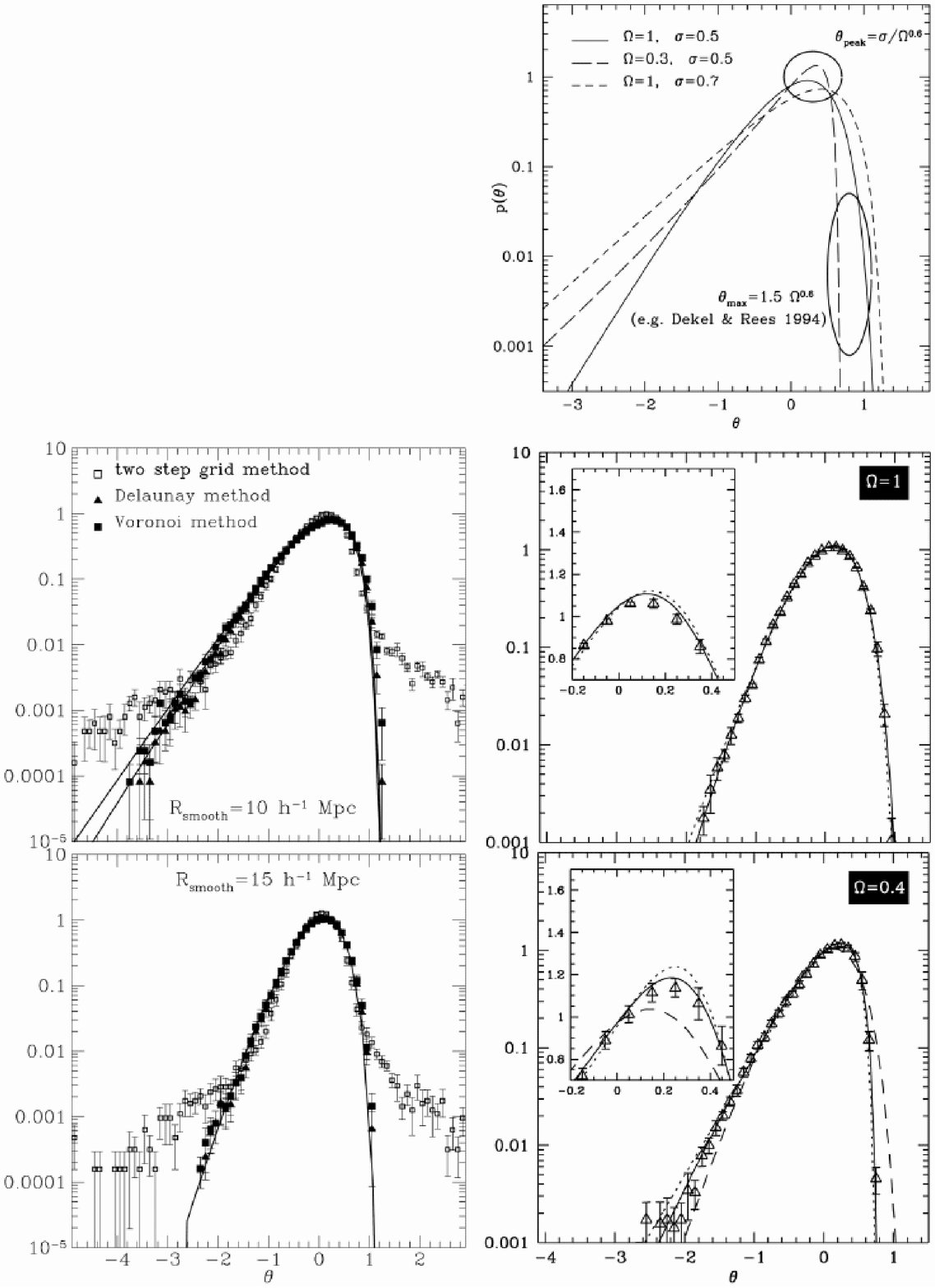}}
    \vskip -0.0truecm
\end{figure}
\begin{figure}
    \caption{DTFE determination of the probability distribution function (pdf) of the 
velocity divergence $\theta$. Top frame illustrates the sensitivity of the $theta$ pdf 
to the underlying cosmology. The superposition of the theoretical pdf curves of 
three cosmologies immediately shows that the location of the maximum of the pdf and 
its sharp cutoff at the positive end are highly sensitive to $\Omega$. The other 
four frames demonstrate DTFE's ability to succesfully trace these marks. 
Lefthand frames: confrontation of the DTFE velocity divergence pdf and that 
determined by more conventional two-step fixed grid interpolation method. Both 
curves concern the same $128^3$ SCDM N-body particle simulation (i.e. $\Omega=1$). 
Top lefthand frame:  tophat filter radius $R_{\rm TH}=10\hmpc$. Bottom lefthand 
panel: $R_{\rm TH}=15\hmpc$. The solid lines represent theoretical predictions of the PDF for 
the measured values of $\sigma_{\theta}$ (Bernardeau 1994). Black triangles are the pdf values 
measured by the DTFE method, the black squares by the equivalent VTFE Voronoi method. 
The open squares show the results obtained by a more conventional two-step fixed grid 
interpolation method. Especially for low and high $\theta$ values the tessellation methods 
prove to produce a much more genuine velocity divergence pdf. From Bernardeau \& van de 
Weygaert 1996. Righthand frames: on the basis of DTFE's ability to trace the velocity divergence pdf 
we have tested its operation in a $\Omega=1$ and a $\Omega=0.4$ CDM N-body simulation. For both 
configurations DTFE manages to trace the pdf both at its high-value edge and near its peak. From 
Bernardeau et al. 1997}
\label{fig:divvpdf} 
\end{figure} 
An impression of the generic behaviour and changing global shape of the resulting non-Gaussian pdf as a function 
of $\Omega_m$ and the amplitude $\sigma_{\theta}$ of the velocity divergence perturbations may be obtained from the 
top frame of fig.~\ref{fig:divvpdf}. The curves correspond to the theoretically predicted velocity divergence pdf 
for three different (matter-dominated) FRW cosmologies. Not only does $\Omega_m$ influence the overall 
shape of the pdf, it also changes the location of the peak -- indicated by $\theta_{\rm max}$ -- as well 
as that of the cutoff at the high positive values of $\theta$. By means of the Edgeworth expansion one may show that 
the velocity divergenc pdf reaches its maximum for a peak value $\theta_{\rm peak}$ \citep{juszk1995,bernkofm1995}, 
\begin{equation}
\theta_{\rm peak}\,=\,-{\displaystyle T_3(\theta) \over \displaystyle 2}\,\sigma_{\theta}\,;\qquad 
\langle \theta^3\rangle\,=\,T_3\,\langle \theta^2 \rangle^2\,.
\end{equation}
\noindent where the coefficient $T_3$ relates the 3rd order moment of the pdf to the second moment \citep[see e.g][]{bernardeau1994}. 
While the exact location of the peak is set by the bias-independent coefficient $T_3$, it does 
not only include a dependence on $\Omega_m$, but also on the shape of the power spectrum, the geometry of the 
window function that has been used to filter the velocity field and on the value of the cosmological 
constant $\Lambda$. To infer the value of $\Omega_m$ extra assumptions need to be invoked. Most direct therefore 
is the determination of $\Omega_m$ via the sharp cutoff of the pdf, related to the expansion velocity of the deepest 
voids in a particular cosmology (see eqn.~\ref{eq:voidmax}).  

While conventional techniques may reproduce the pdf for moderate values of the velocity divergence $\theta$, they 
tend to have severe difficulties in tracing the distribution for the more extreme positive values and the 
highest inflow velocities. An illustration of the tendency to deviate from the analytically predicted distribution 
can be seen in the two lefthand frames of fig.~\ref{fig:divvpdf}, showing the velocity divergence pdf determined 
from a SCDM N-body simulation for two scales of tophat filtered velocity fields ($R=10\hmpc$ and $R=15\hmpc$). The 
open squares depict the velocity divergence pdf determined from an N-body simulation through a two-step grid procedure 
(see sect.~\ref{sec:massvolweight}). Conventional grid interpolation clearly fails by huge margins to recover 
inflow onto the high-density filamentary structures as well as the outflow from voids. 

What came as the first evidence for the success of tessellation based interpolation is the rather 
tight correspondence of the Delaunay tessellation interpolation results with the analytically 
predicted pdf. This finding by \cite{bernwey96} suggested that it would indeed be feasible to 
probe the nonlinear velocity field and infer directly accur4ate estimates of $\Omega_m$. In a follow-up 
study \cite{bernwey97} succesfully tested this on a range of N-body simulations of structure formation, 
showing Delaunay interpolation indeed recovered the right values for $\Omega_m$. The centre and bottom 
righthand frames depict two examples: the pdf's for a $\Omega=1$ and a $\Omega=0.4$ Universe accurately 
traced by the Delaunay interpolated velocity field. 

\subsection{PSCz velocity field}
In a recent study \cite{weyrom2007} applied DTFE towards reconstructing the velocity flow map throughout 
the nearby Universe volume sampled by the PSCz galaxy redshift survey \citep[also see][]{emiliophd2004}. 

\subsubsection{The PSCz sample of the Local Universe}
The IRAS-\pscz ~catalog \citep{saunders2000} is an extension of
the $1.2$-Jy catalog \citep{fisher1995}. It contains $\sim 15
~500$ galaxies with a flux at $60 \mu$m larger than $0.6$-Jy.  For a
full description of the catalog, selection criteria and the procedures
used to avoid stellar contamination and galactic cirrus, we refer the
reader to \cite{saunders2000}.  For our purposes the most
important characteristics of the catalog are the large area sampled
($\sim 84 \%$ of the sky), its depth with a median redshift of $8~500
\kms$, and the dense sampling (the mean galaxy separation at $10~000
\kms$ is $\langle l \rangle = 1~000 \kms$). It implies that PSCz contains   
most of the gravitationally relevant mass concentrations in our local 
cosmic neighbourhood, certainly sufficient to explain and study the 
cosmic flows within a local volume of radius $\sim 120 \hmpc$.  

To translate the redshift space distribution of galaxies in the PSCz catalog into galaxy positions and 
velocities in real space the study based itself on an iterative processing of the galaxy sample by 
\cite{branchini1999} based on the linear theory of gravitational instability~\citep{peebles80}. The 
method involved a specific realization of an iterative technique to minimize 
redshift-space distortions \citep{yahil1991}. While the resulting galaxy velocities relate by 
construction to the linear clustering regime, the reconstruction of the velocity field throughout 
the sample volume does appeal to the capability of DTFE to process a discretely sampled field into 
a continuous volume-filling field and its ability to resolve flows in both high-density 
regions as well as the outflows from underdense voids. 
 
\subsubsection{The PSCz velocity and density field}
\begin{figure} 
  \vskip 1.0truecm
  \centering
    \mbox{\hskip -0.0truecm\includegraphics[height=13.0cm,angle=90.0]{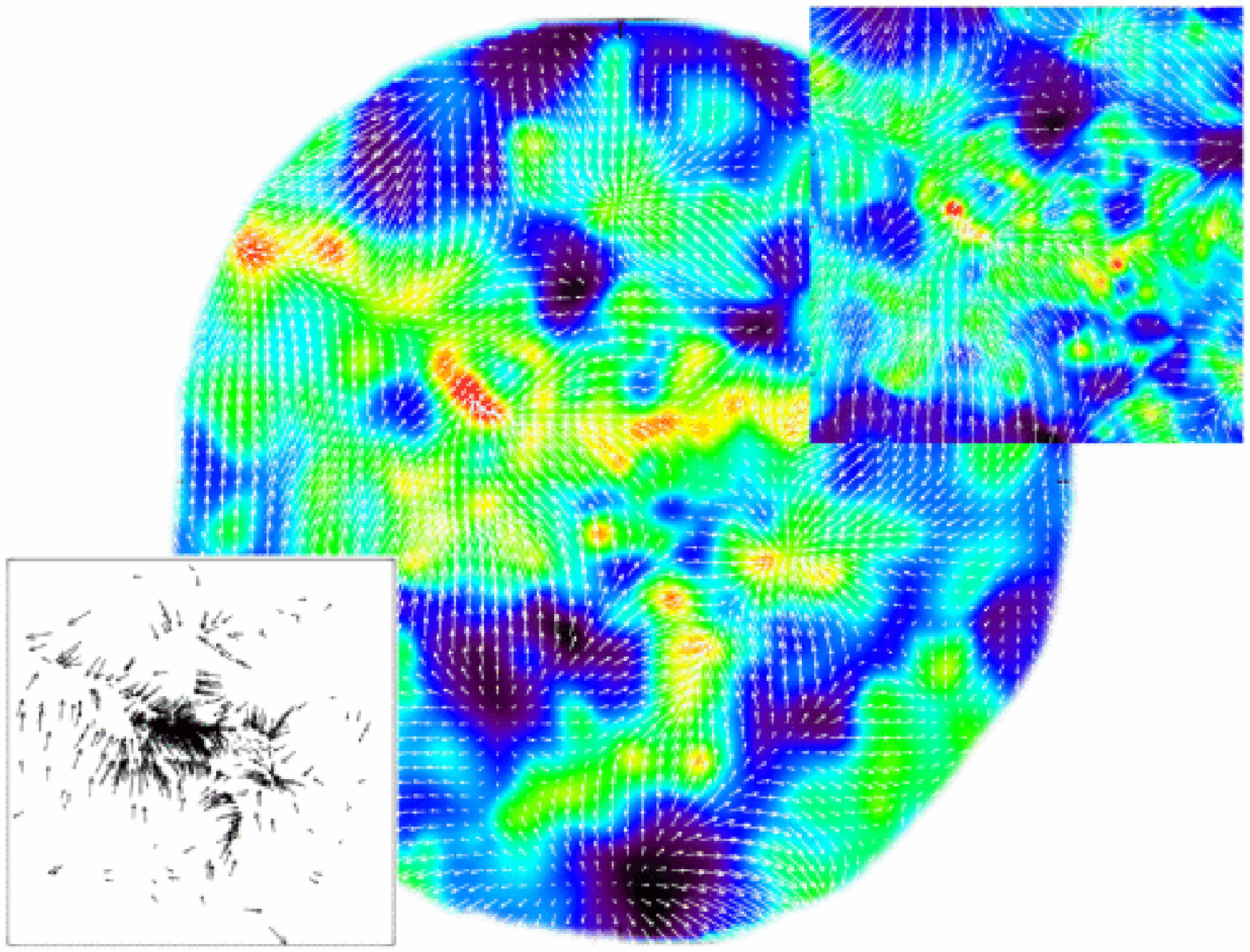}}
    \vskip -0.0truecm
\end{figure}
\begin{figure}
    \caption{Density and velocity field in the local Universe determined by DTFE on the basis 
    of the PSCz galaxy redshift survey. Our Local Group is at the centre of the map. To the left 
    we see the Great Attractor region extending out towards the Shapley supercluster. To the lefthand 
    side we can find the Pisces-Perseus supercluster. The peculiar velocities of the galaxies in the PSCz galaxy 
    redshift catalogue were determined by means of a linearization procedure (Branchini et al. 1999). 
    The resulting galaxy positions and velocities (vectors) of the input sample for the DTFE 
    reconstruction are shown in the bottom lefthand frame.  The density values range from 
    $\sim 4.9$ (red) down to $\sim -0.75$ (darkblue), with cyan coloured regions having 
    a density near the global cosmic average ($\delta \sim 0$). The velocity 
    vectors are scaled such that a vector with a length of $\approx 1/33rd$ of the 
    region's diameter corresponds to $650$ km/s. The density and velocity field have an 
    effective Gaussian smoothing radius of $R_G\sim\sqrt{5} \hmpc$. The top righthand insert 
    zooms in on the Local Supercluster and Great Attractor complex. From: Romano-D\'{\i}az 
    2004.} 
\label{fig:psczvel} 
\end{figure} 
The central circular field of fig.~\ref{fig:psczvel} presents the DTFE velocity field in 
the Supergalactic Plane. For comparison, the bottom lefthand insert shows the discrete sample 
of galaxy velocities which formed the input for the reconstruction. The velocity field is shown 
by means of the projected velocity vectors within the Z-supergalactic plane, 
superposed upon the corresponding DTFE density contourmaps inferred from the same PSCz galaxy sample.
The length of the velocity arrows can be inferred from the arrow in the lower lefthand corner, which corresponds 
to a velocity of $650$ km/s. The Local Group is located at the origin of the map. 

The map of fig.~\ref{fig:psczvel} shows the success of DTFE in converting 
a sample of discretely sampled galaxy velocities and locations into a sensible 
volume-covering flow and density field. The processed DTFE velocity field reveals intricate details along the whole 
volume. The first impression is that of the meticulously detailed DTFE flow field, marked by sharply defined flow 
regions over the whole supergalactic plane. Large scale bulk flows, distorted flow patterns such as shear, expansion 
and contraction modes of the velocity field are clear features revealed through by the DTFE technique. DTFE recovers 
clearly outlined patches marked by strong bulk flows, regions with characteristic shear flow patterns around 
anisotropically shaped supercluster complexes, radial inflow towards a few massive clusters and, perhaps most 
outstanding, strong radial outflows from the underdense void regions. 

Overall, there is a tight correspondence with the large scale structures in the 
underlying density distribution. While the density field shows features down to a scale of 
$\sqrt{5} \hmpc$, the patterns in the flow field clearly have a substantially larger 
coherence scale, nearly all in excess of $10 \hmpc$. The DTFE velocity flow sharply 
follows the elongated ridge of the Pisces-Perseus supercluster. In addition we 
find the DTFE velocity field to contain markedly sharp transition regions between 
void expansion and the flows along the body of a supercluster.

The local nature of the DTFE interpolation guarantees a highly 
resolved flow field in high density regions. Massive bulk motions are concentrated 
near and around the massive structure extending from the Local Supercluster (center map) 
towards the Great Attractor region and the Shapley concentration. The DTFE map nicely 
renders this pronounced bulk flow towards the Hydra-Centaurus region and shows 
that it dominates the general motions at our Local Group and Local Supercluster. 
The top righthand insert zooms in on the flow in and around this region. The most massive 
and coherent bulk flows in the supergalactic plane appear to be connected 
to the Sculptor void and the connected void regions (towards the lefthand side of the 
figure). They are the manifestation of the combination of gravitational attraction by the 
heavy matter concentration of the Pavo-Indus-Telescopium complex (at [SGX,SGY]$\approx[-40,-10] \hmpc$), 
the more distant ``Hydra-Centaurus-Shapley ridge'', and the effective push by the Sculptor void region. 
Conspicuous shear flows can be recognized along the ridge defined by the Cetus wall towards 
the Pisces-Perseus supercluster ([SGX,SGY]$\approx[20,-40] \hmpc$. A similar strong shear 
flow is seen along the extension of the Hydra-Centaurus supercluster towards the Shapley 
concentration. 

The influence of the Coma cluster is beautifully outlined by the strong and 
near perfect radial infall of the surrounding matter, visible at the top-centre of 
figure~\ref{fig:psczvel}. Also the velocity field near the Perseus cluster, 
in the Pisces-Perseus supercluster region, does contain a strong radial inflow 
component. 

Perhaps most outstanding are the radial outflow patterns in and around voids. 
In particular its ability to interpolate over the low-density and thus sparsely sampled 
regions is striking: the voids show up as regions marked by a near-spherical outflow. The 
intrinsic suppression of shot noise effects through the adaptive spatial interpolation procedure 
of DTFE highlights these important components of the Megaparsec flow field and emphasizes the dynamical 
role of voids in organizing the matter distribution of the large scale 
Cosmic Web.  By contrast, more conventional schemes, such as TSC or SPH \citep[see][]{schaapwey2007a}, 
meet substantial problems in defining a sensible field reconstruction in low density 
regions without excessive smoothing and thus loss of resolution. 

\section{DTFE meets reality: 2dFGRS and the Cosmic Web}
To finish the expos\'e on the potential of the Delaunay Tessellation Field Estimator, 
we present a reconstruction of the foamy morphology of the galaxy distribution in the 
2dF Galaxy Redshift Survey (2dFGRS). DTFE was used to reconstruct the projected galaxy surface 
density field as well as the full three-dimensional galaxy density field. 

\begin{figure*}
\begin{center}
\vskip -0.0truecm
\mbox{\hskip -0.8truecm\includegraphics[width=13.0cm,height=19.0cm]{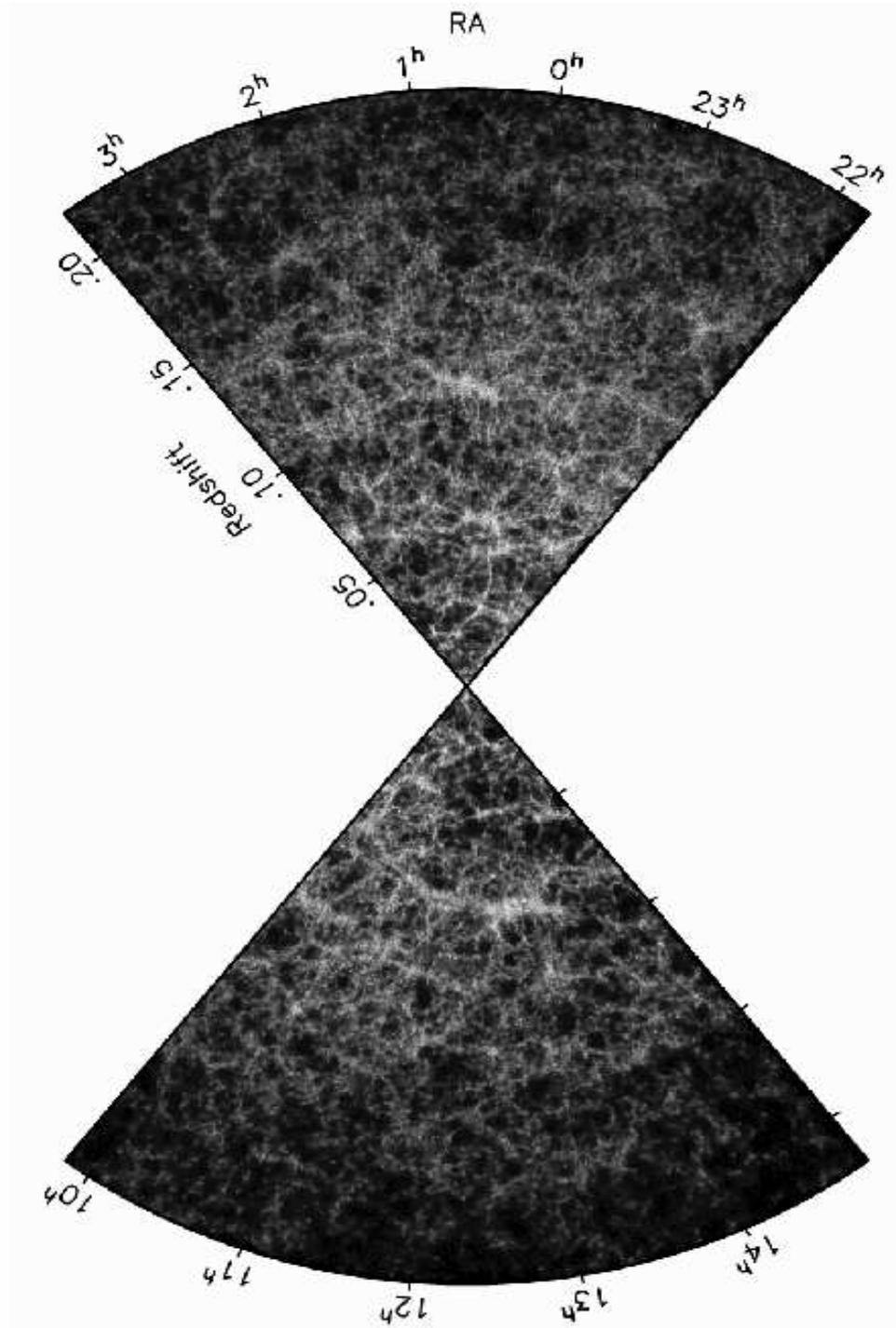}}
\vskip 0.0truecm
\caption{DTFE galaxy surface density reconstructions of the projected galaxy distribution 
in the two 2dF slices (north and south). For comparison see the galaxy distribution in fig.~\ref{fig:2dfgaldist}. 
A description may be found in the text (sect.~\ref{sec:2dfsurface}). From Schaap 2007 and 
Schaap \& van de Weygaert 2007d.}
\label{fig:2dfsurfdens}
\end{center}
\end{figure*}

\subsection{the 2dF Galaxy Redshift Survey}
The 2dFGRS is one of the major spectroscopic surveys in which the
spectra of $245 \, 591$ objects have been obtained, with the scope of
obtaining a representative picture of the large scale distribution of
galaxies in the nearby universe \citep{colless2003}. It is a
magnitude-limited survey, with galaxies selected down to a limiting
magnitude of \mbox{$b_J\sim19.45$} from the extended APM Galaxy Survey
\citep{maddox1990a,maddox1990b,maddox1990c}. The galaxy redshifts of galaxies were measured 
with the 2dF multifibre spectrograph on the Anglo-Australian Telescope, 
capable of simultaneously observing 400 objects in a $2^{\circ}$ diameter field. 
The galaxies were confined to three regions, together covering an area of approximately 
$1500$ squared degrees. These regions include two declination strips, each consisting of
overlapping $2^{\circ}$ fields, as well as a number of `randomly'
distributed $2^{\circ}$ control fields. One strip (the SGP strip) is
located in the southern Galactic hemisphere and covers about
\mbox{$80^{\circ}\times15^{\circ}$} close to the South Galactic Pole
(\mbox{$21^h40^m<\alpha<03^h40^m$}, \mbox{$-37.5^{\circ}<\delta<-22.5^{\circ}$}). 
The other strip (the NGP strip) is located in the northern Galactic hemisphere and 
covers about \mbox{$75^{\circ}\times10^{\circ}$}(\mbox{$09^h50^m<\alpha<14^h50^m$},
\mbox{$-7.5^{\circ}<\delta<+2.5^{\circ}$)}.  Reliable redshifts were obtained for 
$221\,414$ galaxies. 

These data have been made publically available in the form of the 2dFGRS final data release
(available at {\tt http://msowww.anu.edu.au/2dFGRS/}).

\begin{figure*}
\begin{center}
\vskip 0.75truecm
\mbox{\hskip -0.3truecm\includegraphics[width=13.0cm]{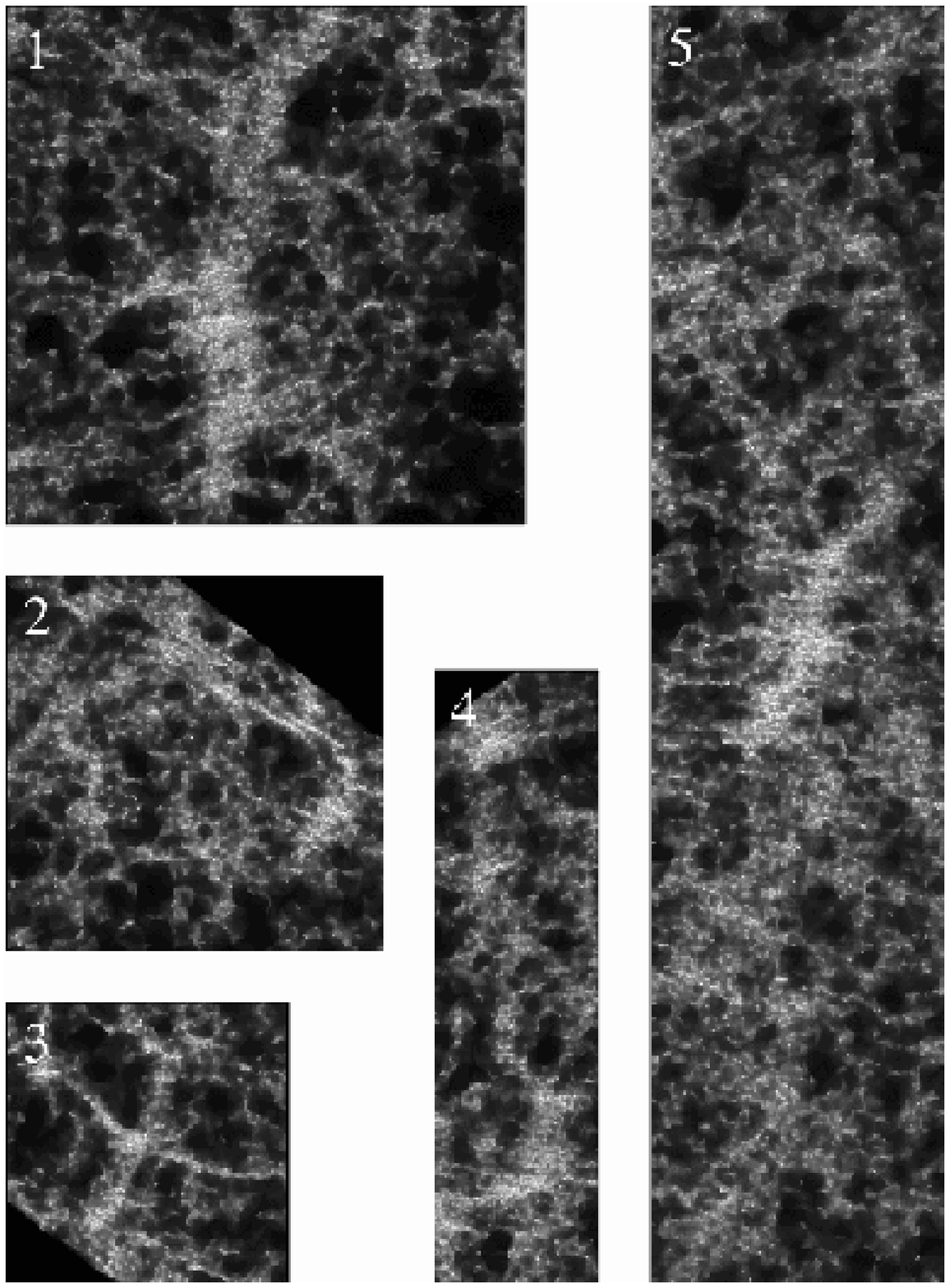}}
\vskip 0.0truecm
\end{center}
\end{figure*}
\begin{figure*}
\caption{DTFE galaxy surface density in selected regions in the 2dFGRS galaxy surface
density field. For the density field in the total 2dFGRS region see fig.~\ref{fig:2dfsurfdens}. 
For a discussion see sect.~\ref{sec:2dfindividual}. Frame 1 zooms in on the 
Great Wall in the southern (SGP) 2dF slice, frame 5 on the one in the northern (NGP) slice. 
Note the sharply rendered ``fingers of God'' marking the sites of clusters of 
galaxies. Perhaps the most salient feature is the one seen in frame 3, the cross-like 
configuration in the lower half of the NGP region. From Schaap 2007 and 
Schaap \& van de Weygaert 2007d.}
\label{fig:2dfdetails}
\end{figure*}

\subsection{Galaxy surface density field reconstructions}
\label{sec:2dfsurface}
The galaxy distribution in the 2dFGRS is mainly confined to the two 
large strips, NGP and SGP. Since their width is reasonably thin, a good impression 
of the spatial patterns in the galaxy distribution may be obtained 
from the 2-D projection shown in fig.~\ref{fig:2dfgaldist}. 

We have reconstructed the galaxy surface density field in redshift
space in the 2dFGRS NGP and SGP regions. 
All density field reconstructions are DTFE reconstructions on the basis 
of the measured galaxy redshift space positions. As no corrections 
were attempted to translate these into genuine positions in 
physical space, the density reconstructions in this section 
concern redshift space. In order to warrant a direct comparison 
with the galaxy distribution in fig.~\ref{fig:2dfgaldist} the results 
shown in this section were not corrected for any observational selection 
effect, also not for the survey radial selection function. For 
selection corrected density field reconstructions we refer to the 
analysis in \cite{schaapphd2007,schaapwey2007c}.

Fig.~\ref{fig:2dfsurfdens} shows the resulting DTFE reconstructed density field. 
DTFE manages to reveal the strong density contrasts in the large scale density 
distribution. The resolution is optimal in that the smallest interpolation 
units are also the smallest units set by the data.  At the same time the
DTFE manages to bring out the fine structural detail of the intricate
and often tenuous filamentary structures. Particularly noteworthy are the 
thin sharp edges surrounding voidlike regions. 

\begin{figure*}
\begin{center}
\vskip -0.0truecm
\mbox{\hskip -0.5truecm\includegraphics[width=13.0cm]{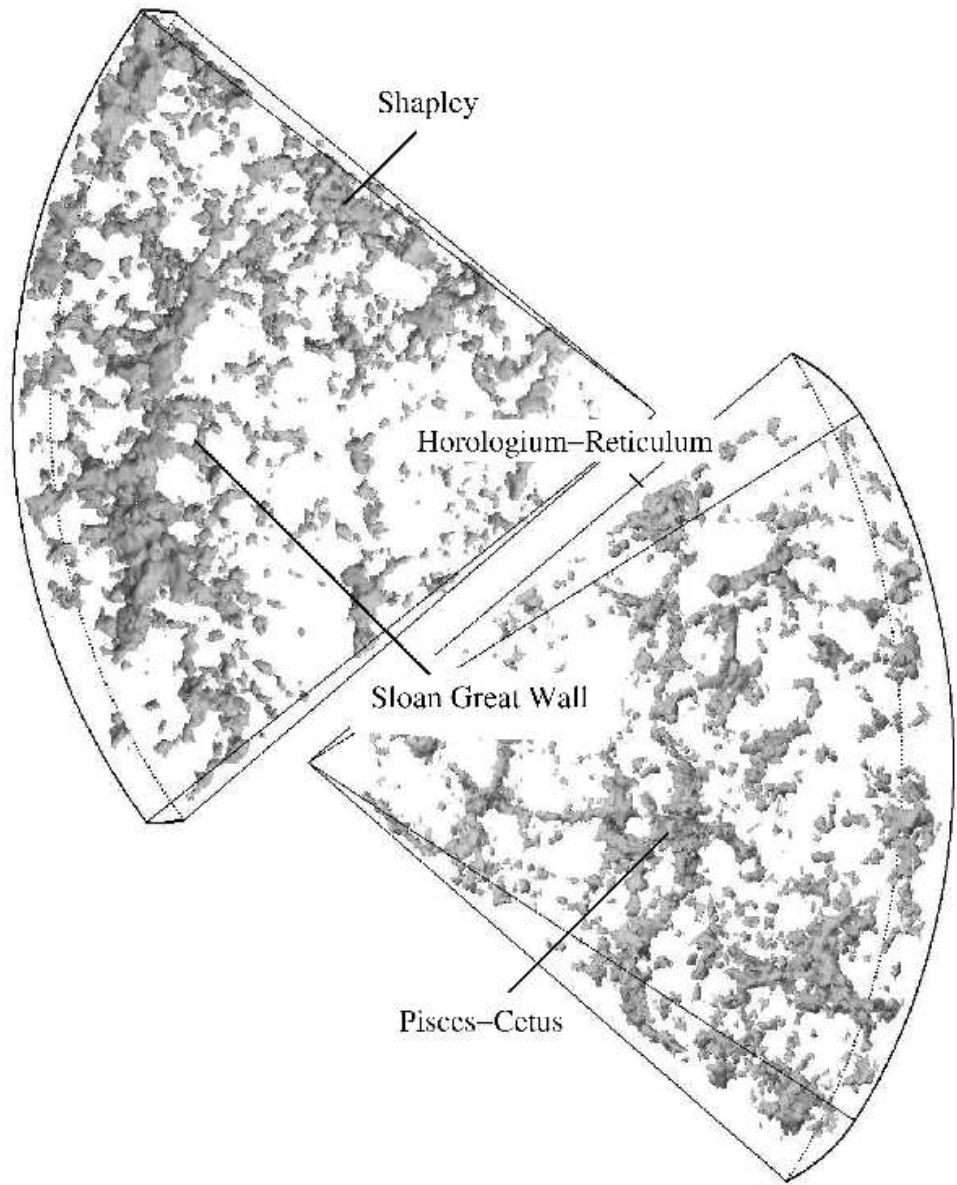}}
\vskip 0.0truecm
\caption{Isodensity surface of the galaxy distribution in the north (top) and 
south region (bottom) of the 2dFGRs. The densities are determined by means of the 
DTFE technology, subsequently Gaussian smoothed on a scale of $2\hmpc$. Several well-known 
structures are indicated. From Schaap 2007 and Schaap \& van de Weygaert 2007d.}
\label{fig:2dfdtfe3d}
\end{center}
\end{figure*}

\subsubsection{Individual Structures}
\label{sec:2dfindividual}
The impressive resolution and shape sensitivity of the DTFE reconstruction 
becomes particularly visible when zooming in on structural details in 
the Cosmic Web. Figure~\ref{fig:2dfdetails} zooms in on a few interesting 
regions in the map of fig.~\ref{fig:2dfgaldist}. Region~1 focuses on the major 
mass concentration in the NGP region, the Sloan Great Wall. 
Various filamentary regions emanate from the high density core. In region~2 
a void-like region is depicted. The DTFE renders the void as a low density area 
surrounded by various filamentary and wall-like features. Two fingers of God are 
visible in the upper right-hand corner of region~2, which show up as very sharply
defined linear features. Many such features can be recognized in high density environments. 
Note that the void is not a totally empty or featureless region. The void is marked 
by substructure and contains a number of smaller subvoids, reminding of the 
hierarchical evolution of voids \citep{dubinski1993,shethwey2004}. Region~3 
is amongst the most conspicuous structures encountered in the 2dF survey. The 
cross-shaped feature consists of four tenuous filamentary structures emanating from 
a high density core located at the center of the region. Region~4 zooms in on some of 
the larger structures in the SGP region. Part of the Pisces-Cetus supercluster is 
visible near the bottom of the frame, while the concentration at the top of this region 
is the upper part of the Horologium-Reticulum supercluster. Finally, region~5 
zooms in on the largest structure in the SGP region, the Sculptor
supercluster. 

\subsubsection{DTFE artefacts}
Even though the DTFE offers a sharp image of the cosmic web, the reconstructions 
also highlight some artefacts. At the highest resolution we can directly discern 
the triangular imprint of the DTFE kernel. Also a considerable amount of noise is 
visible in the reconstructions. This is a direct consequence of the high resolution 
of the DTFE reconstruction. Part of this noise is natural, a result of the statistical 
nature of the galaxy formation process. An additional source of noise is due to the fact 
that the galaxy positions have been projected onto a two-dimensional plane. Because DTFE 
connects galaxies which lie closely together in the projection, it may involve objects which 
in reality are quite distant. A full three-dimensional reconstruction followed 
by projection or smoothing on a sufficiently large scale would alleviate the problem. 

\subsection{Three-dimensional structure 2dFGRS}
We have also reconstructed the full three-dimensional galaxy density
field in the NGP and SGP regions of the 2dFGRS. The result is shown
in Fig.~\ref{fig:2dfdtfe3d}. It shows the three-dimensional rendering 
of the NGP (left) and SGP slices (right) out to a redshift $z=0.1$. 
The maps demonstrate that DTFE is able of recovering the three-dimensional 
structure of the cosmic web as well as its individual elements. Although less 
obvious than for the two-dimensional reconstructions, the
effective resolution of the three-dimensional reconstructions is also
varying across the map. 

The NGP region is dominated by the large supercluster at a redshift of about 0.08, 
the Sloan Great Wall. The structure near the upper edge at a redshift of 0.05 to 0.06 
is part of the upper edge of the Shapley supercluster. In the SGP region several
known superclusters can be recognized as well. The supercluser in the center
of this region is part of the Pisces-Cetus supercluster. The huge
concentration at a redshift of about 0.07 is the upper part of the enormous 
Horologium-Reticulum supercluster.

\section{Extensions, Applications and Prospects}
In this study we have demonstrated that DTFE density and velocity fields are optimal in the sense of defining a continuous and unbiased 
representation of the data while retaining all information available in the point sample.

In the present review we have emphasized the prospects for the analysis of weblike structures or, more general, any 
discretely sampled complex spatial pattern. In the meantime, DTFE has been applied in a number of studies of cosmic structure formation. 
These studies do indeed suggest a major improvement over the more conventional analysis tools. Evidently, even though density/intensity field 
analysis is one of the primary objectives of the DTFE formalism one of its important features is its ability to extend its 
Delaunay tessellation based spatial interpolation to any corresponding spatially sampled physical quantity. The true potential of DTFE 
and related adaptive random tessellation based techniques will become more apparent as further applications and extensions will come to 
the fore. The prospects for major developments based on the use of tessellations are tremendous. As yet we may identify a diversity of 
astrophysical and cosmological applications which will benefit substantially from the availability of adaptive random tessellation 
related methods. A variety of recent techniques have recognized the high dynamic range and adaptivity of tessellations to the spatial 
and morphological resolution of the systems they seek to analyze.

The first major application of DTFE concerns its potential towards uncovering morphological and statistical characteristics 
of spatial patterns. A second major class of applications is that of issues concerning the dynamics of many particle systems. 
They may prove to be highly useful for the analysis of phase space structure and evolution of gravitational 
systems in their ability to efficiently trace density concentrations and singularities in higher dimensional space. 
Along similar lines a highly promising avenue is that of the application of similar formalisms within the context 
of computational hydrodynamics. The application of Delaunay tessellations as a fully self-adaptive grid may finally 
open a Lagrangian grid treatment of the hydrodynamical equations describing complex multiscale systems such as 
encountered in cosmology and galaxy formation. Various attempts along these lines have already been followed and 
with the arrival of efficient adaptive Delaunay tessellation calculations they may finally enable a practically 
feasible implementation. A third and highly innovative application, also using both the {\it adaptive} and {\it random} 
nature of Delaunay tessellations, is their use as Monte Carlo solvers for systems of complex equations describing complex 
physical systems. Particularly interesting has been the recent work by \cite{ritzericke2006}. They managed to exploit the 
random and adaptive nature of Delaunay tessellations as a stochastic grid for Monte Carlo lattice treatment of 
radiative transfer in the case of multiple sources. In the following we will describe a few of these applications 
in some detail. 

\subsection{Gravitational Dynamics}
Extrapolating the observation that DTFE is able to simultaneously handle spatial density and velocity 
field \citep[e.g.][]{bernwey96,emiliophd2004,weyrom2007}, and encouraged by the success of Voronoi-based methods 
in identifying dark halos in N-body simulations \citep{neyrinck2005}, \cite{arad2004} used DTFE to assess the six-dimensional 
phase space density distribution of dark halos in cosmological $N$-body simulations. While a fully six-dimensional analysis may be 
computationally cumbersome \citep{ascalbin2005}, and not warranted because of the symplectic character of phase-space, 
the splitting of the problem into a separate spatial and velocity-space three-dimensional tessellation may indeed hold 
promise for an innovative analysis of the dynamical evolution of dark halos. 

\subsubsection{Gravitational Lensing and Singularity Detection}
A related promising avenue seeks to use the ability of DTFE to trace sharp density contrasts. This impelled \cite{bradac2004} 
to apply DTFE to gravitational lensing. They computed the surface density map for a galaxy from the projection of the DTFE 
volume density field and used the obtained surface density map to compute the gravitational lensing pattern around the object. 
Recently, \cite{li2006} evaluated the method and demonstrated that it is indeed a promising tool for tracing 
higher-order singularities. 

\subsection{Computational Hydrodynamics}
Ultimately, the ideal fluid dynamics code would combine the advantages of the Eulerian as well as of the Lagrangian approach. 
In their simplest formulation, Eulerian algorithms cover the volume of study with a fixed grid and compute the fluid transfer 
through the faces of the (fixed) grid cell volumes to follow the evolution of the system. Lagrangian formulations, on the other hand, 
compute the system by following the ever changing volume and shape of a particular individual element of gas\footnote{Interestingly, 
the {\it Lagrangian} formulation is also due to Euler \citep{eul1862} who employed this formalism in a letter to Lagrange, who later 
proposed these ideas in a publication by himself \citep{lag1762}}. Their emphasis on mass resolution makes Lagrangian codes usually 
better equipped to follow the system into the highest density regions, at the price of a decreased resolution in regions of a lower 
density. 

As we may have appreciated from the adaptive resolution and morphological properties of Delaunay tessellations in DTFE and 
the more general class of {\it Natural Neighbour} interpolations the recruitment of Delaunay tessellations may define 
an appropriate combination of Eulerian and Lagrangian description of fluid dynamical systems. 

\subsubsection{Particle Hydrodynamics}
A well-known strategy in computational hydrodynamics is to follow the Lagrangian description by discretizing the 
fluid system in terms of a (large) number of fluid particles which encapsulate, trace and sample the relevant 
dynamical and thermodynamical characteristics of the entire fluid \citep[see][for a recent review]{koumoutsakos2005}. Smooth 
Particle Hydrodynamics (SPH) codes \citep{lucy1977,gingold1977,monaghan1992} have found widespread use in many areas of science 
and have arguably become the most prominent computational tool in cosmology \citep[e.g.][]{katz1996,springel2001,springel2005}. SPH particles 
should be seen as discrete balls of fluid, whose shape, extent and properties is specified according to user-defined 
criteria deemed appropriate for the physical system at hand. Notwithstanding substantial virtues of SPH -- amongst which 
one should include its ability to deal with systems of high dynamic range, its adaptive spatial resolution, and its 
flexibility and conceptual simplicity -- it also involves various disadvantages. 

One straightforward downside concerns its comparatively bad performance in low-density regions. 
In a highly topical cosmological issue such as that of the {\it reionization of the Universe} upon the formation 
of the first galaxies and stars it may therefore not be able to appropriately resolve the void patches 
which may be relatively important for the transport of radiation. Even in high density regions it may 
encounter problems in tracing singularities. As a result of its user-defined finite size it may not 
succeed in a sufficiently accurate outlining of shocks. The role of the user-dependent and artificial 
particle kernel has also become apparent when comparing the performance of SPH versus the more natural 
DTFE kernels (see fig.~\ref{fig:eta4panel}, fig.~\ref{fig:dtfescaling} and sect.~\ref{sec:dtfescaling}). 

In summary: SPH involves at least four user-defined aspects which affect its performance,
\begin{enumerate}
\item{} SPH needs an arbitrary user-specified kernel function W. 
\item{} SPH needs a smoothing length h;\\ 
\ \ \ even though the standard choice is a length adapting to the particle density it does imply a finite extent. 
\item{} SPH kernel needs a shape; a spherical shape is usually presumed.
\item{} SPH needs an (unphysical) artificial viscosity to stabilize solutions 
and to be able to capture shocks.
\end{enumerate}

Given the evidently superior density and morphology tracing abilities of DTFE and related methods based upon 
adaptive grids \cite{pelu2003} investigated the question what the effect would be of replacing a regular 
SPH kernel by an equivalent DTFE based kernel. In an application to a TreeSPH simulation of the (neutral) ISM 
they managed to identify various density and morphological environments where the natural adaptivity of 
DTFE proves to yield substantially better results. They concluded with various suggestions for the formulation of  
an alternative particle based adaptive hydrodynamics code in which the kernel would be defined by DTFE. 

A closely related and considerably advanced development concerns the formulation of a complete and self-consistent  
particle hydrodynamics code by Espa{\~nol}, Coveney and collaborators \citep{espanol1998,flekkoy2000,serrano2001,
serrano2002,fabritiis2003}. Their {\it (Multiscale) Dissipative Particle Hydrodynamics} code is entirely formulated on the basis of 
{\it Voronoi fluid particles}. Working out the Lagrangian equations for a fluid system they demonstrate 
that the subsequent compartimentalization of the system into discrete thermodynamic systems - {\it fluid 
particles} - leads to a set of equations whose solution would benefit best if they are taken to be 
defined by the Voronoi cells of the corresponding Voronoi tessellation. In other words, the geometrical 
features of the Voronoi model is directly connected to the physics of the system by interpretation of the 
Voronoi volumes as coarse-grained ``fluid clusters''. Not only does their formalism 
capture the basic physical symmetries and conservation laws and reproduce the continuum for (sampled) 
smooth fields, but it also suggests a striking similarity with turbulence. Their 
formalism has been invoked in the modeling {\it mesoscale} systems, simulating the molecular 
dynamics of complex fluid and soft condense matter systems which are usually are marked fluctuations 
and Brownian motion. While the absence of long-range forces such as gravity simplifies the description 
to some extent, the {\it Voronoi particle} descriptions does provide enough incentive for looking into 
possible implementations within an astrophysical context. 

\subsubsection{Adaptive Grid Hydrodynamics}
For a substantial part the success of the DTFE may be ascribed to the use of Delaunay tessellations as 
optimally covering grid. This implies them to also be ideal for the use in {\it moving \& adaptive grid} implementations 
of computational hydrodynamics. 

\begin{figure*}
\begin{center}
\mbox{\hskip -0.1truecm\includegraphics[width=11.7cm]{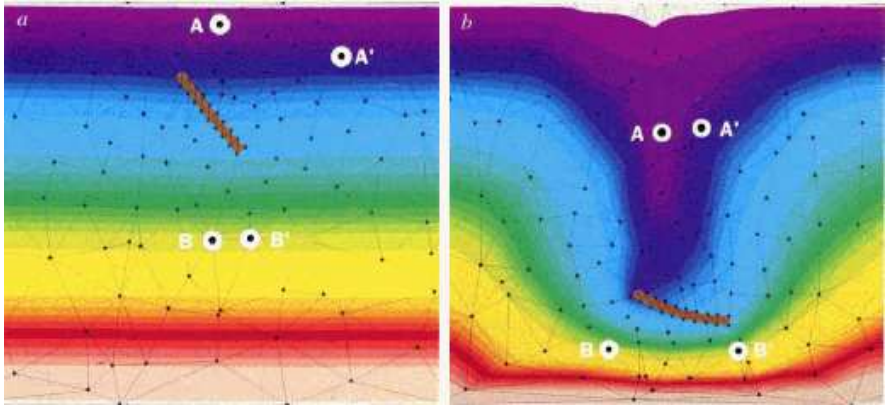}}
\vskip 0.0truecm
\caption{Application of the Natural Neighbour scheme solution of partial differential equations on highly irregular 
evolving Delaunay grids, described by Braun \& Sambridge 1995. It involves the natural-element method (NEM) solution 
of the Stokes flow problem in which motion is driven by the fall of an elasto-plastic plate denser than the viscous fluid. 
The problem is solved by means of a natural neighbour scheme, the Delaunay grid is used as unstructured computation mesh. The Stokes 
equation is solved at the integration points in the linear fluid, the equations of force balance at the integration points 
located within the plate. The solution is shown at two timesteps (1 and 10000 of the integration). Image courtesy 
of M. Sambridge also see Braun \& Sambridge 1995.}
\label{fig:nnsbmhydro}
\end{center}
\end{figure*}

Unstructured grid methods originally emerged as a viable alternative to structured or block-structured grid techniques 
for discretization of the Euler and Navier-Stokes equations for structures with complex geometries 
\citep[see][for an extensive review]{mavripilis1997}. The use of unstructured grids provides greater flexibility for 
discretizing systems with a complex shape and allows the adaptive update of mesh points and connectivities in order 
to increase the resolution. A notable and early example of the use of unstructured grids may be found General Relativity 
in the formulation of Regge calculus \citep{regge1961}. Its applications includes quantum gravity formulations on the 
basis of Voronoi and Delaunay grids \citep[e.g.][]{wmiller1997}.

One class of unstructured grids is based upon the use of specific geometrical elements (triangles in 2-D, tetrahedra in 3-D) to 
cover in a non-regular fashion a structure whose evolution one seeks to compute. It has become a prominent technique in a large 
variety of technological applications, with those used in the design of aerodynamically optimally shaped cars and the design 
of aeroplanes are perhaps the most visible ones. The definition and design of optimally covering and suitable meshes is a 
major industry in computational science, including issues involved with the definition of optimal Delaunay meshes 
\citep[see e.g][]{amenta1998,amenta1999,shewchuk2002,chengdey2004,dey2004,alliez2005a,alliez2005b,boissonnat2006}. A second 
class of unstructured grids is based upon the use of structural elements of a mixed type with irregular connectivity. The 
latter class allows the definition of a self-adaptive grid which follows the evolution of a physical system. It is in this 
context that one may propose the use of Voronoi and Delaunay tessellations as adaptive hydrodynamics lattices. 

Hydrocodes with Delaunay tessellations at their core warrant a close connection to the underlying matter 
distribution and would represent an ideal compromise between an Eulerian and Lagrangian description, 
guaranteeing an optimal resolution and dynamic range while taking care of an appropriate coverage of 
low-density regions. Within astrophysical context there have been a few notable attempts to develop {\it moving grid} 
codes. Although these have shown their potential \citep{gnedin1995,pen1998}, their complexity and the implicit 
complications raised by the dominant presence of the long-range force of gravity have as yet prevented their 
wide-range use. 

It is here that Delaunay tessellations may prove to offer a highly promising alternative. The advantages of a moving 
grid fluid dynamics code based on Delaunay tessellations has been explicitly addressed in a detailed and 
enticing study by \cite{whitehurst1995}. His two-dimensional FLAME Lagrangian hydrocode used a first-order 
solver and proven to be far superior to all tested first-order and many second-order Eulerian codes. The adaptive 
nature of the Langrangian method and the absence of preferred directions in the grid proved to be key 
factors for the performance of the method. \cite{whitehurst1995} illustrated this with impressive examples 
such as the capturing of shock and collision of perfectly spherical shells. The related higher-order 
Natural Neighbour hydrocodes used by \cite{braunsambridge1995}, for a range of geophysical problems, 
and \cite{sukumarphd1998} are perhaps the most convincing examples and applications of Delaunay grid 
based hydrocodes. The advantages of Delaunay grids in principle apply to any algorithm invoking them, in particular 
also for three-dimensional implementations (of which we are currently unaware). 

\begin{figure*}
\begin{center}
\mbox{\hskip -0.5truecm\includegraphics[width=12.3cm]{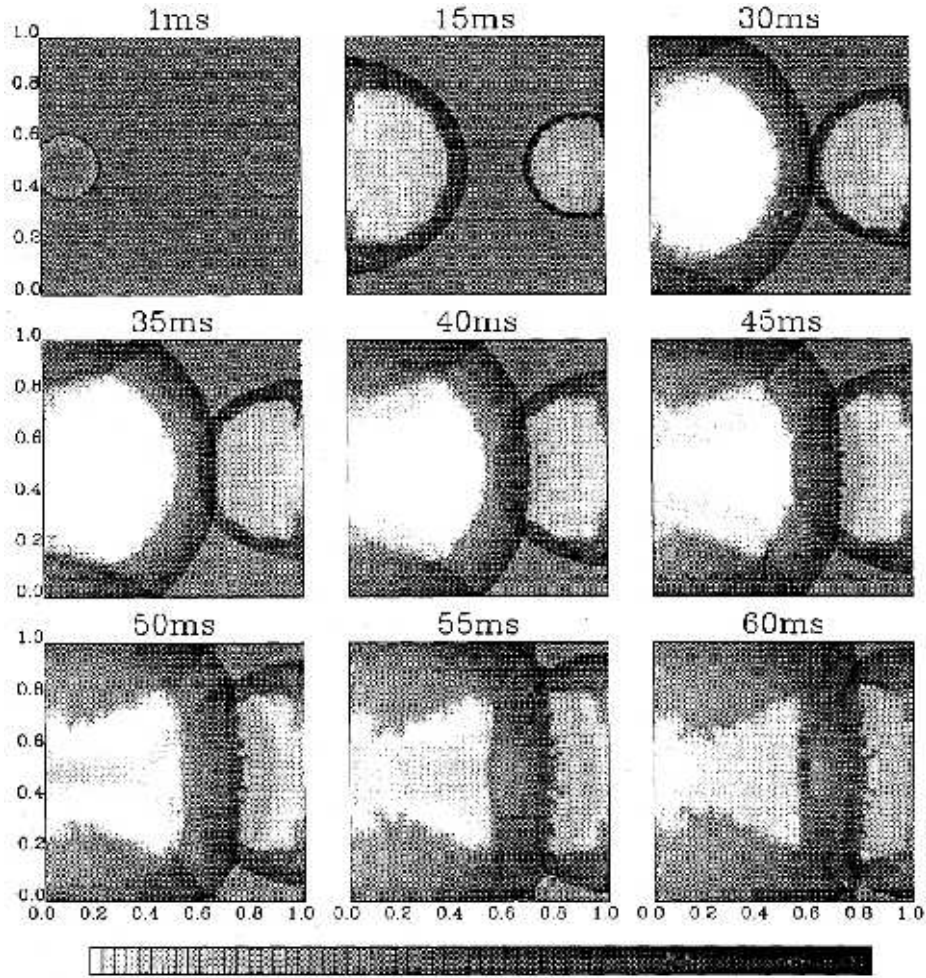}}
\vskip 0.0truecm
\caption{Nine non-linear gray-scale images of the density evolution of the 2D interacting blast waves test of the 
FLAME Delaunay grid hydrocode of Whitehurst (1995). The circular shape of the shockfronts is well represented; the 
contact fronts are clearly visible. Of interest is the symmetry of the results, which is not enforced and so is 
genuine, and the instability of the decelerated contact fronts. From Whitehurst 1995.}
\label{fig:blastwhitehurst}
\end{center}
\vskip -0.5truecm
\end{figure*}

\subsubsection{Kinetic and Dynamic Delaunay Grids}
If anything, the major impediment towards the use of Delaunay grids in evolving (hydro)dynamical systems is the 
high computational expense for computing a Delaunay grid. In the conventional situation one needs to 
completely upgrade a Delaunay lattice at each timestep. It renders any Delaunay-based code prohibitively 
expensive. 

In recent years considerable attention has been devoted towards developing kinetic and dynamic Delaunay 
codes which manage to limit the update of a grid to the few sample points that have moved so far that 
the identity of their Natural Neighbours has changed. The work by Meyer-Hermann and collaborators \citep{schaller2004,beyer2005,
beyerhermann2006} may prove to represent a watershed in having managed to define a parallel code for the kinetic and dynamic 
calculation of Delaunay grids. Once the Delaunay grid of the initial point configuration has been computed, 
subsequent timesteps involve a continuous dynamical upgrade via local Delaunay simplex upgrades as points move 
around and switch the identity of their natural neighbours. 

The code is an elaboration on the more basic step of inserting one point into a Delaunay grid and 
evaluating which Delaunay simplices need to be upgraded. Construction of Delaunay triangulations 
via incremental insertion was introduced by \cite{fortune1992} and \cite{edelsbrunner1996}. Only sample 
points which experience a change in identity of their natural neighbours need an update of their Delaunay 
simplices, based upon {\it vertex flips}. The update of sample points that have retained the 
same natural neighbours is restricted to a mere recalculation of the position and shape of 
their Delaunay cells.

\subsection{Random Lattice Solvers and \texttt{SimpleX}}
The use of Delaunay tessellations as adaptive and dynamic meshes for evaluating the 
fluid dynamical equations of an evolving physical system emphasizes their adaptive nature. 
Perhaps even more enticing is the use of the random nature of the these meshes. Taking 
into account their spatial adaptivity as well as their intrinsic stochastic nature, 
they provide an excellent template for following physical systems along the lines 
of Monte Carlo methods. 

\begin{figure*}
\begin{center}
\mbox{\hskip -0.1truecm\includegraphics[width=11.9cm]{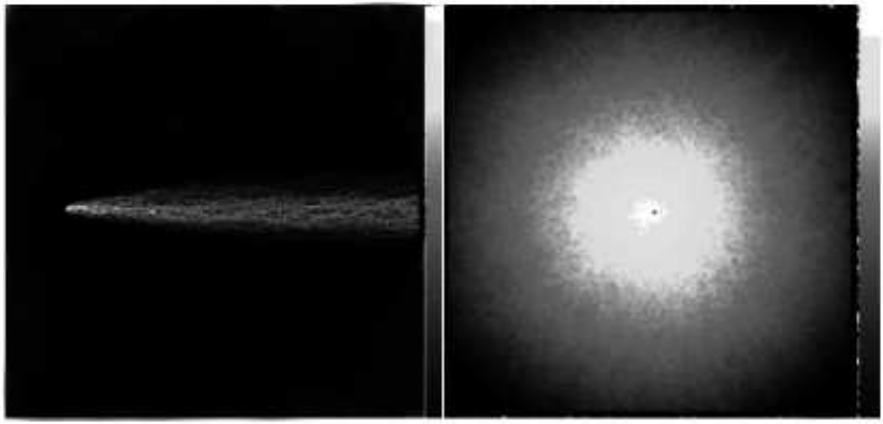}}
\vskip 0.0truecm
\caption{The result of two simple \texttt{SimpleX} radiative transfer tests on a 2D Poisson-Delaunay random 
lattice with $N=5\times10^4$ points. Both are logarithmic plots of the number of particles at 
each site. Left: illustration of the conservation of momentum by means of the transport of particles 
with constant momentum vectors. Right: Illustration of a scattering transport process. Image 
courtesy of J. Ritzerveld, also see Ritzerveld 2007.}
\label{fig:simplex1}
\end{center}
\end{figure*}

Generally speaking, Monte Carlo methods determine the characteristics of many body systems as 
statistical averages over a large number of particle histories, which are computed with the 
use of random sampling techniques. They are particularly useful for transport problems 
since the transport of particles through a medium can be described stochastically as a random 
walk from one interaction event to the next. The first to develop a computer-oriented Monte Carlo 
method for transport problems were \cite{metropolis1949}. Such transport problems may in general 
be formulated in terms of a stochastic Master equation which may evaluated by means 
of Monte Carlo methods by simulating large numbers of different particle trajectories or 
histories \citep[see][for a nice discussion]{ritzerveldphd2007}. 

\begin{figure*}
\begin{center}
\mbox{\hskip -0.1truecm\includegraphics[width=11.9cm]{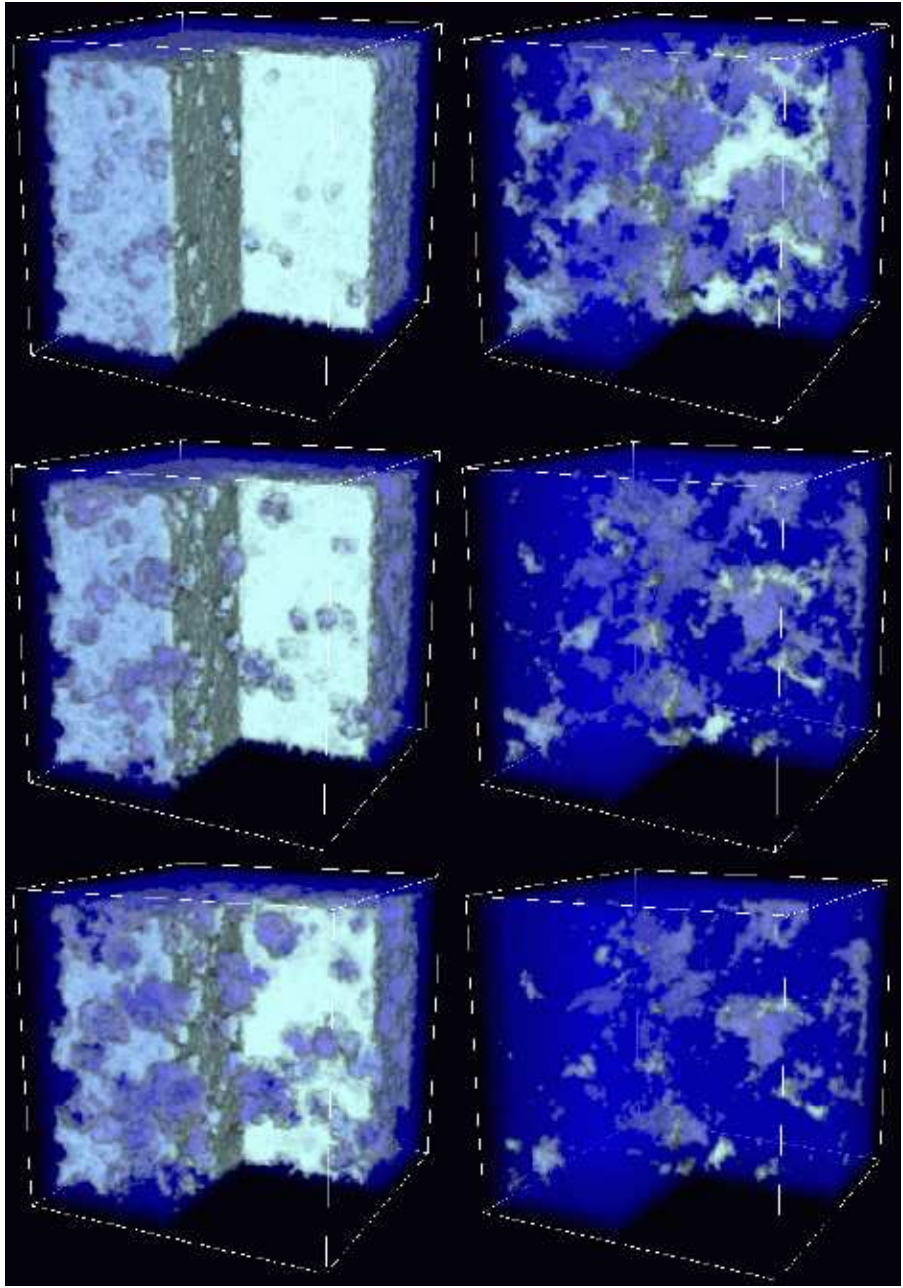}}
\vskip 0.0truecm
\end{center}
\caption{A volume rendering of the result of using the \texttt{SimpleX} method to transport ionizing photons 
through the intergalactic medium at and around the epoch of reionization. The \texttt{SimpleX} method was 
applied to a PMFAST simulation of the large scale structure in a $\Lambda$CDM universe. The results of 6 different 
timesteps are plotted, in which the white corresponds to hydrogen to the hydrogen that is still neutral 
(opaque), the blue to the already ionized hydrogen (transparent). Image courtesy of J. Ritzerveld, also see Ritzerveld 2007.}
\label{fig:simplexreion}
\end{figure*}

While the asymmetry of regular meshes and lattices for Monte Carlo calculations introduces 
various undesirable effects, random lattices may alleviate or solve the problematic lack of 
symmetry. Lattice Boltzmann studies were amongst the first to recognize this 
\citep[see e.g.][]{ubertini2005}. In a set of innovative publications \cite{christ1982a,
christ1982b,christ1982c} applied random lattices, including Voronoi and Delaunay tessellations, 
to solve (QCD) field theoretical equations. 

Recently, a highly innovative and interesting application to (astrophysical) radiative 
transfer problems followed the same philosophy as the Random Lattice Gauge theories. 
\cite{ritzericke2006} and \cite{ritzerveldphd2007} translated the problem of radiative transfer 
into one which would be solved on an appropriately defined (Poisson)-Delaunay grid. It leads to
the formulation of the \texttt{SimpleX} radiation transfer technique which translates the transport 
of radiation through a medium into the {\it deterministic} displacement of photons from one 
vertex to another vertex of the {\it stochastic} Delaunay mesh.
The perfect random and isotropic nature of the Delaunay grid assures an unbiased Monte Carlo sampling of the 
photon paths. The specification of appropriate absorption and scattering coefficients at the nodes 
completes the proper Monte Carlo solution of the Master equation for radiative transfer. 

One of the crucial aspects of \texttt{SimpleX} is the sampling of the underlying density field 
$\rho({\bf x})$ such that the Delaunay grid edge lengths $L_{\rm D}$ represent a proper random 
sample of the (free) paths $\lambda_{\gamma}$ of photons through a scattering and absorbing medium. 
In other words, \texttt{SimpleX} is built on the assumption that the (local) mean Delaunay edge 
length $\langle L_{\rm D}\rangle$ should be proportional to the (local) mean free path of the photon. 
For a medium of density $\rho({\bf x})$ in d-dimensional space (d=2 or d=3), with absorption/scattering 
coefficient $\sigma$, sampled by a Delaunay grid generated by a point sample with local number 
density $n_{\rm X}(\bf x)$, 
\begin{eqnarray}
\langle L_{\rm D} \rangle \,\propto\, \lambda_{\gamma}\ \ \ \Longleftarrow\ \ \ 
\begin{cases}
\langle L_{\rm D} \rangle\,=\,\zeta(d)\, n_{\rm X}^{-1/d}\\
\ \\
\lambda_{\gamma}\,=\,{\displaystyle 1 \over \displaystyle \rho({\bf x})\,\sigma}
\end{cases}
\end{eqnarray}
where $\zeta(d)$ is a dimension dependent constant. By sampling of the underlying density 
field $\rho({\bf x})$ by a point density 
\begin{eqnarray}
n_{\rm X}({\bf x})\,\propto \rho({\bf x})^d 
\end{eqnarray} 
\texttt{SimpleX} produces a Delaunay grid whose edges are guaranteed to form a representative stochastic 
sample of the photon paths through the medium. 

To illustrate the operation of \texttt{SimpleX} Fig.~\ref{fig:simplex1} shows the outcome of a 
two-dimensional \texttt{SimpleX} test calculations. One involves a test of the ability of 
\texttt{Simplex} to follow a beam of radiation, the second one its ability to follow the 
spherical spread of radiation emitted by an isotropically emitting source. The figure nicely 
illustrates the success of \texttt{SimpleX}, meanwhile providing an impression of the ebffect 
and its erratic action at the scale of Delaunay grid cells.

The \texttt{SimpleX} method has been applied to the challenging problem of the reionization of the 
intergalactic medium by the first galaxies and stars. A major problem for most radiative transfer 
techniques is to deal with multiple sources of radiation. While astrophysical problems may often be a
pproximated by a description in terms of a single source, a proper evaluation of reionization 
should take into account the role of a multitude of sources. \texttt{SimpleX} proved its ability to 
yield sensible answers in a series of test comparisons between different radiative transfer 
codes applied to aspects typical for reionization \citep{illiev2006}. The first results of the 
application of \texttt{SimpleX} to genuine cosmological circumstances, by coupling it to a cosmological 
SPH code, do yield highly encouraging results \citep{ritzerveld2007b}. Fig.~\ref{fig:simplexreion} 
is a beautiful illustration of the outcome of one of the reionization simulations.  

If anything, \texttt{SimpleX} proves that the use of random Delaunay grids have the potential of 
representing a genuine breakthrough for addressing a number of highly complex 
astrophysical problems. 

\subsection{Spatial Statistics and Pattern Analysis}
Within its cosmological context DTFE will meet its real potential in more sophisticated applications tuned towards uncovering 
morphological characteristics of the reconstructed spatial patterns. 

A potentially interesting application would be its implementation in the SURFGEN machinery. SURFGEN seeks to provide locally defined 
topological measures, cq. local Minkowski functionals, of the density field \citep{sahni1998,jsheth2003,shandarin2004}. A recent 
new sophisticated technique for tracing the cosmic web is the {\it skeleton} formalism developed by \cite{novikov2006}, based upon 
the framework of Morse theory \citep{colombi2000}. The {\it skeleton } formalism seeks to identify filaments in the web by identifying 
ridges along which the gradient of the density field is extremal along its isocontours \citep[also see][] {sousbie2006}. The use 
of unbiased weblike density fields traced by DTFE would undoubtedly sharpen the ability of the skeleton formalism to 
trace intricate weblike features over a wider dynamical range. 

Pursuing an alternative yet related track concerns the direct use of tessellations themselves in outlining topological properties of 
a complex spatial distribution of points. {\it Alpha-shapes} of a discrete point distribution are subsets of Delaunay triangulations which 
are sensitive tracers of its topology and may be exploited towards inferring Minkowski functionals and Betti numbers \citep{edelsbrunner1983,edelsbrunner1994,
edelsbrunner2002}. A recent and ongoing study seeks their application to the description of the 
Cosmic Web \citep{vegter2007} (also see sec~\ref{sec:alphashape}). 

Two major extensions of DTFE already set out to the identification of key aspects of the cosmic web. Specifically focussed on the 
hierarchical nature of the Megaparsec matter distribution is the detection of weblike anisotropic features over a range of spatial 
scales by the Multiscale Morphology Filter (MMF), introduced and defined by \cite{aragonmmf2007}. The {\it Watershed Void Finding} (WVF) 
algorithm of \cite{platen2007} is a void detection algorithm meant to outline the hierarchical nature of the cosmic void population. 
We will shortly touch upon these extensions in order to illustrate the potential of DTFE. 

\subsection{Alpha-shapes}
\label{sec:alphashape}
{\it Alpha shape} are a description of the (intuitive) notion of the shape of a discrete point set. {\it Alpha-shapes} of a discrete 
point distribution are subsets of a Delaunay triangulation and were introduced by Edelsbrunner and collaborators \citep{edelsbrunner1983,
mueckephd1993,edelsbrunner1994,edelsbrunner2002}. Alpha shapes are generalizations of the convex hull of a point set and are concrete geometric objects which are uniquely defined for a particular point set. Reflecting the topological structure of a point distribution, it 
is one of the most essential concepts in the field of Computational Topology \citep{dey1998,vegter2004,zomorodian2005}. 

\begin{figure*}
\begin{center}
\mbox{\hskip -0.1truecm\includegraphics[width=12.2cm]{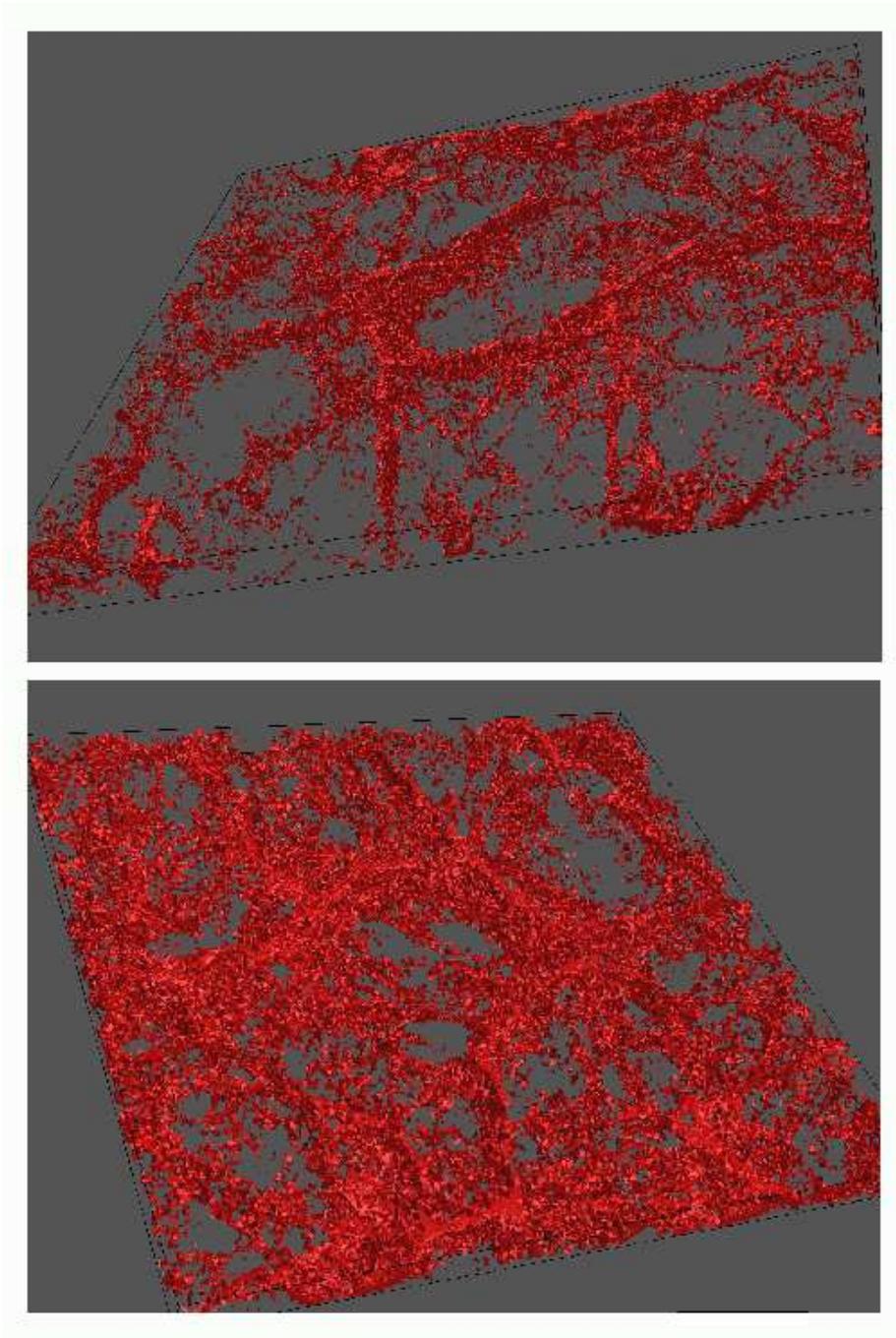}}
\vskip 0.0truecm
\caption{Examples of {\it alpha shapes} of the LCDM GIF simulation. Shown are central slices through 
two alpha shapes (top: low alpha; bottom: high alpha). The image shows the sensitivity of alpha shapes 
to the topology of the matter distribution. From: Vegter et al. 2007. Courtesy: Bob Eldering.}
\label{fig:gifalpha}
\end{center}
\end{figure*}

If we have a point set $S$ and its corresponding Delaunay triangulation, we may identify all {\it Delaunay simplices} 
-- tetrahedra, triangles, edges, vertices -- of the triangulation. For a given non-negative value of $\alpha$, the 
{\it alpha complex} of a point sets consists of all simplices in the Delaunay triangulation which have 
an empty circumsphere with squared radius less than or equal to $\alpha$, $R^2\leq \alpha$. Here ``empty'' means that the 
open sphere does not include any points of $S$. For an extreme value $\alpha=0$ the alpha complex merely consists of 
the vertices of the point set. The set also defines a maximum value $\alpha_{\rm max}$, such that for $\alpha \geq 
\alpha_{\rm max}$ the alpha shape is the convex hull of the point set. 

The {\it alpha shape} is the union of all simplices of the alpha complex. Note that it implies that although the alpha shape 
is defined for all $0\leq \alpha < \infty$ there are only a finited number of different alpha shapes for any one point set. 
The alpha shape is a polytope in a fairly general sense, it can be concave and even disconnected. Its components can be 
three-dimensional patches of tetrahedra, two-dimensional ones of triangles, one-dimensional strings of edges and 
even single points. The set of all real numbers $\alpha$ leads to a family of shapes capturing the intuitive notion of 
the overall versus fine shape of a point set. Starting from the convex hull of a point set and gradually decreasing $\alpha$ the shape 
of the point set gradually shrinks and starts to develop cavities. These cavities may join to form tunnels and 
voids. For sufficiently small $\alpha$ the alpha shape is empty. 

Following this description, one may find that alpha shapes are intimately related to the topology of a point set. 
As a result they form a direct and unique way of characterizing the topology of a point distribution. A complete 
quantitative description of the topology is that in terms of Betti numbers $\beta_{\rm k}$ and these may indeed 
be directly inferred from the alpha shape. The first Betti number $\beta_0$ specifies the number of 
independent components of an object. In ${\mathbb R}$ $\beta_1$ may be interpreted as the number of independent 
tunnels, and $\beta_2$ as the number of independent enclose voids. The $k^{th}$ Betti number effectively counts 
the number of independent $k$-dimensional holes in the simplicial complex. 

Applications of alpha shapes have as yet focussed on biological systems. Their use in characterizing the topology 
and structure of macromolecules. The work by Liang and collaborators \citep{edelsbrunner1998,liang1998a,liang1998b,liang1998c} 
uses alpha shapes and betti numbers to assess the voids and pockets in an effort to classify complex protein structures, a highly challenging 
task given the 10,000-30,000 protein families involving 1,000-4,000 complicated folds. Given the interest in the topology 
of the cosmic mass distribution \citep[e.g.][]{gott1986,mecke1994,schmalzing1999}, it is evident that {\it alpha shapes} 
also provide a highly interesting tool for studying the topology of the galaxy distribution and N-body simulations of cosmic 
structure formation. Directly connected to the topology of the point distribution itself it would discard the need of 
user-defined filter kernels. 

In a recent study \cite{vegter2007} computed the alpha shapes for a set of GIF simulations of cosmic structure 
formation (see fig.~\ref{fig:gifalpha}). On the basis of a calibration of the inferred Minkowski functionals and 
Betti numbers from a range of Voronoi clustering models their study intends to refine the knowledge of 
the topological structure of the Cosmic Web. 

\subsection{the Multiscale Morphology Filter}
\label{sec:mmf}
The multiscale detection technique -- MMF -- is used for characterizing the different morphological 
elements of the large scale matter distribution in the Universe \cite{aragonmmf2007}. The method is ideally suited 
for extracting catalogues of clusters, walls, filaments and voids from samples of galaxies in redshift surveys or particles 
in cosmological N-body simulations. 

The multiscale filter technology is particularly oriented towards recognizing and identifying features in a 
density or intensity field on the basis of an assessment of their coherence along a range of spatial scales and with the 
virtue of providing a generic framework for characterizing the local morphology of the density field and enabling the 
selection of those morphological features which the analysis at hand seeks to study. The Multiscale Morphology Filter (MMF) 
method has been developed on the basis of visualization and feature extraction techniques in computer vision and medical research 
\citep{florack1992}. The technology, finding its origin in computer vision research, has been optimized within the context of feature 
detections in medical imaging. \cite{frangi1998} and \cite{sato1998} presented its operation for the specific situation 
of detecting the web of blood vessels in a medical image. This defines a notoriously complex pattern of elongated tenuous features 
whose branching make it closely resemble a fractal network. 

The density or intensity field of the sample is smoothed over a range of multiple scales. Subsequently, this signal is processed through 
a morphology response filter. Its form is dictated by the particular morphological feature it seeks to extract, and depends on the local 
shape and spatial coherence of the intensity field. The morphology signal at each location is then defined to be the one with the maximum 
response across the full range of smoothing scales.  The MMF translates, extends and optimizes this technology towards the recognition 
of the major characteristic structural elements in the Megaparsec matter distribution. It yields a unique framework for the combined 
identification of dense, compact bloblike clusters, of the salient and moderately dense elongated filaments and of tenuous planar walls. 
Figure~\ref{fig:mmf} includes a few of the stages involved in the MMF procedure.

Crucial for the ability of the method to identify anisotropic features such as filaments and walls is the use of a morphologically 
unbiased and optimized continuous density field retaining all features visible in a discrete galaxy or particle distribution. 
Accordingly, DTFE is used to process the particle distribution into a continuous density field (top centre frame, fig.~\ref{fig:mmf}). 
The morphological intentions of the MMF method renders DTFE a key element for translating the particle or galaxy distribution into a 
representative continuous density field $f_{\tiny{\textrm{DTFE}}}$. 

\begin{table}
\label{tab:morphmask}
\centering
\begin{large}
\begin{tabular} {|p{2.65cm}|p{3.75cm}|p{5.0cm}|}
\hline
\hline
\ &&\\
\hskip 0.5truecm Structure & \hskip 1.0truecm $\lambda$ ratios & \hskip 1.2truecm $\lambda$ constraints \\
\  &&\\
\hline
\ &&\\
\hskip 0.5truecm Blob \hskip 0.5truecm &  \hskip 0.6truecm $\lambda_1 \simeq \lambda_2 \simeq \lambda_3$ \hskip 0.5truecm & \hskip 0.4truecm $\lambda_3 <0\,\,;\,\, \lambda_2 <0 \,\,;\,\, \lambda_1 <0 $ \hskip 0.5truecm  \\
\hskip 0.5truecm Line  \hskip 0.5truecm    &  \hskip 0.6truecm $\lambda_1 \simeq \lambda_2 \gg    \lambda_3$ \hskip 0.5truecm & \hskip 0.4truecm $\lambda_3 <0 \,\,;\,\, \lambda_2 <0  $  \\
\hskip 0.5truecm Sheet \hskip 0.5truecm &  \hskip 0.6truecm $\lambda_1 \gg\lambda_2 \simeq \lambda_3$ \hskip 0.5truecm & \hskip 0.4truecm $\lambda_3 <0 $    \\
\ &&\\
\hline
\hline
\end{tabular}
\end{large}
\vskip 0.25truecm
\caption{Behaviour of the eigenvalues for the characteristic morphologies. The
         lambda conditions describe objects with intensity higher that their
		 background (as clusters, filaments and walls). For voids we must reverse the
	     sign of the eigenvalues.  From the constraints imposed by the $\lambda$ 
         conditions we can describe the blob morphology  as a subset of the line 
         which is itself a subset of the wall.}
\end{table}

In the implementation of \cite{aragonmmf2007} the scaled representations of the data are obtained by repeatedly smoothing the 
DTFE reconstructed density field $f_{\tiny{\textrm{DTFE}}}$ with a hierarchy of spherically symmetric Gaussian filters 
$W_{\rm G}$ having different widths $R$:
\begin{equation}
f_{\rm S}({\vec x}) =\, \int\,{\rm d}{\vec y}\,f_{\tiny{\textrm{DTFE}}}({\vec y})\,W_{\rm G}({\vec y},{\vec x})\nonumber
\end{equation}
where $W_{\rm G}$ denotes a Gaussian filter of width $R$. A pass of the Gaussian smoothing filter attenuates structure on 
scales smaller than the filter width. The Scale Space itself is constructed by stacking these variously smoothed data sets, 
yielding the family $\Phi$ of smoothed density maps $f_n$:
\begin{equation}
\label{eq:scalespace}
\Phi\,=\,\bigcup_{levels \; n} f_n 
\end{equation}
In essence the {\it Scale Space} structure of the field is the $(D+1)$ dimensional space defined by the $D$-dimensional density 
or intensity field smoothed over a continuum of filter scales $R_G$. As a result a data point can be viewed at any of the scales 
where scaled data has been generated.  

\begin{figure*}
\begin{center}
\vskip 0.5truecm
\mbox{\hskip -0.7truecm\includegraphics[height=14.0cm,angle=90.0]{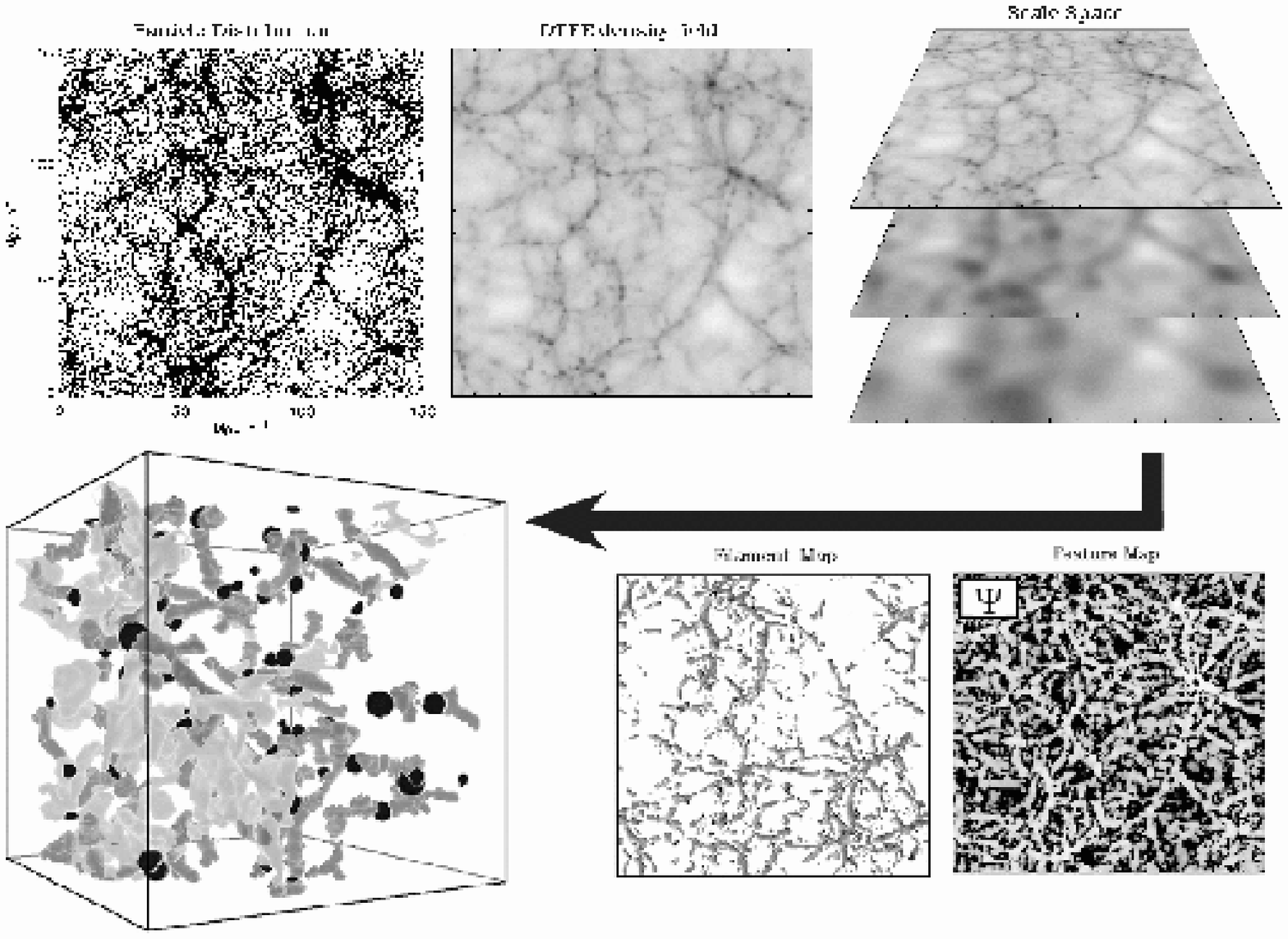}}
\vskip 0.0truecm
\end{center}
\end{figure*}
\begin{figure*}
\caption{Schematic overview of the Multiscale Morphology Filter (MMF) to isolate and 
extract elongated filaments (dark grey), sheetlike walls (light grey) and clusters (black 
dots) in the weblike pattern of a cosmological N-body simulation. The first stage is the 
translation of a discrete particle distribution (top lefthand frame) into a DTFE density 
field (top centre). The DTFE field is filtered over a range of scales (top righthand stack 
of filtered fields). By means of morphology filter operations defined on the basis of the 
Hessian of the filtered density fields the MMF successively selects the regions which 
have a bloblike (cluster) morphology, a filamentary morphology and a planar morphology, 
at the scale at which the morphological signal is optimal. This produces a feature map 
(bottom lefthand). By means of a percolation criterion the physically significant 
filaments are selected (bottom centre). Following a sequence of blob, filament and wall 
filtering finally produces a map of the different morphological features in the particle 
distribution (bottom lefthand). The 3-D isodensity contours in the bottom lefthand 
frame depict the most pronounced features. From Arag\'on-Calvo et al. 2007.}
\label{fig:mmf}
\end{figure*}

The crux of the concept is that the neighbourhood of a given point will look different at each scale. While there are potentially 
many ways of making a comparison of the scale dependence of local environment, \cite{aragonmmf2007} chose to calculate the 
Hessian matrix in each of the smoothed replicas of the original data to describe the the local "shape" of the density field 
in the neighbourhood of that point. In terms of the Hessian, the local variations around a point $\vec{x}_0$ of the density 
field $f(\vec{x})$ may be written as the Taylor expansion
\begin{equation}\label{eq:taylor_exp_1}
f(\vec{x}_0 + \vec{s})\,=\,f(\vec{x}_0)\,+\,\vec{s}^T \nabla f(\vec{x}_0)\,+\,
\frac{1}{2}\vec{s}^T \mathcal{H} (\vec{x}_0) \vec{s} + ...
\end{equation} 
where 
\begin{equation}\label{eq:hessian_1} 
  \mathcal{H}\,=\,\left ( \begin{array}{ccc}
                f_{xx} & f_{yx} & f_{zx} \\
                f_{xy} & f_{yy} & f_{zy} \\
                f_{xz} & f_{yz} & f_{zz}
		\end{array} \right )
\end{equation} 
is the Hessian matrix. Subscripts denote the partial derivatives of $f$ with respect to the named variable.  There are many 
possible algorithms for evaluating these derivatives. In practice, the Scale Space procedure evaluates the Hessian directly 
for a discrete set of filter scales by smoothing the DTFE density field by means of Mexican Hat filter, 
\begin{eqnarray}
\frac{\partial^2}{\partial x_i \partial x_j} f_S({\vec x})&\,=\,&f_{\tiny{\textrm{DTFE}}}\,\otimes\,\frac{\partial^2}{\partial x_i \partial x_j} W_{\rm G}(R_{\rm S})\nonumber \\
&\,=\,&\int\,{\rm d}{\vec y}\,f({\vec y})\,\,\frac{(x_i-y_i)(x_j-y_j)-\delta_{ij}R_{\rm S}^2}{R_{\rm S}^4}\,W_{\rm G}({\vec y},{\vec x})
\end{eqnarray} 
with ${x_1,x_2,x_3}={x,y,z}$ and $\delta_{ij}$ the Kronecker delta. In other words, the scale space representation of the Hessian matrix for each level $n$ is evaluated by means of a convolution with the second derivatives of the Gaussian filter, also known as the {\it Marr or ''Mexican Hat''} Wavelet. 

The eigenvalues of the Hessian matrix at a point encapsulate the information on the local shape of the field. Eigenvalues are denoted 
as $\lambda_{a}(\vec{x})$ and arranged so that 
$ \lambda_1 \ge \lambda_2 \ge \lambda_3 $:
\begin{eqnarray}
\qquad \bigg\vert \; \frac{\partial^2 f_n({\vec x})}{\partial x_i \partial x_j}  - \lambda_a({\vec x})\; \delta_{ij} \; \bigg\vert  
&=& 0,  \quad a = 1,2,3 \\
\mathrm{with} \quad \lambda_1 &>& \lambda_2  >  \lambda_3 \nonumber
\end{eqnarray}
The $\lambda_{i}(\vec{x})$ are coordinate independent descriptors of the behaviour of the density field in the locality of the point 
$\vec{x}$ and can be combined to create a variety of morphological indicators. They quantify the rate of change of the field gradient in 
various directions about each point. A small eigenvalue indicates a low rate of change of the field values in the corresponding 
eigen-direction, and vice versa. The corresponding eigenvectors show the local orientation of the morphology characteristics.   

Evaluating the eigenvalues and eigenvectors for the renormalised Hessian $\tilde {\mathcal{H}}$ of each dataset in a Scale Space shows 
how the local morphology changes with scale. First regions in scale space are selected according to the appropriate morphology 
filter, identifying bloblike, filamentary and wall-like features at a range of scales. The selections are made according 
to the eigenvalue criteria listed in table~\ref{tab:morphmask}. Subsequently, a sophisticated machinery of filters 
and masks is applied to assure the suppression of noise and the identification of the locally relevant scales for the 
various morphologies.
 
Finally, for the cosmological or astrophysical purpose at hand the identified 
spatial patches are tested by means of an {\it erosion} threshold criterion. Identified blobs should have a 
critical overdensity corresponding to virialization, while identified filaments should fulfil a percolation requirement 
(bottom central frame). By successive repetition for the identification of blobs, filaments and sheets -- each with their 
own morphology filter -- the MMF has dissected the cosmological density field into the corresponding features. The box 
in the bottom lefthand frame shows a segment from a large cosmological N-body simulation: filaments are coloured 
dark grey, the walls light grey and the clusters are indicated by the black blobs.

Once these features have been marked and identified by MMF, a large variety of issues may be adressed. An important 
issue is that of environmental influences on the formation of galaxies. The MMF identification of filaments 
allowed \citep{aragonmmf2007} to show that galaxies in filaments and walls do indeed have a mass-dependent alignment. 

\subsection{the Watershed Void Finder}
The Watershed Void Finder (WVF) is an implementation of the {\it Watershed 
Transform} for segmentation of images of the galaxy and matter distribution 
into distinct regions and objects and the subsequent identification of voids 
\citep{platen2005,platen2007}. The watershed transform is a
concept defined within the context of mathematical morphology. The basic 
idea behind the watershed transform finds its origin in geophysics. It 
delineates the boundaries of the separate domains, the {\it basins}, into which yields of 
e.g. rainfall will collect. The analogy with the cosmological context is 
straightforward: {\it voids} are to be identified with the {\it basins}, 
while the {\it filaments} and {\it walls} of the cosmic web are the ridges 
separating the voids from each other.

\begin{figure*}
\begin{center}
\mbox{\hskip -0.1truecm\includegraphics[width=11.9cm]{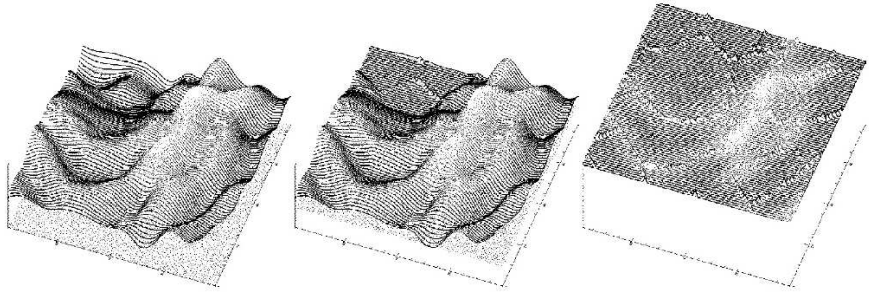}}
\vskip 0.0truecm
\caption{Three frames illustrating the principle of the watershed
       transform. The lefthand frame shows the surface to be segmented. 
       Starting from the local minima the surrounding basins of the surface 
       start to flood as the water level continues to rise (dotted plane 
       initially below the surface). Where two basins meet up near a ridge 
       of the density surface, a ``dam'' is erected (central frame). Ultimately, 
       the entire surface is flooded, leaving a network of dams 
       defines a segmented volume and delineates the corresponding 
       cosmic web (righthand frame). From: Platen, van de Weygaert \& Jones 2007.}
\label{fig:wvfcart}
\end{center}
\end{figure*}

The identification of voids in the cosmic matter distribution is hampered by 
the absence of a clearly defined criterion of what a void is. Unlike overdense 
and virialized clumps of matter voids are not genuinely defined physical 
objects. The boundary and identity of voids is therefore mostly a matter of 
definition. As a consequence there is a variety of voidfinding algorithms 
\citep{kauffair1991,elad1996,plionbas2002,hoyvog2002,shandfeld2006,colberg2005b,
patiri2006a}. A recent study \citep{colpear2007} contains a balanced comparison of 
the performance of the various void finders with respect to a small region taken 
from the Millennium simulation \cite{springmillen2005}

\begin{figure*}
\begin{center}
\mbox{\hskip -0.1truecm\includegraphics[width=11.9cm]{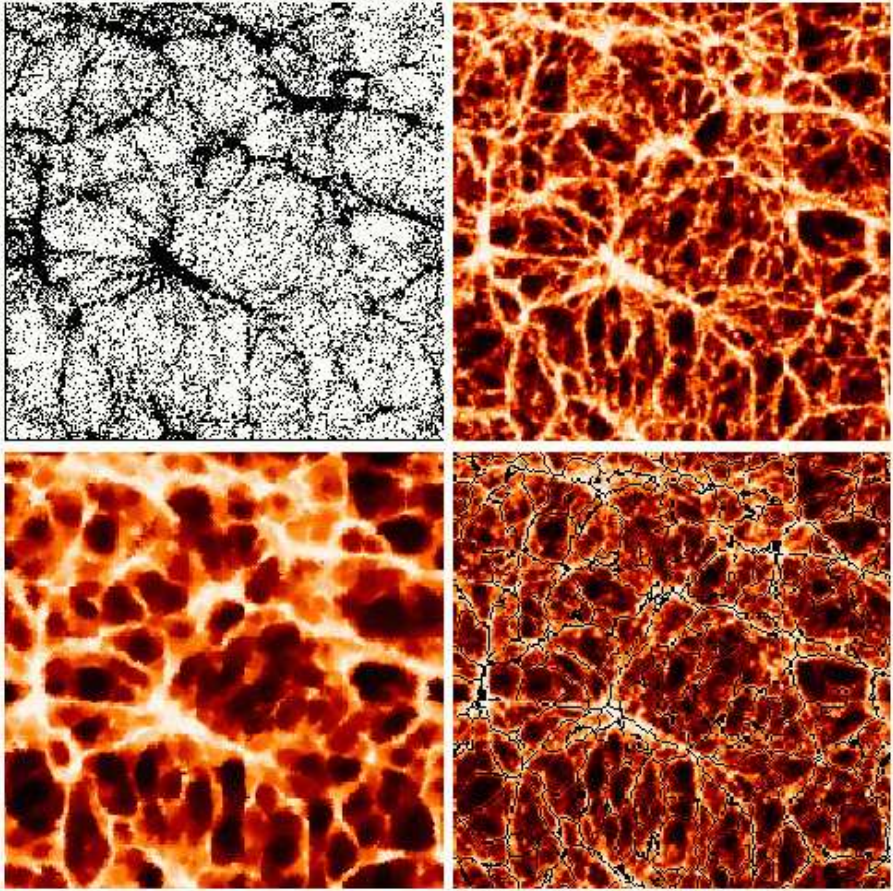}}
\vskip 0.0truecm
\caption{A visualization of several intermediate steps of the
  Watershed Void Finding (WVF) method. The top lefthand frame shows the
  particles of a slice in the LCDM GIF simulation. The corresponding 
  DTFE density field is shown in the top righthand frame. The 
  next, bottom lefthand, frame shows the resulting $n$-th order 
  median-filtered image. Finally, the bottom righthand frame shows 
  the resulting WVF segmentation computed on the basis of the median 
  filtered image. From: Platen, van de Weygaert \& Jones 2007.}
\label{fig:wvf}
\end{center}
\end{figure*}

\subsubsection{Watersheds}
\label{sec:watershed}
With respect to the other void finders the watershed algorithm seeks to define  
a natural formalism for probing the hierarchical nature of the void distribution in 
maps of the galaxy distribution and in N-body simulations of cosmic 
structure formation. The WVF has several advantages \citep[see e.g.][]{meyerbeucher1990}, 
Because it is identifies a void segment on the basis of the crests in a density field 
surrounding a density minimum it is able to trace the void boundary even though 
it has a distored and twisted shape. Also, because the contours around well chosen 
minima are by definition closed the transform is not sensitive to local protrusions between 
two adjacent voids. The main advantage of the WVF is that for an ideally smoothed 
density field it is able to find voids in an entirely parameter free fashion. 

The word {\it watershed} finds its origin in the analogy of the procedure 
with that of a landscape being flooded by a rising level of water. 
Figure~\ref{fig:wvfcart} illustrates the concept. Suppose 
we have a surface in the shape of a landscape. The surface is pierced at the location of 
each of the minima. As the waterlevel rises a growing fraction of the landscape will be flooded
by the water in the expanding basins. Ultimately basins will meet at
the ridges corresponding to saddlepoints in th e density field. These
define the boundaries of the basins, enforced by means of a
sufficiently high {\it dam}. The final result of the completely immersed landscape 
is a division of the landscape into individual cells, separated 
by the {\it ridge dams}.

\subsubsection{Formalism}
The WVF consists of eightfold crucial steps. The two essential first steps 
relate directly to DTFE. The use of DTFE is essential to infer a continuous 
density field from a given N-body particle distribution of galaxy redshift 
survey. For the success of the WVF it is of utmost importance that the density 
field retains its morphological character, ie. the hierarchical nature, the weblike 
morphology dominated by filaments and walls, and the presence of voids. In particular 
in and around low-density void regions the raw density field is characterized by 
a considerable level of noise. In an essential second step noise gets suppressed 
by an adaptive smoothing algorithm which in a consecutive sequence of repetitive 
steps determines the median of densities within the {\it contiguous Voronoi cell} 
surrounding a point. The determination of the median density of the natural 
neighbours turns out to define a stable and asymptotically converging smooth 
density field fit for a proper watershed segmentation. 

Figure~\ref{fig:wvf} is an illustration of four essential stages in the WVF procedure. 
Starting from a discrete point distribution (top left), the continuous density field is 
determined by the DTFE procedure (top right). Following the natural smoothing by {\it Natural Neighbour Median} 
filtering (bottom left), the watershed formalism is able to identify the void segments in the density field 
(bottom right). 

\section{Concluding Remarks}
This extensive presentation of tessellation-based machinery for the analysis of weblike patterns in the 
spatial matter and galaxy distribution intends to provide the interested reader with a framework to 
exploit and extend the large potential of Voronoi and Delaunay Tessellations. Even though conceptually 
not too complex, they do represent an intriguing world by themselves. Their cellular geometry paves the 
way towards a natural analysis of the intricate fabric known as Cosmic Web. 

\section{Acknowledgments}
RvdW wishes to thank Vicent Mart\'{\i}nez, Enn Saar and Maria Pons for their invitation and hospitality 
during this fine and inspiring September week in Valencia, and their almost infinite 
patience regarding my ever shifting deadlines. We owe thanks to O. Abramov, S. Borgani, M. Colless, 
A. Fairall, T. Jarrett, M. Juri\'C, R. Landsberg, O. L\'opez-Cruz, A. McEwen, M. Ouchi, J. Ritzerveld, M. Sambridge, 
V. Springel and N. Sukumar for their figures used in this manuscript. The work described in this lecture is the 
result of many years of work, involving numerous collaborations. In particular we wish 
to thank Vincent Icke, Francis Bernardeau, Dietrich Stoyan, Sungnok Chiu, Jacco Dankers, Inti Pelupessy, 
Emilio Romano-D\'{\i}az,  Jelle Ritzerveld, Miguel Arag\'on-Calvo, Erwin Platen, Sergei Shandarin, 
Gert Vegter, Niko Kruithof and  Bob Eldering for invaluable contributions and discussions, covering nearly 
all issues touched upon in this study. In particularly grateful we are to Bernard Jones, for his enthusiastic and crucial 
support and inspiration, and the many original ideas, already over two decades, for all that involved 
tessellations, DTFE, the Universe, and much more .... RvdW is especially grateful to Manolis and Menia for 
their overwhelming hospitality during the 2005 Greek Easter weeks in which a substantial fraction of this 
manuscript was conceived ... What better setting to wish for than the view from the Pentelic mountain overlooking 
the cradle of western civilization, the city of Pallas Athena. Finally, this work could not have 
been finished without the patience and support of Marlies ... 
\printindex

\end{document}